\documentclass[preprint2,tighten]{aastex62_reep}

\usepackage{epstopdf}
\usepackage{subfigure}  
\usepackage{amsmath}
\usepackage{hyperref}
\usepackage{graphbox}

\usepackage{longtable}

\usepackage{fancyhdr}   
\fancypagestyle{specialfooter}{%
 \fancyhf{}
 
 \fancyfoot[C]{Distribution Statement A.  Approved for public release.  Distribution unlimited.}  }

\shorttitle{Effect of Geometry on Hydrodynamic Response}
\shortauthors{Reep et al.}

\begin{document}

\title{Geometric Assumptions in Hydrodynamic Modeling of Coronal and Flaring Loops}

\author[0000-0003-4739-1152]{Jeffrey W. Reep}
\affiliation{Space Science Division, Naval Research Laboratory, Washington, DC 20375}

\author[0000-0001-5503-0491]{Ignacio Ugarte-Urra}
\affiliation{Space Science Division, Naval Research Laboratory, Washington, DC 20375}

\author[0000-0001-6102-6851]{Harry P. Warren}
\affiliation{Space Science Division, Naval Research Laboratory, Washington, DC 20375}

\author[0000-0001-9642-6089]{Will T. Barnes}
\affiliation{National Research Council Research Associate Residing at the Naval Research Laboratory, Washington, DC 20375}
\affiliation{NASA Goddard Space Flight Center, Heliophysics Sciences Division, Greenbelt, MD 20771}
\affiliation{Department of Physics, American University, Washington, DC 20016}

\email{jeffrey.reep@nrl.navy.mil}

\begin{abstract}
In coronal loop modeling, it is commonly assumed that the loops are semi-circular with a uniform cross-sectional area.  However, observed loops are rarely semi-circular, and extrapolations of the magnetic field show that the field strength decreases with height, implying that the cross-sectional area expands with height.  We examine these two assumptions directly to understand how they affect the hydrodynamic and radiative response of short, hot loops to strong, impulsive electron beam heating events.  Both the magnitude and rate of area expansion impact the dynamics directly, and an expanding cross-section significantly lengthens the time for a loop to cool and drain, increases upflow durations, and suppresses sound waves.  The standard $T \sim n^{2}$ relation for radiative cooling does not hold with expanding loops, which cool with relatively little draining.  An increase in the eccentricity of loops, on the other hand, only increases the draining timescale, and is a minor effect in general.  Spectral line intensities are also strongly impacted by the variation in the cross-sectional area since they depend on both the volume of the emitting region as well as the density and ionization state.  With a larger expansion, the density is reduced, so the lines at all heights are relatively reduced in intensity and, because of the increase of cooling times, the hottest lines remain bright for significantly longer.  Area expansion is critical to accurate modeling of the hydrodynamics and radiation, and observations are needed to constrain the magnitude, rate, and location of the expansion or lack thereof.
\end{abstract}


\nopagebreak

\section{Introduction}
\label{sec:intro}
\thispagestyle{specialfooter}   

Field-aligned hydrodynamic loop modeling of both quiescent regions and flares typically makes simplifying assumptions about the geometry, namely that the loops are semi-circular in shape with constant cross-section along their lengths.  These assumptions, however, have not been critically examined nor are they necessarily correct.  Observations of loops with imagers such as the Extreme Ultraviolet Imager \citep[EUVI,][]{wuelser2004} or the Atmospheric Imaging Assembly (AIA, \citealt{lemen2012}) onboard the Solar Dynamics Observatory (SDO, \citealt{pesnell2012}) very often show loops that are not circular \citep[e.g.][]{aschwanden2009}, often appearing more elliptical, which will alter the gravitational acceleration parallel to the loop.  Furthermore, since the magnetic field strength decreases from the photosphere into the corona, by the conservation of magnetic flux, we expect that the cross-sectional area of loops must expand.  This expansion of the area would affect the flows along the loop drastically, which can then impact the temperatures, densities, and therefore the entire resultant spectrum of emission from the loop.  

In \citet{mikic2013}, the effect of a non-uniform cross-sectional area was shown to affect the transport of energy by thermal conduction, which in turn impacts the occurrence of thermal non-equilibrium (TNE).  One manifestation of TNE is the occurrence of periodic coronal rain events known to occur in active regions \citep{auchere2018,pelouze2020,pelouze2022}.  Rain is also seen prominently in solar flares \citep{jing2016,scullion2016}, but does not occur in hydrodynamic simulations of impulsive heating \citep{reep2020}.  A non-uniform area is likely an important ingredient to the production of rain, but currently poorly understood in both quiescent and flare contexts.  

Imaging observations of coronal loops widths, however, do not show significant expansion across the length of coronal loops \citep{klimchuk2000,klimchuk2020}.  This lack of significant expansion is found for both flaring and non-flaring loops, and does not depend on loop length \citep{watko2000}.  As noted by \citet{watko2000}, this seemingly contradicts the observation that the magnetic field strength decreases from photosphere to corona \citep{gary2001}, which would necessitate an increase in the area expansion due to conservation of magnetic flux.  Comparison of observed and modeled spectral line intensities observed across the solar atmosphere suggests that the area does expand in both quiescent \citep{warren2010} and flaring contexts \citep{reep2022}, but this is far from settled.  In this work, we do not seek to explain this discrepancy, and we simply work with the assumption that the area may expand.

As we will show, the cooling time of a coronal loop depends directly on the area expansion, and therefore this is a possible explanation for a commonly reported problem that modeled loops cool too fast relative to observations.  In both quiescent loops and flaring loops, it has been noted that the observed cooling times are too long compared to model predictions (whether in relative or absolute terms).  For example, \citet{ugarte2006} found that there was a discrepancy between the time delays observed in active regions measured with Yohkoh's Soft X-ray Telescope (SXT, \citealt{tsuneta1991}) and the 195\,\AA\ channel of the Transition Region and Coronal Explorer (TRACE, \citealt{handy1999}) when compared to hydrodynamic simulations.    \citet{warren2010b} found that simulations could not reproduce the long cooling phase of post-flare loop arcades observed by Hinode's X-Ray Telescope (XRT, \citealt{golub2007}).  Many similar flare observations have suggested that the long-lasting cooling phase of post-flare loops may be due to a number of factors, such as gradual phase heating \citep{kopp1976,cargill1982,cargill1983,czaykowska1999,qiu2016} or turbulent suppression of thermal conduction \citep{karpen1987,bian2016,zhu2018}, among other possibilities.  Area expansion is one such possibility that requires further exploration.

In this work, therefore, we examine the effects of the two geometric assumptions -- area expansion and loop ellipticity -- and how they might affect conclusions drawn from hydrodynamic simulations.  We show that an area expansion can significantly impact both the hydrodynamic evolution of loops and the intensities of lines across the solar atmosphere, and further that the rate and location of expansion also impact the evolution.  Ellipticity, on the other hand, is relatively minor, only affecting the draining time of loops.

\section{Hydrodynamic Modeling}
\label{sec:hydro}

In this work, we use the HYDrodynamics and RADiation code (\texttt{HYDRAD}\footnote{\url{https://github.com/rice-solar-physics/HYDRAD}}, \citealt{bradshaw2003,bradshaw2013}) to solve the field-aligned hydrodynamic equations of a two-fluid plasma confined to a magnetic flux tube.  \texttt{HYDRAD} assumes that the plasma consists of electrons and hydrogen (neutral or ionized), and trace elements are included as an extra term to the electron density, in the radiative loss calculation, and as an increase in the effective mass.  \texttt{HYDRAD} uses full adaptive mesh refinement to spatially resolve conserved quantities in sufficient detail.

In the presence of a non-uniform cross-sectional area $A$, the conservation of mass, momentum, and energy are then given by:
\begin{equation}
    \frac{\partial \rho}{\partial t} + \frac{1}{A} \frac{\partial}{\partial s} \Big(A \rho v\Big) = 0 
    \label{eqn:mass}
\end{equation}    
\begin{align}
    \frac{\partial}{\partial t} \Big(\rho v\Big) + 
    \frac{1}{A}\frac{\partial}{\partial s} \Big(A \rho v^{2}\Big) &= \nonumber \\ 
    \frac{1}{A}\frac{\partial}{\partial s} \Big(\frac{4A \eta}{3} \frac{\partial v}{\partial s}\Big) &- \rho g_{\parallel} - \frac{\partial}{\partial s}\Big(P_{e} + P_{H}\Big)
    \label{eqn:momentum}
\end{align}
\begin{align}
    \frac{\partial E_{e}}{\partial t} + \frac{1}{A} \frac{\partial}{\partial s}\Big[A(E_{e} + P_{e}) v\Big] &= \nonumber \\
    - \frac{1}{A} \frac{\partial}{\partial s} \Big(A F_{ce}\Big) &+ v \frac{\partial P_{e}}{\partial s} - n_{e}n_{H}\Lambda(T) \nonumber \\
    + \frac{k_{B} n_{H}}{\gamma-1} \nu_{ie} \Big(T_{H} &- T_{e}\Big)  + H_{e}
\end{align}

\begin{align}
    \frac{\partial E_{H}}{\partial t} + \frac{1}{A} \frac{\partial}{\partial s}\Big[A(E_{H} + P_{H}) v\Big] &= \nonumber \\
    - \frac{1}{A} \frac{\partial}{\partial s} \Big(A F_{ci}\Big) &+ v \frac{\partial P_{e}}{\partial s} +\rho v g_{\parallel} \nonumber \\ 
    + \frac{1}{A}\frac{\partial}{\partial s} \Big(\frac{4A \eta v}{3} \frac{\partial v}{\partial s}\Big) \nonumber \\
    + \frac{k_{B} n_{H}}{\gamma-1} \nu_{ie} \Big(T_{e} &- T_{H}\Big)  + H_{H}
\end{align}

\noindent In these equations, $s$ is the field-aligned coordinate, $t$ time, $\rho$ the mass density, $n$ the number density, $v$ the bulk-flow velocity, $E_{e} = \frac{P_{e}}{\gamma - 1}$ the electron energy and $P_{e}$ pressure, and $E_{H} = \frac{P_{H}}{\gamma-1} + \frac{1}{2} \rho v^{2}$ the hydrogen energy and pressure.  $\eta$ is the dynamic ion viscosity, where electron viscosity is neglected, $F_c$ is the thermal conduction, $\nu_{ie}$ the collision frequency between ions and electrons, $g_{\parallel}$ the parallel acceleration due to gravity, and $\Lambda(T)$ is the optically thin emissivity.  Finally, the terms $H_{e}$ and $H_{H}$ are the heating rates applied to either the electrons or hydrogen.  

In most of this paper, we use loops of total length $2L = 50$\, Mm, and modify their cross-sectional areas $A(s)$, their eccentricity through $g_{\parallel}(s)$, or both.  The initial coronal temperature was approximately 0.5 MK, and coronal density was approximately $8 \times 10^{7}$\,cm$^{-3}$, which is energetically negligible compared to the input heat.  There was no coronal background heating.

In this work, we use the so-called VAL C chromospheric density and temperature profile \citep{vernazza1981}, with an approximation to the optically thick radiative losses there \citep{carlsson2012}.  There is a chromospheric background heating term that exactly balances the initial state of the chromospheric radiative losses, intended to simply maintain the chromosphere in the absence of heating events.  We use impulsive heating by an electron beam, but the major results of this paper are in the cooling phase and so do not depend strongly on the details of the heating \citep{winebarger2004}.  The heating function is then approximated by the equations in \citet{emslie1978}, with modifications for non-uniform ionization by \citet{hawley1994}.  Recent work by \citet{allred2020} has shown that this approximation is relatively inaccurate with particularly strong heating, and a solution to the Fokker-Planck equation is more suitable in general, though the approximation works well when the only force is due to Coulomb collisions of the non-thermal electrons.  

\section{Expanding Cross-Sectional Area}
\label{sec:area}

The impact of expanding cross-sections has been touched upon only sparsely in the literature of field-aligned hydrodynamic simulations.  \citet{emslie1992} examined the impact of an expanding area on the evaporative flows driven by an impulsive electron beam and how it affects observed blue-shifts of \ion{Ca}{19} emission.  \citet{mikic2013} explored how the cross-sectional area affects thermal conduction and thermal non-equilibrium (TNE) in coronal loops, and \citet{froment2018} performed an in-depth parameter space survey to understand the occurrence of TNE or coronal condensations, but found that TNE can occur in nearly any geometry.  \citet{reep2020} showed that an expanding area itself is not sufficient to cause coronal condensations (rain) in impulsively heated loops.  \citet{cargill2022} examined how loops evolved with modestly expanding cross-sectional areas, finding higher peak temperatures and denser loops when compared to loops with uniform area.

In this section, therefore, we examine how an expanding area can impact both the hydrodynamic evolution and the resultant irradiance at various temperatures.  Since it is not well known where the expansion occurs, or the rate at which it occurs, we consider two forms of expansion: one where the cross-section increases gradually and continuously, and one where the expansion only occurs in the transition region.  We examine area expansions factors of more than a factor of 100 from footpoint to apex, consistent with the range implied from measurements of the magnetic field strength (see Appendix \ref{app:area}).

\subsection{Continuous Area Expansion}
\label{subsec:continuous}

\begin{figure}
\centering
\includegraphics[align=c,width=0.48\textwidth]{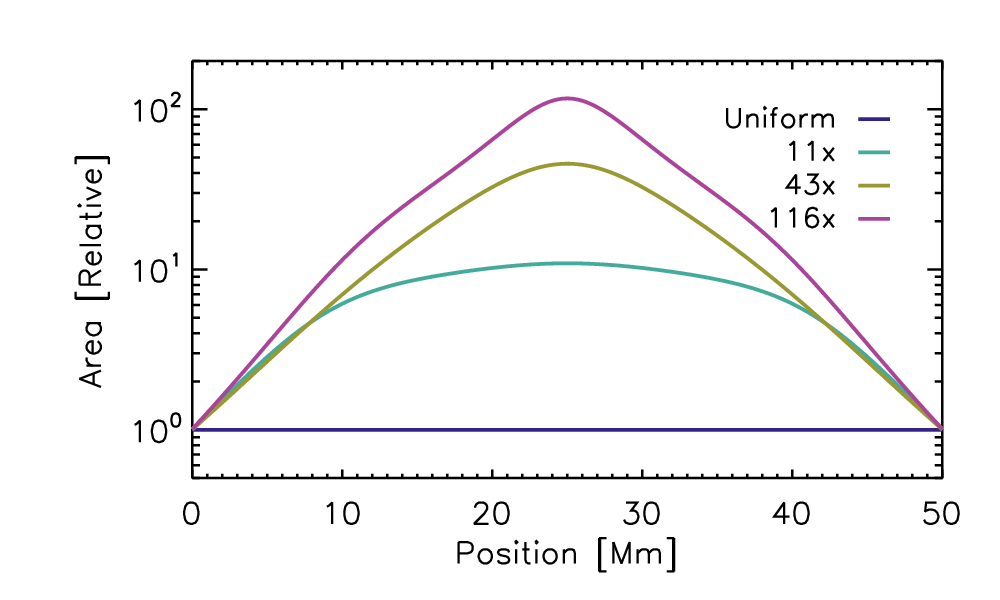}
\includegraphics[align=c,width=0.48\textwidth]{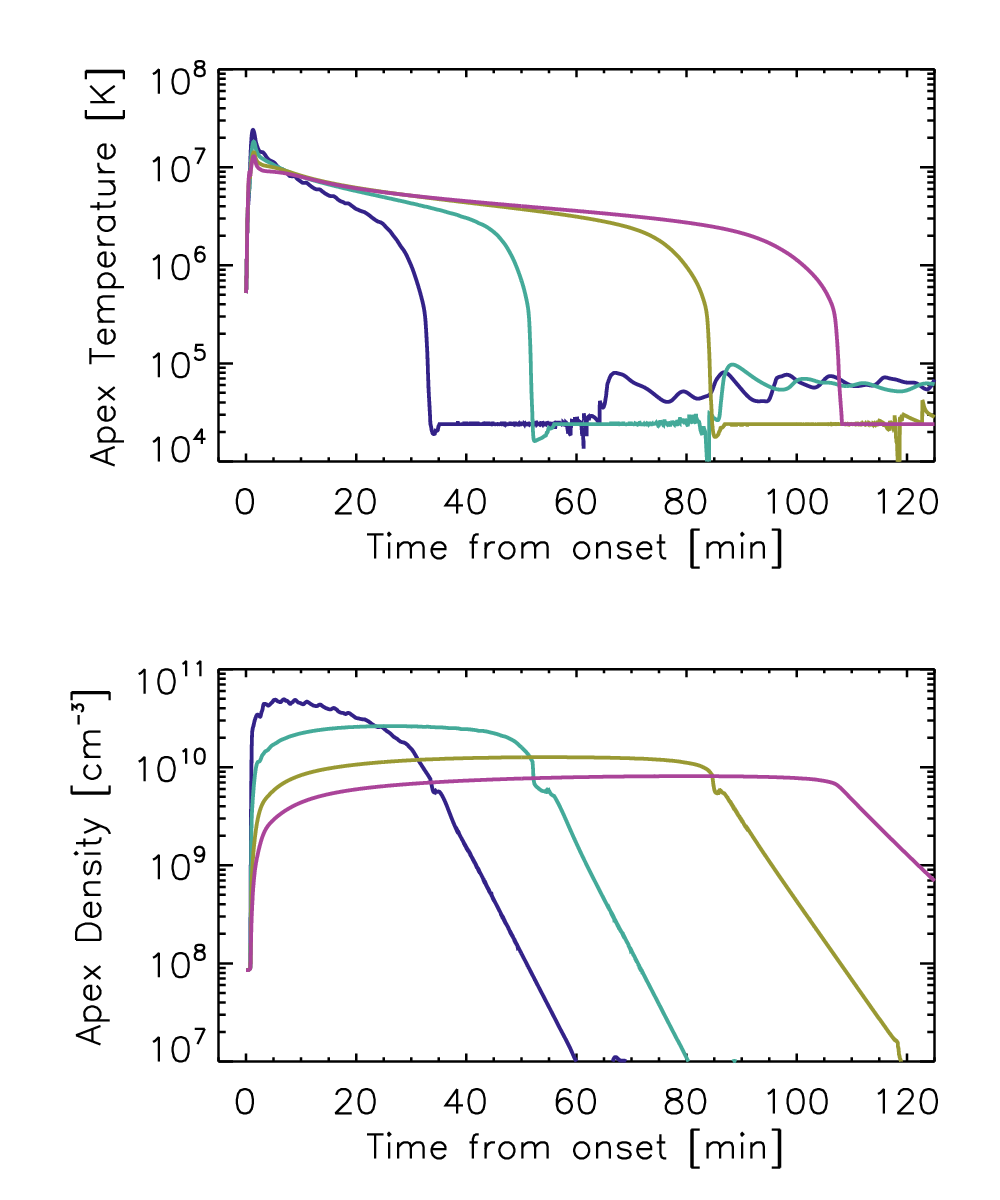}
\caption{A comparison of the area profiles (top) used in this section and the evolution of the apex temperatures (center) and densities (bottom).  We contrast four cases with four different expansion factors: uniform area, and expansions of 11, 43, and 116 from footpoint to apex of the loop.  Note that the area expansion is relative, since the dynamics do not depend on the actual area magnitude, only the relative change. \label{fig:area_comp}}
\end{figure}

In order to better understand the effect of a non-uniform cross-sectional area, we examine four simulations with different cross-sectional area profiles, but otherwise identical parameters.  We use the magnetic field strength $B(s)$ function from \citet{mikic2013}, which produces a continuous expansion over the length of the loop:
\begin{equation}
    B(s) = B_{0} + B_{1}\ (\exp(-s/l) + \exp(\frac{-(L-s)}{l})
\end{equation}
where $L$ is the loop length and $l$ a scale height. As those authors do, we then assume that the area expansion $A(s) \propto \frac{1}{B(s)}$ to calculate the (relative) area expansion.  We choose values such that the the expansion from footpoint to apex is 1 (uniform), 11, 43, and 116, which span the approximate range of expected expansion factors (see Appendix \ref{app:area}), and produces a continuous expansion along $s$ from footpoint to apex.  The profiles $A(s)$ are shown at top in Figure \ref{fig:area_comp}.  Note that the expansion used in the simulations is only a relative expansion, and does not require a physical value.  The larger area expansions assumed here, additionally, require coronal magnetic field strengths that are likely small enough that the low $\beta$ assumption typical of coronal loop modeling might not apply at all times, particularly with strong heating.  This would likely necessitate both a positionally and temporally varying cross-sectional area (see also the discussion in the appendix of \citealt{reep2022}).

We heat the loops with an electron beam with peak flux $F = 10^{10.3}$ erg s$^{-1}$ cm$^{-2}$, lasting for 100 s, with a low energy cut-off of 15 keV and spectral index of 5.  The heating parameters are not crucially important to examine the differences in hydrodynamic response over long time scales, and the overall evolution would be similar for other parameters.  The center and bottom plots in Figure \ref{fig:area_comp} show the evolution of the apex temperatures and densities in these four simulations.  The total times for the loop to cool and drain after heating onset are both significantly increased with an increasing area, as found in previous studies such as \citet{antiochos1976,antiochos1978}.  The small volume of the TR limits both the amount of energy that can be radiated away and the amount of plasma that can be evaporated into the loop. 

There are a few important features to notice from this figure.  The time of peak density, roughly when the dominant cooling mechanism switches from conduction to radiation \citep{cargill1994}, is delayed with a larger area expansion, so that there are extremely long conductive cooling phases (see also \citealt{antiochos1978}).  Additionally, after the density peaks, it remains relatively constant, while the temperature slowly falls through radiation.  That is, the $T \sim n^{2}$ relation for the radiative cooling phase (\textit{e.g.} \citealt{serio1991,jakimiec1992,cargill1995,bradshaw2005}) does not hold here, and the loop drains only after the temperature falls below $\approx 10^{5}$ K.  This behavior is fundamentally different to a uniform loop, that there is essentially no enthalpy/radiative cooling phase with large expansions, and as a result it is not correct to assume that emission cooler than 3--4 MK is from the radiative phase.

Following \citet{bradshaw2010}, we have more generally the relation $T \sim n^{\delta}$, where $\delta = \gamma - 1 + \frac{\tau_{\nu}}{\tau_{R}}$, with $\gamma = 5/3$, $\tau_{R}$ the radiative time scale, and a coronal draining timescale $\tau_{\nu} = \frac{L_{c}}{v}$, depending on the (coronal) loop length $L_{c}$ and a downflow speed $v$.  With large area expansions, this enters the static radiative cooling regime, or $\delta \rightarrow \infty$, implying that the draining velocity becomes significantly reduced ($\tau_{\nu} \rightarrow \infty$ as $v \rightarrow 0$).  We look at the impact of area expansion on the velocities now.

\begin{figure*}
\centering
\includegraphics[width=0.24\linewidth]{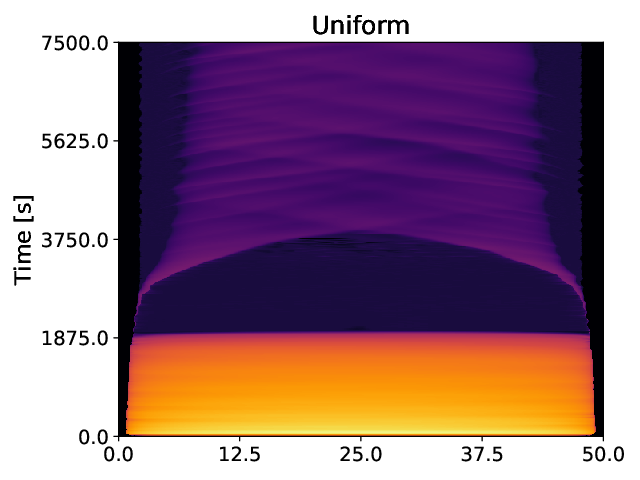}
\includegraphics[width=0.24\linewidth]{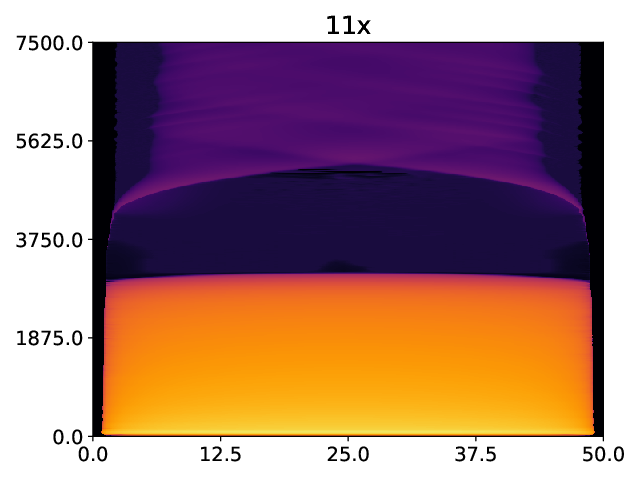}
\includegraphics[width=0.24\linewidth]{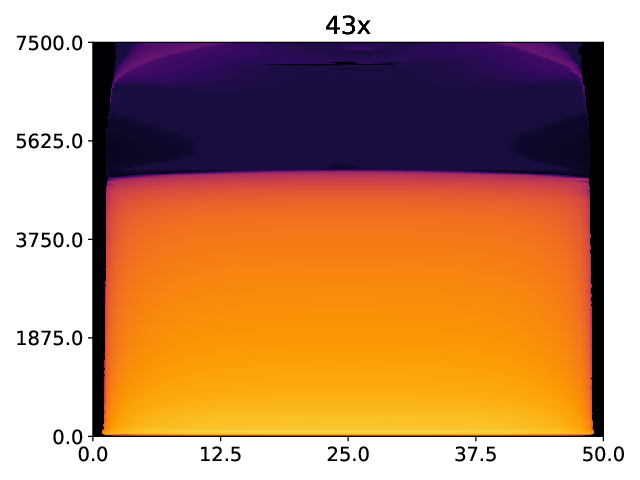}
\includegraphics[width=0.24\linewidth]{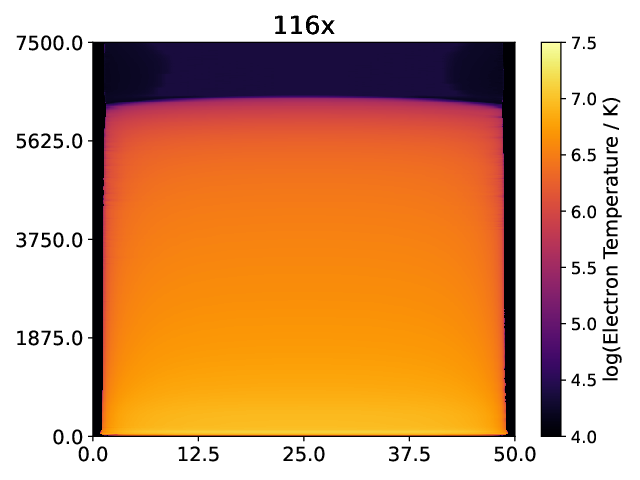}
\includegraphics[width=0.24\linewidth]{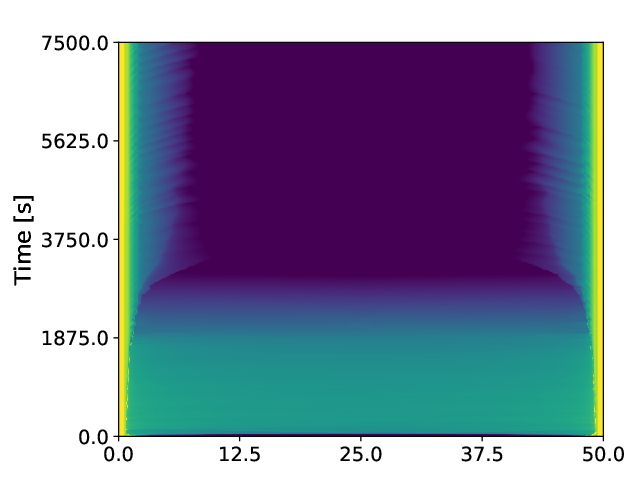}
\includegraphics[width=0.24\linewidth]{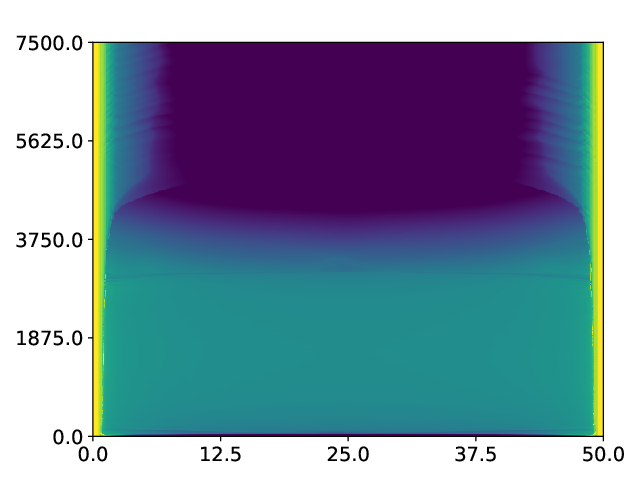}
\includegraphics[width=0.24\linewidth]{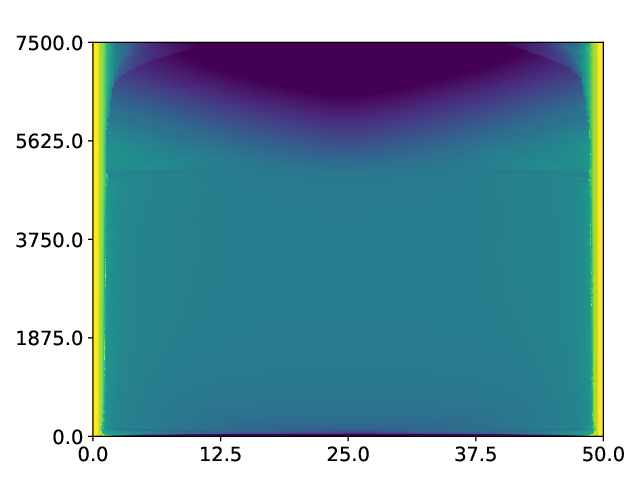}
\includegraphics[width=0.24\linewidth]{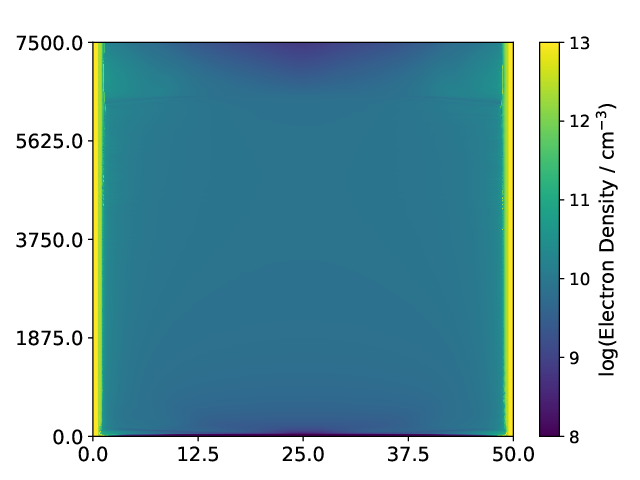}
\includegraphics[width=0.24\linewidth]{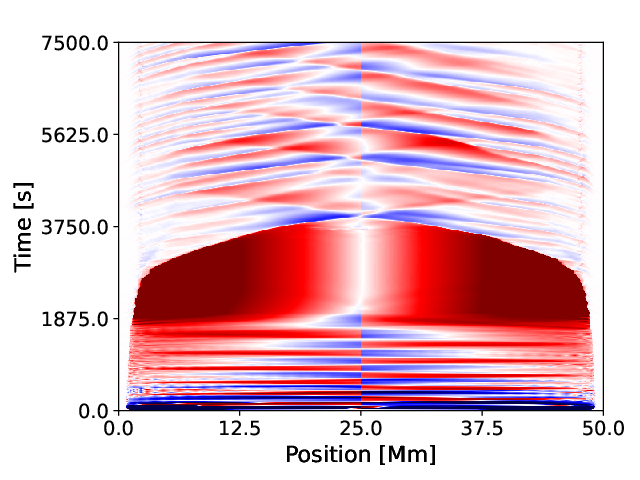}
\includegraphics[width=0.24\linewidth]{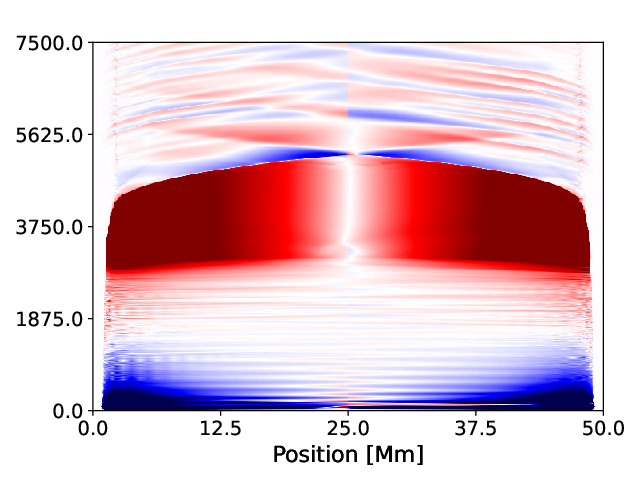}
\includegraphics[width=0.24\linewidth]{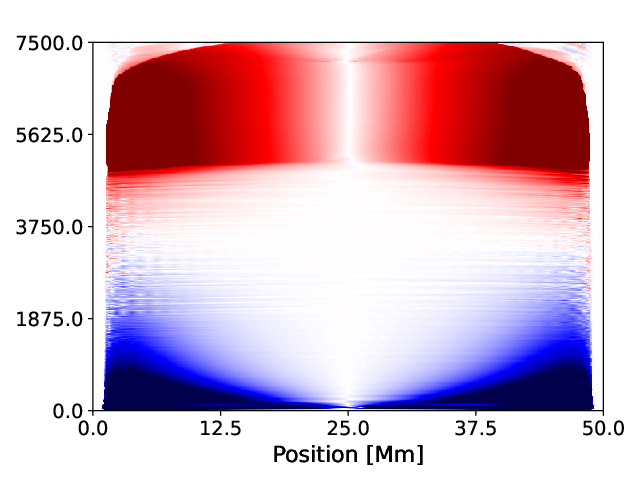}
\includegraphics[width=0.24\linewidth]{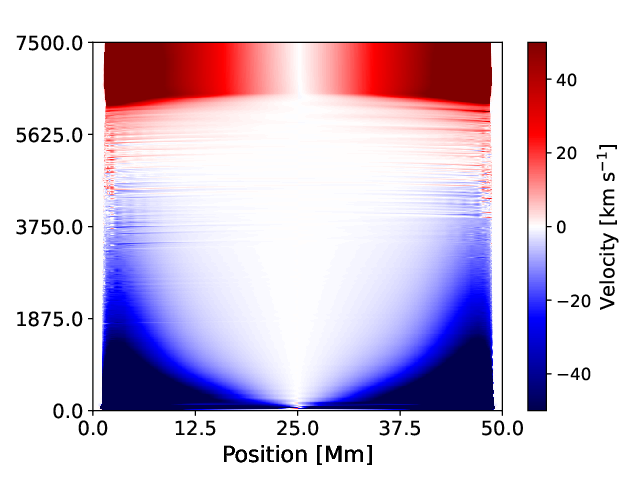}
\caption{The evolution of the electron temperature (top), density (center), and bulk flow velocity (bottom) in four circular loops with varying cross-sectional area.  From left, the total area expansion is 1 (uniform), 11, 43, and 116, corresponding to the profiles in Figure \ref{fig:area_comp}.  The x-axis shows the position along the loop, while the y-axis shows the evolution in time.  In the velocity plots, blue indicates a flow towards the apex, while red indicates a flow away from the apex, and does not necessarily correspond to a Doppler shift.  The cooling time is significantly longer with increasing area.  Evaporative upflows last for significantly longer with a larger area expansion, and sound waves are suppressed.  \label{fig:area_hydro}}
\end{figure*}

Next, we examine the overall evolution of the loops.  Figure \ref{fig:area_hydro} shows the hydrodynamic evolution along the loops (x-axis) with time (y-axis) for the electron temperature (top), electron density (center), and bulk flow velocity (bottom).  The four cases are shown in order from left to right (1x, 11x, 43x, and 116x).  In the velocity plots, blue indicates a flow toward the loop apex and red away from the apex, which does not necessarily correspond to a Doppler shift.  Once again, it is apparent that the cooling and draining times are both significantly increased with increasing area expansion.  As in \citet{reep2020}, we find that there are no coronal condensations (rain events), indicating that area expansion alone is not sufficient to produce coronal rain.  The velocity plots show two interesting features.  First, as the area expansion is increased, the duration of upflows is also increased, lasting significantly longer than the assumed heating duration.  \citet{reep2018} examined the relation between upflow duration and heating duration, finding that they are approximately equal, but that study only examined uniform area loops.  This should be reexamined in detail.  Second, the sound waves shown at bottom left, seen as alternating up- and downflows, are suppressed with larger area expansions.  As evaporative material flows up into an area of the loop with larger cross-section, the speed slows, effectively causing a damping of the sound waves.  Importantly, the speed of downflows is significantly reduced during the radiative cooling phase, and so the draining timescale $\tau_{\nu}$ increases with the area expansion, or $\delta \propto A_{\text{max}}$.

We now turn to the effect on radiative emission.  In each simulation, we calculate the emission measure for each grid cell at each time step, EM$ = \int n_{e}(s)\ n_{H}(s)\ dV = \int n_{e}(s)\ n_{H}(s)\ A(s)\ ds$, which we then combine with the emissivities from the CHIANTI atomic database (version 10, \citealt{dere1997, delzanna2021}).  We then sum up the emission along the full length of the loop at each time step to create a time series.  Note that this assumes the spectral lines are optically thin, which may not be true in general, particularly for the cooler lines.  

\begin{figure*}
    \centering
    \includegraphics[width=0.32\textwidth]{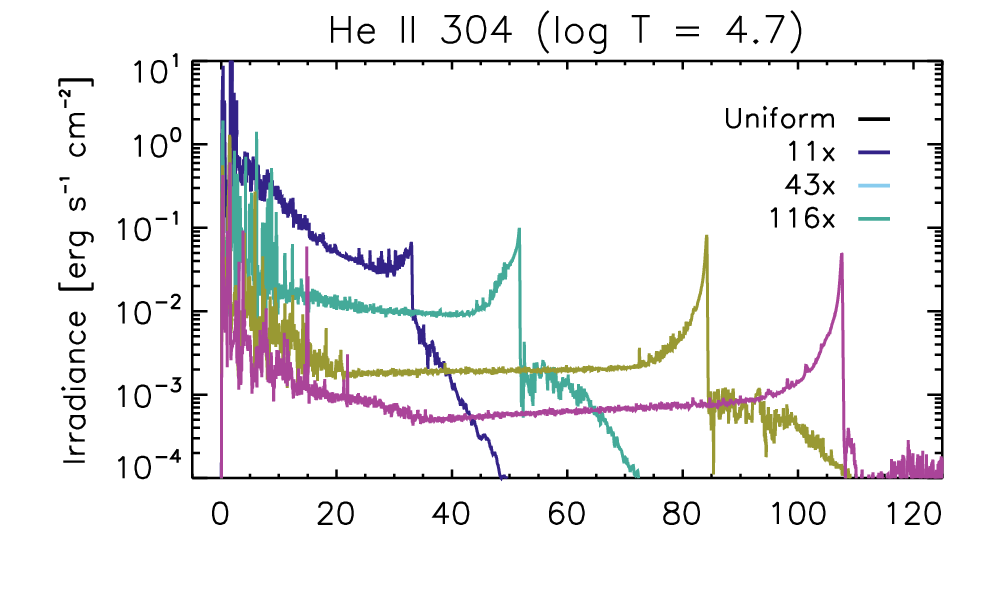}
    \includegraphics[width=0.32\textwidth]{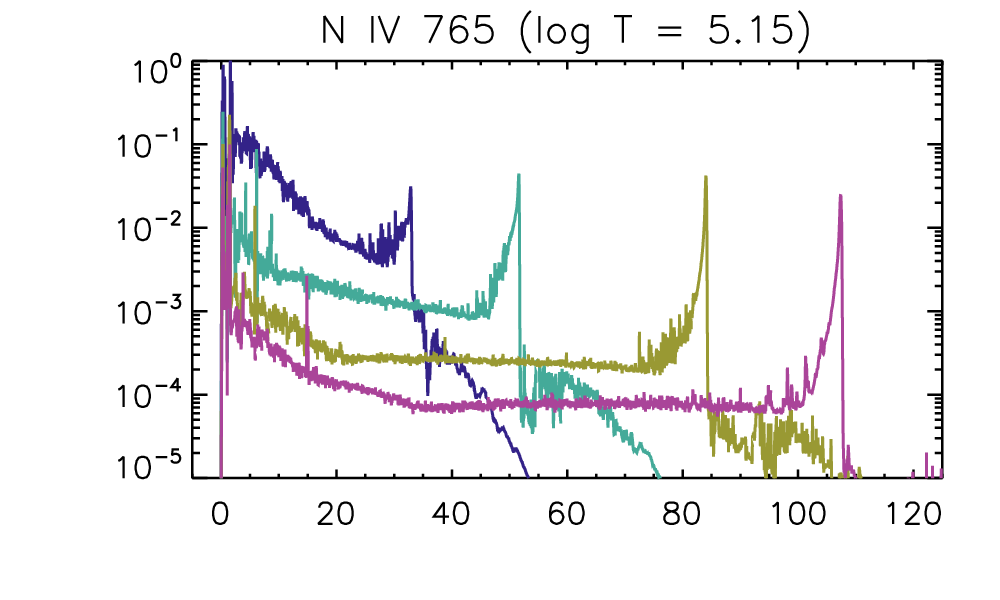}
    \includegraphics[width=0.32\textwidth]{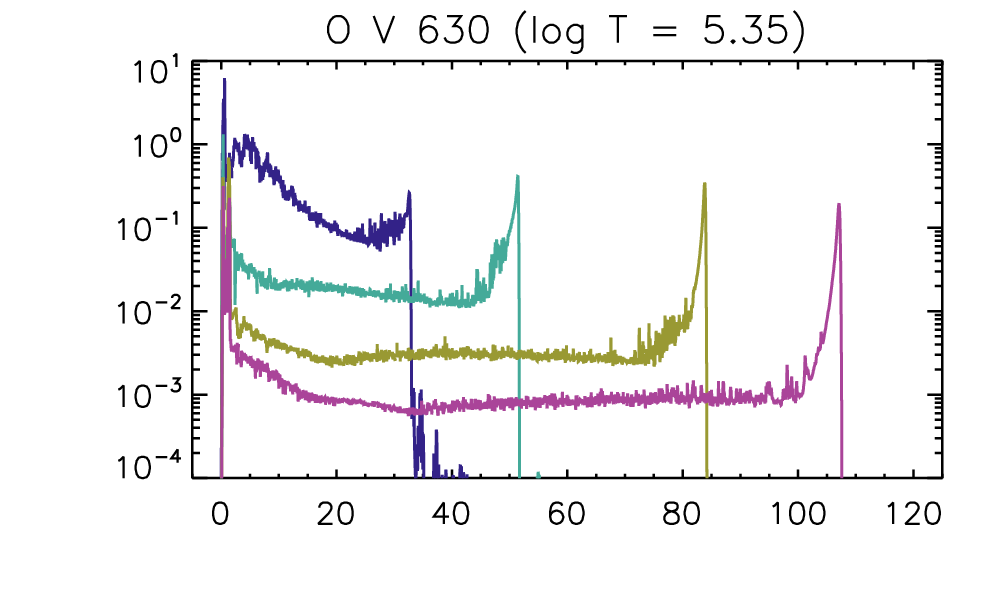}
    \includegraphics[width=0.32\textwidth]{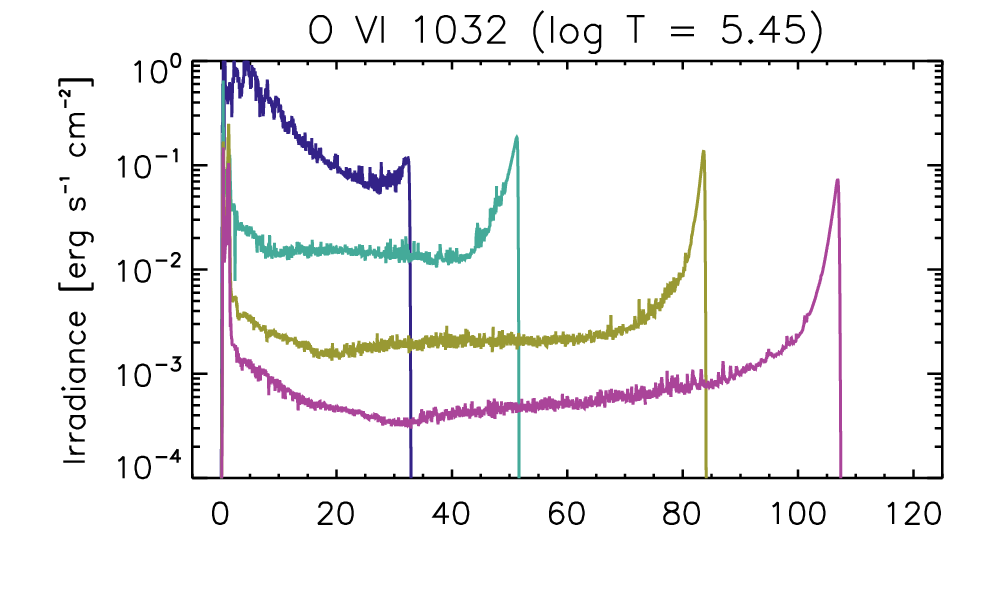}
    \includegraphics[width=0.32\textwidth]{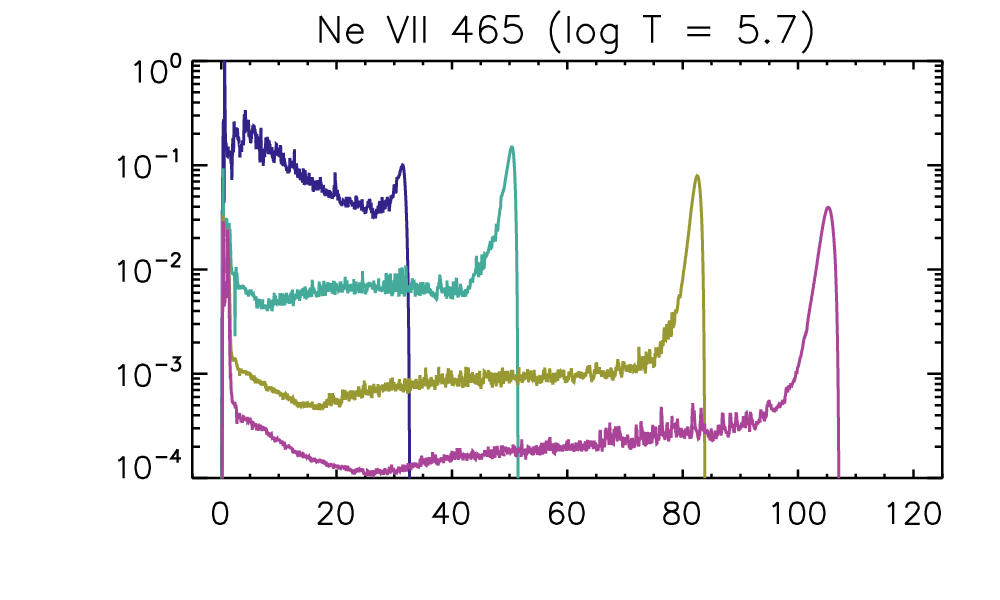}
    \includegraphics[width=0.32\textwidth]{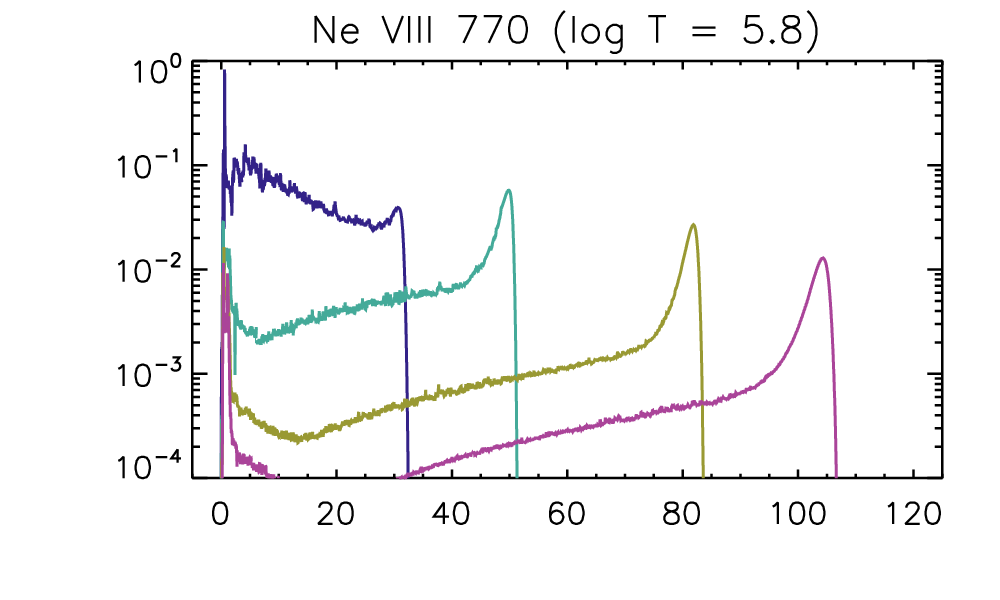}
    \includegraphics[width=0.32\textwidth]{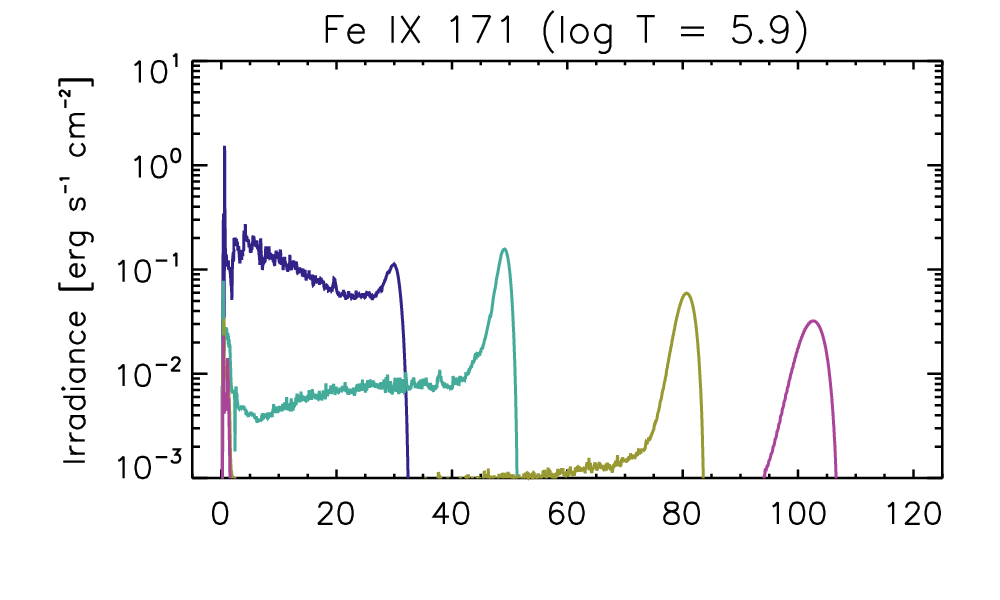}
    \includegraphics[width=0.32\textwidth]{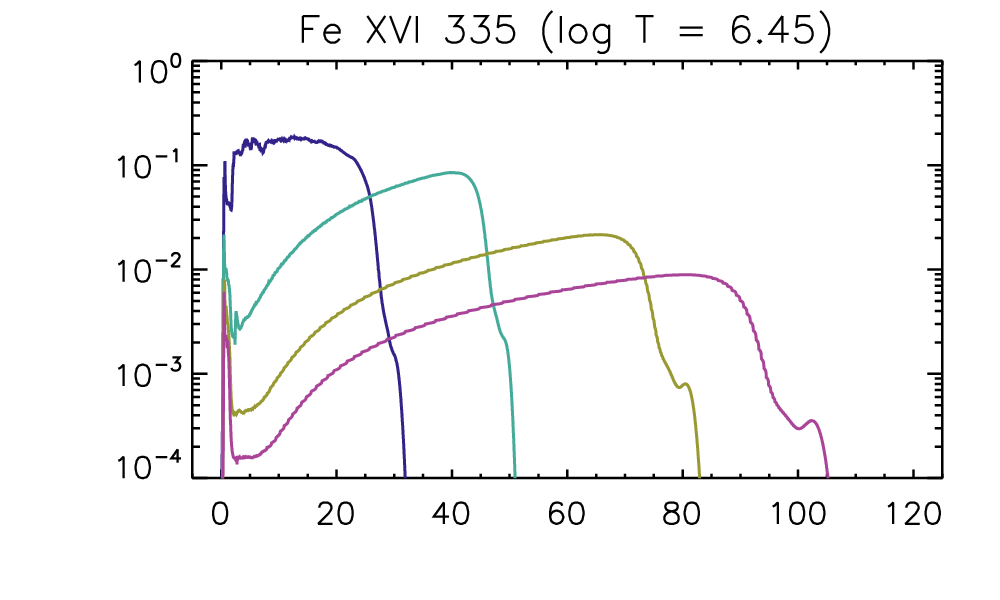}
    \includegraphics[width=0.32\textwidth]{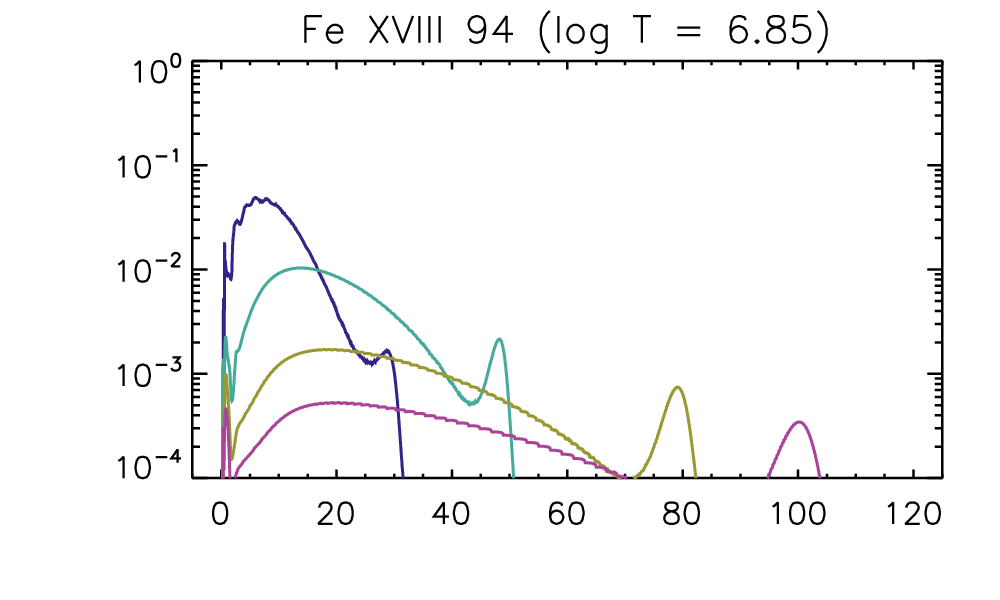}
    \includegraphics[width=0.32\textwidth]{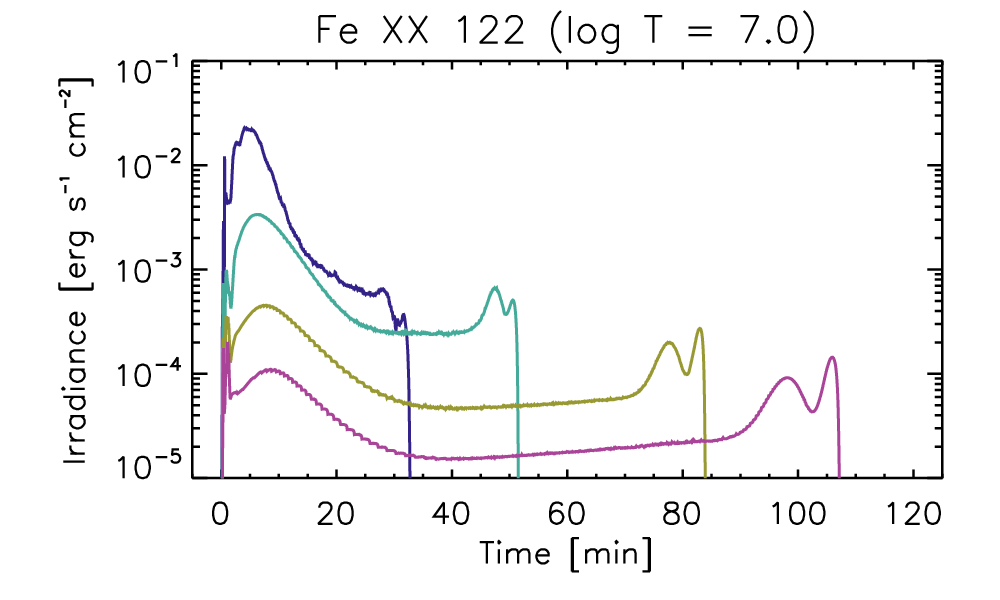}
    \includegraphics[width=0.32\textwidth]{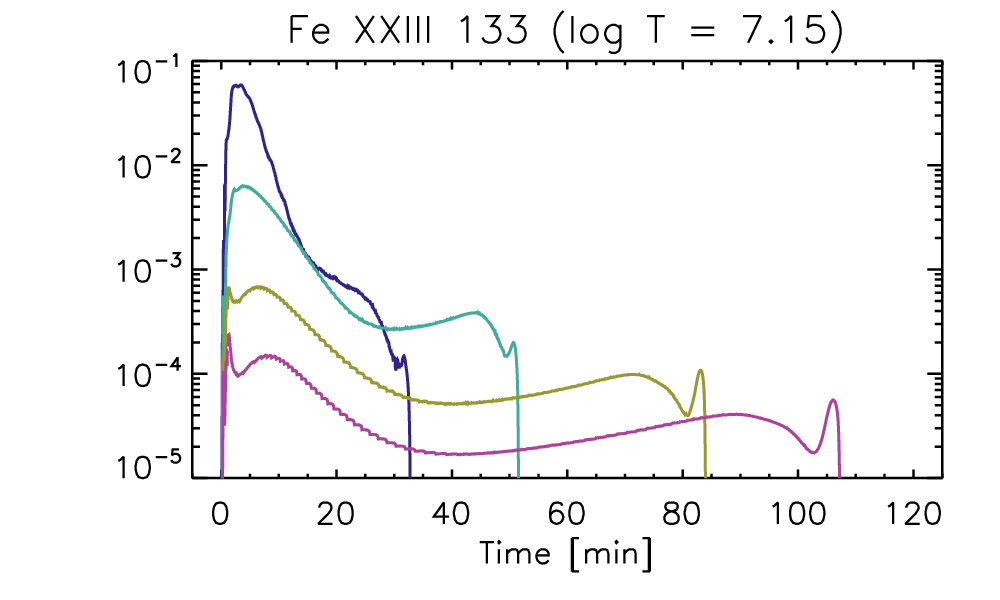}
    \includegraphics[width=0.32\textwidth]{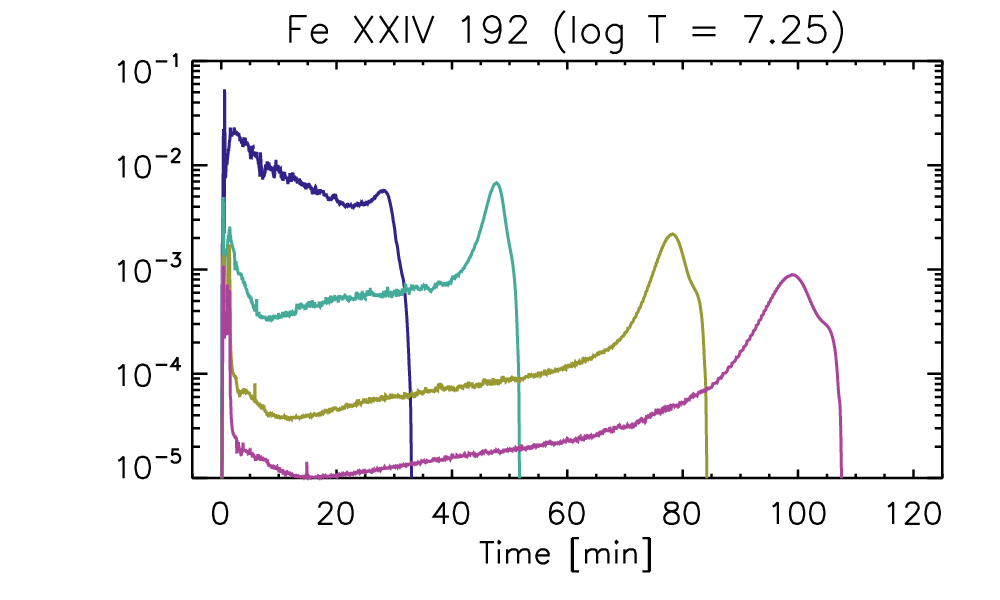}
    \caption{Synthetic irradiance time series for 12 spectral lines, as labeled (arranged by formation temperature, from coolest to hottest), as might be seen by SDO/EVE for each of the four cross-sectional area cases in Figure \ref{fig:area_comp}.  The total volumes (=$\int_{L} A(s)\ ds$) have been normalized to be equal.  The area expansion drastically affects the intensities of all lines at all times, both directly through the volume at a given height and indirectly through the effects on the hydrodynamics.  \label{fig:area_irr}}
\end{figure*}

In Figure \ref{fig:area_irr}, we show a comparison of irradiance light curves synthesized from these four simulations for 12 spectral lines as might be measured by the Extreme Ultraviolet Variability Experiment (EVE, \citealt{woods2012}) onboard SDO, as labeled in the plots, ranging from \ion{He}{2} 304\,\AA\ at $\log{T} = 4.7$ through \ion{Fe}{24} 192\,\AA\ at $\log{T} = 7.25$.  We have assumed that the total volume, $V = \int_{L} A(s)\ ds$, in each simulation is equal, and we have assumed photospheric abundances \citep{asplund2009}.  It is clear that the non-uniform cross-sectional area strongly impacts the irradiance in all of these lines.  First, the emission is reduced with a larger total expansion while the loop remains hot.  Second, the increase in the cooling and draining times that was shown in Figure \ref{fig:area_hydro} is apparent in these plots, where the emission from \textit{e.g.} \ion{Fe}{16} remains relatively bright for a longer time with increasing area.  Third, we note that the peak irradiance in the transition region lines (\ion{N}{4}, \ion{O}{5}, \ion{O}{6}, \ion{Ne}{7}, \ion{Ne}{8}, and \ion{Fe}{9}) is similar for each of the four area expansion factors.  This spike in intensity occurs when the coronal segment of the loop cools through the line's formation temperature, and since the total loop volumes are normalized, the resultant intensities are similar.  Finally, since the peak density is higher with a smaller area (see the apex densities in Figure \ref{fig:area_comp}), the hottest lines are brightest with smaller area expansions (\textit{e.g.} \ion{Fe}{22} 122\,\AA\ and \ion{Fe}{23} 133\,\AA), and all lines are brighter at early times with a smaller area expansion.  

These plots suggest that we may be able to infer the approximate expansion factor from observed cooling times.  That is, because the cross-sectional area expansion correlates well with the cooling time of the loop, and because the intensity spikes sharply when the loop cools through the formation temperature of a line, we can approximate the expansion factor by measuring the cooling time.  The time to cool from, \textit{e.g.}, \ion{Fe}{18} 94\,\AA\ to \ion{He}{2} 304\,\AA\ increases slightly with area expansion (1.2 minutes in the uniform loop versus 3 minutes in the 116x loop).  Of course, SDO/AIA channels have a relatively broad contribution so this sort of comparison requires caution.  

\subsection{Footpoint Localized Expansion}
\label{subsec:tr_exp}

Imaging observations often show little expansion in the coronal portion of loops \citep{klimchuk2000,klimchuk2020}, which is at odds with the decrease of the magnetic field strength from photosphere through corona that would necessitate expansion.  One possible explanation for this discrepancy is that area expansion occurs mostly or only in the transition region (TR), and that any expansion in the corona is limited.  We take an agnostic position in this work, and therefore additionally present a comparison of how expansion localized near the base of the loop would affect the dynamics.

We quantify footpoint localized expansion using a hyperbolic tangent function, chosen \textit{ad hoc} to simply examine this possibility.  The magnetic field strength $B(s)$ near the footpoint is:
\begin{equation}
    B(s) = \frac{B_{\text{max}} + B_{\text{min}}}{2} - \frac{B_{\text{max}} - B_{\text{min}}}{2} \tanh\Big(\frac{s - s_{0}}{2 \sigma L}  \Big)
    \label{eq:tr}
\end{equation}
\noindent where where we have defined $s_{0}$ as an offset for the initial location of the transition region, $\sigma$ is a scale height of the expansion, and $L$ is the loop length.  We assume the coronal segment of the loop has a constant magnetic field strength.  

\begin{figure}
\centering
\includegraphics[align=c,width=0.48\textwidth]{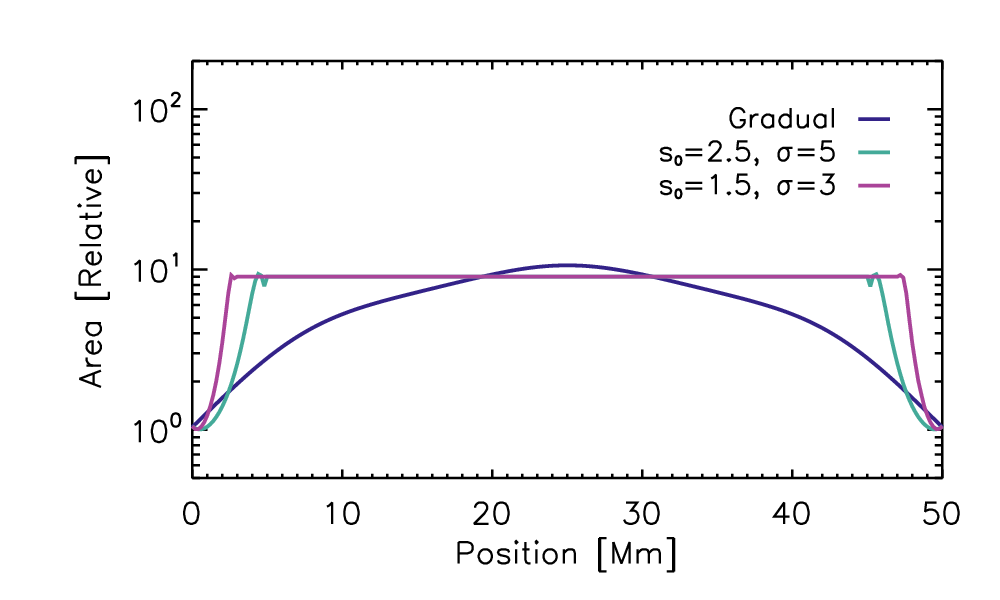}
\includegraphics[align=c,width=0.48\textwidth]{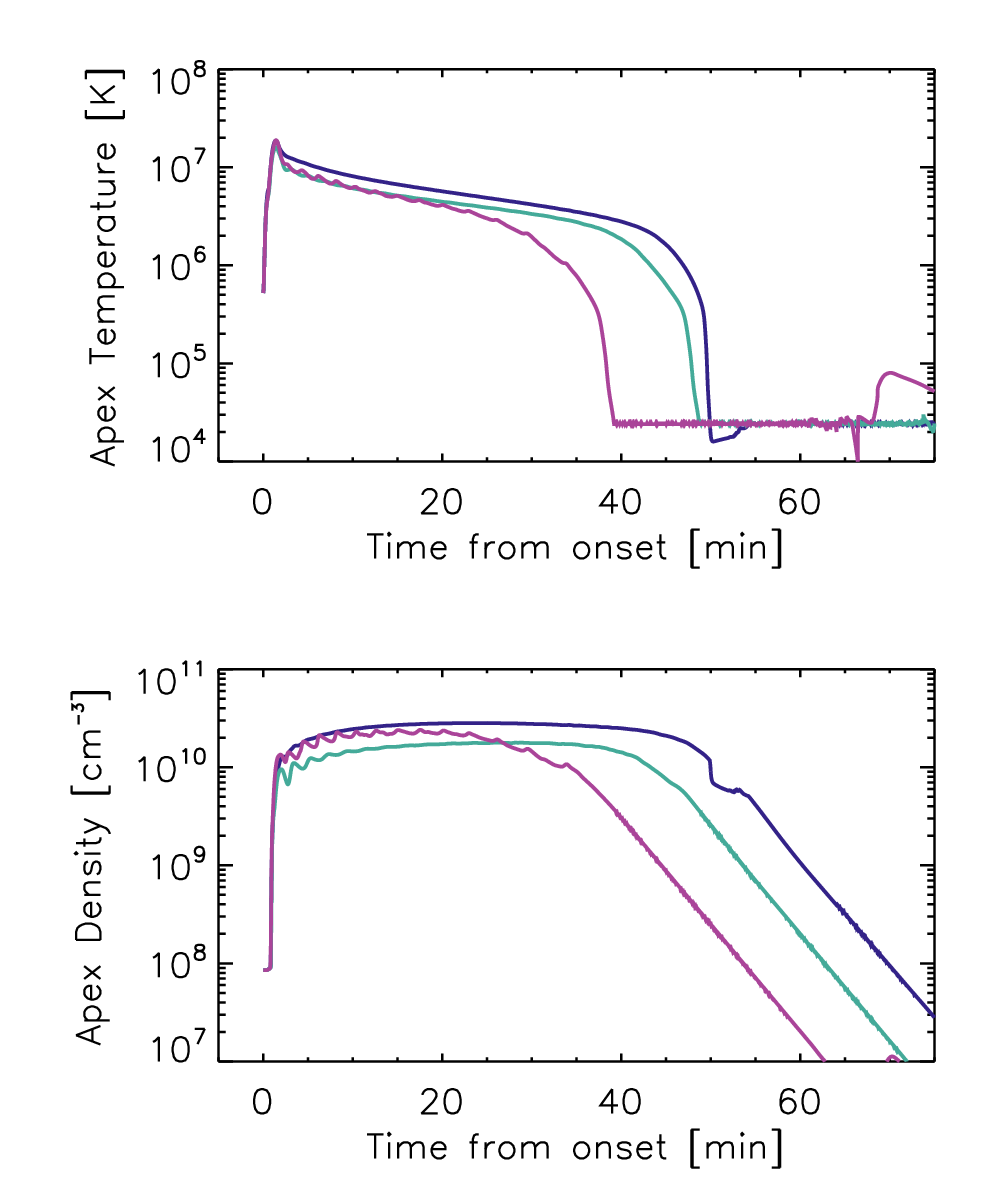}
\caption{Similar to Figure \ref{fig:area_comp}, a comparison of the area profiles (top) used in this section and the evolution of the apex temperatures and densities (bottom).  We contrast three cases: gradual expansion (black), footpoint localized expansions with $s_{0} = 2.5$ Mm and $\sigma = 5$ Mm (blue), and $s_{0} = 1.5$ Mm and $\sigma = 3$ Mm (red).  The cooling and draining times are reduced with expansion more confined near the footpoint.  \label{fig:TR_comp}}
\end{figure}

We examine three cases in this section to better understand this.  We use a gradual expansion case, where the area increases continuously from the footpoint to the apex of the loop to a peak value of 10.  We also compare this to two footpoint localized expansion cases, one with $s_{0} = 2.5$ Mm and $\sigma = 5$ Mm in Equation \ref{eq:tr}, and one with $s_{0} = 1.5$ Mm and $\sigma = 3$ Mm.  In all three cases, we assume a loop length of 50 Mm, and use the same heating parameters as in the previous section.  The area profiles are shown in Figure \ref{fig:TR_comp} at top, along with the evolution of the apex temperatures and densities (bottom).  When the expansion occurs closer to the footpoint, the cooling and draining times are both reduced.  This is because a larger portion of the TR reaches the maximum cross-section, the more localized the expansion is to the footpoint, more effectively transporting away energy through radiation and flows.  We therefore see the gradually expanding case takes the longest to cool and drain, while the profile localized furthest down the loop drains and cools the quickest.  The peak temperature reaches a similar value in all three cases, and all three peak at a similar time.  

\begin{figure*}
\centering
\includegraphics[width=0.32\linewidth]{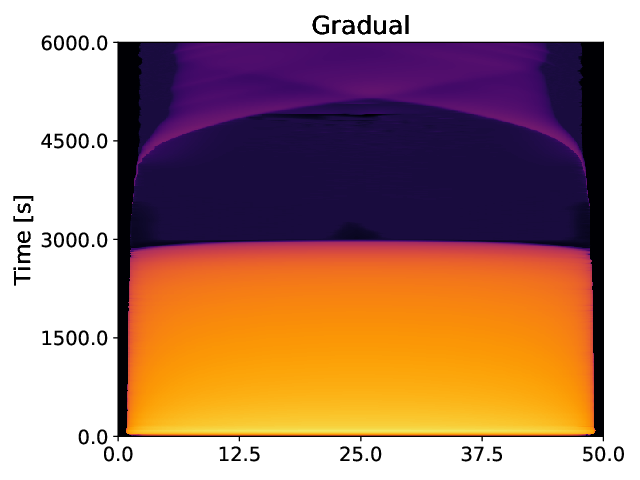}
\includegraphics[width=0.32\linewidth]{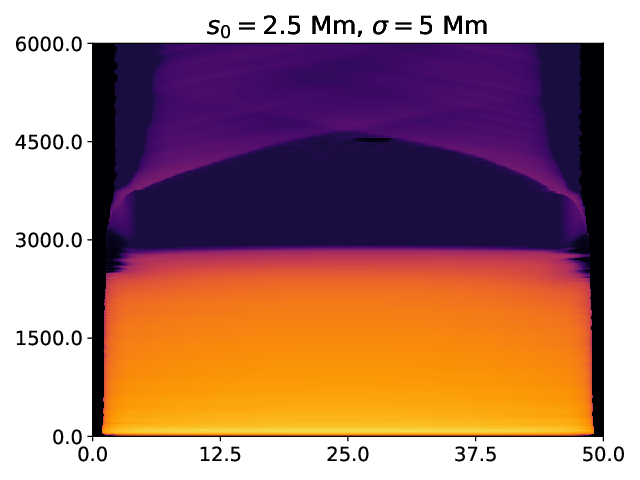}
\includegraphics[width=0.32\linewidth]{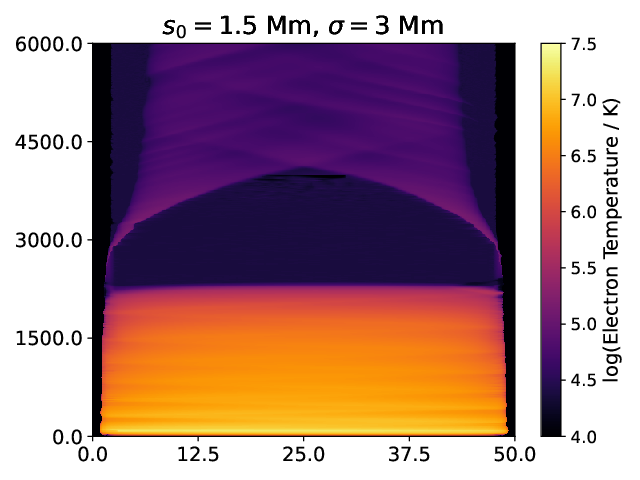}
\includegraphics[width=0.32\linewidth]{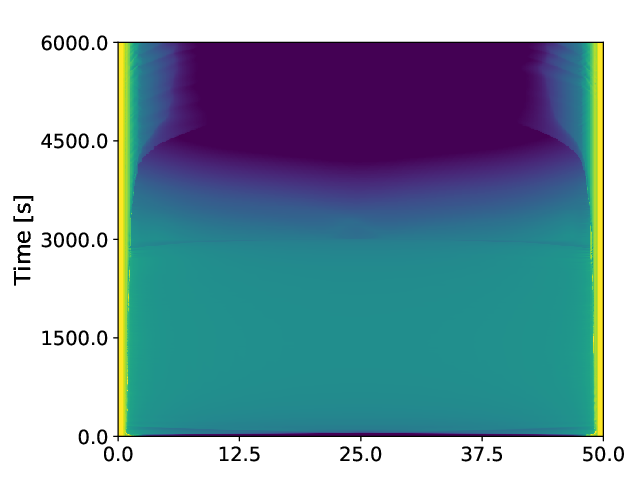}
\includegraphics[width=0.32\linewidth]{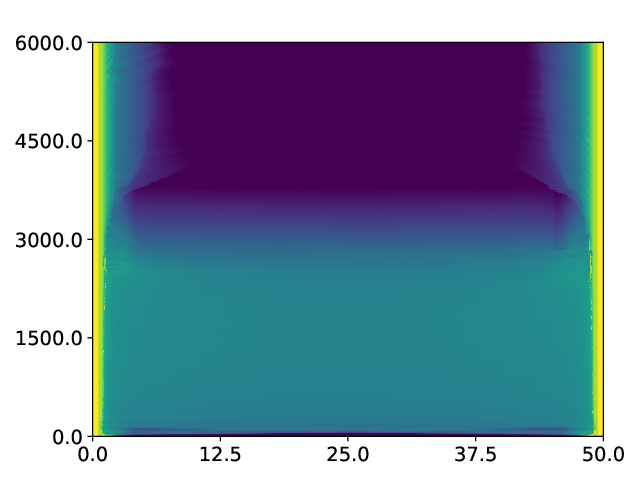}
\includegraphics[width=0.32\linewidth]{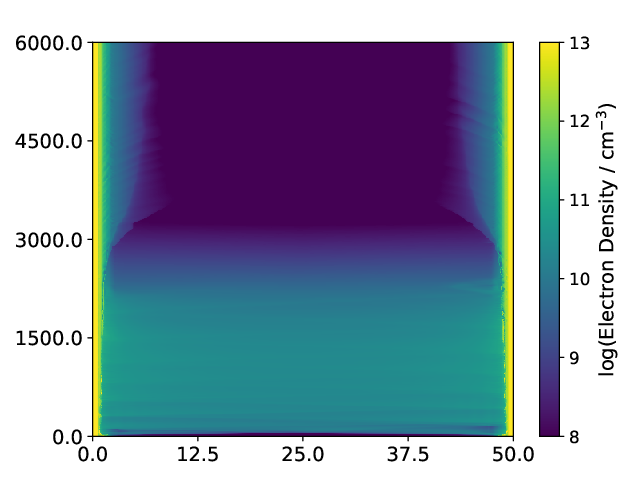}
\includegraphics[width=0.32\linewidth]{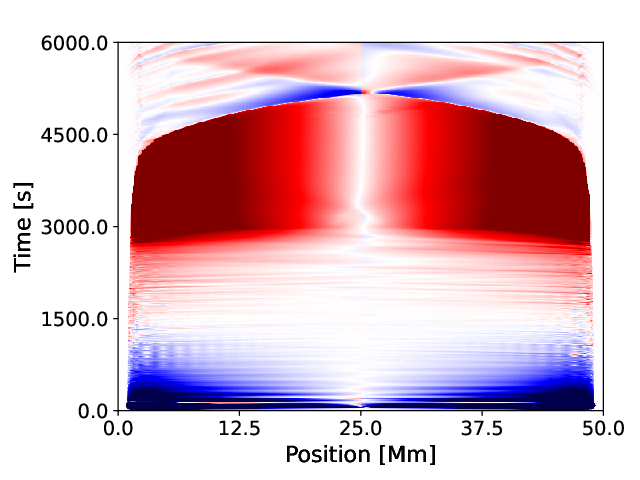}
\includegraphics[width=0.32\linewidth]{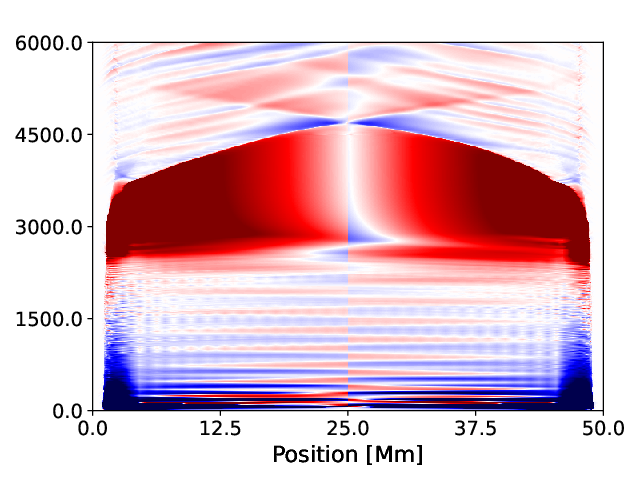}
\includegraphics[width=0.32\linewidth]{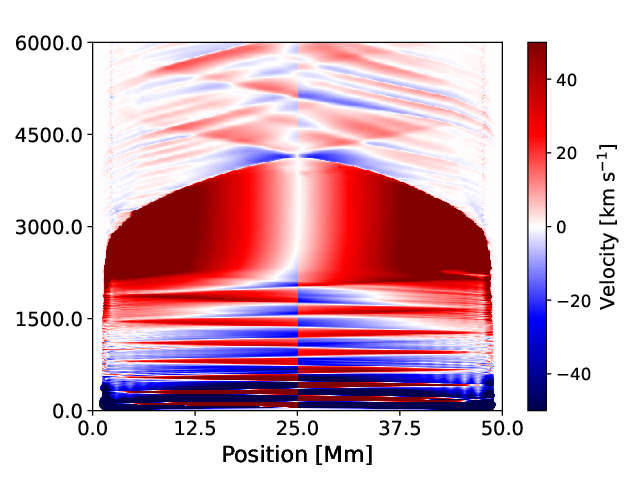}
\caption{Similar to Figure \ref{fig:area_hydro}, comparing three cases, gradual expansion (left), along with TR-confined expansions with $s_{0} = 2.5$ Mm and $\sigma = 5$ Mm (center), and $s_{0} = 1.5$ Mm and $\sigma = 3$ Mm (right).  The cooling and draining times are reduced when the expansion occurs lower in the atmosphere.  The up-flows become similarly confined near the location of expansion, and behaves more like a uniform-area case (compare Fig. \ref{fig:area_hydro}).  \label{fig:TR_hydro} } 
\end{figure*}

To get a fuller picture, we also present the full hydrodynamic evolution in Figure \ref{fig:TR_hydro}.  The gradual case is shown at left, while the case with $s_{0} = 2.5$ Mm and $\sigma = 5$ Mm is in the center, and $s_{0} = 1.5$ Mm and $\sigma = 3$ Mm is at right.  The change to the cooling time is apparent.  Additionally, the flow profiles are somewhat modified.  When the expansion is confined near the footpoints, the up-flow speeds quickly dampen as the area rapidly expands ($A\ v = \text{const}$).  In contrast, when the area expands gradually, its effect on the velocity is less pronounced, and the up-flow speeds remain strong well into the coronal portion of the loop.  Similarly, the suppression of sound waves which we noted in Figure \ref{fig:area_hydro} occurs in the gradual expansion case, but is mostly absent when the expansion is confined near the footpoints.  Once again, no coronal condensation events occur.

\begin{figure*}
    \centering
    \includegraphics[width=0.32\textwidth]{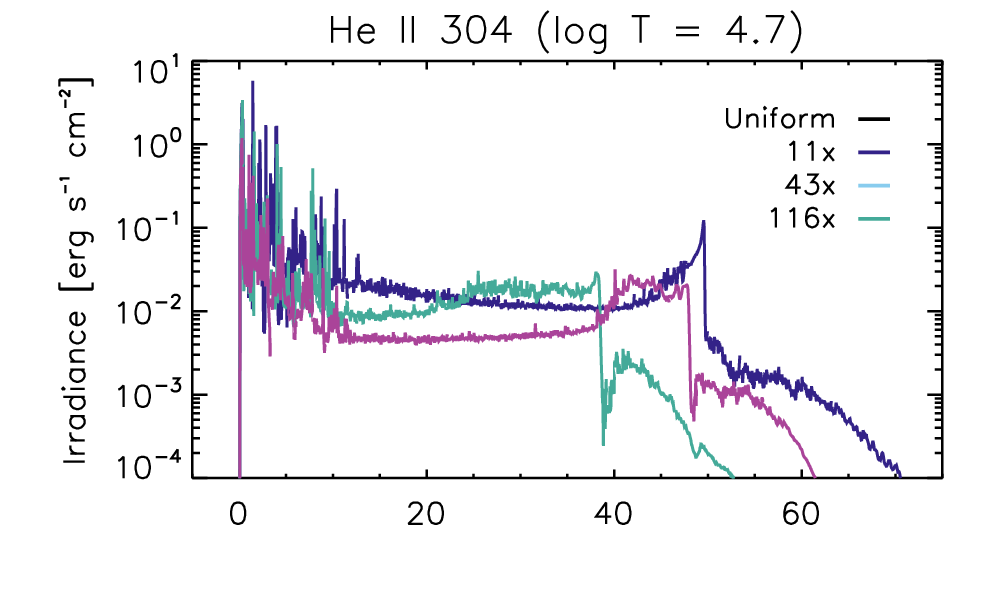}
    \includegraphics[width=0.32\textwidth]{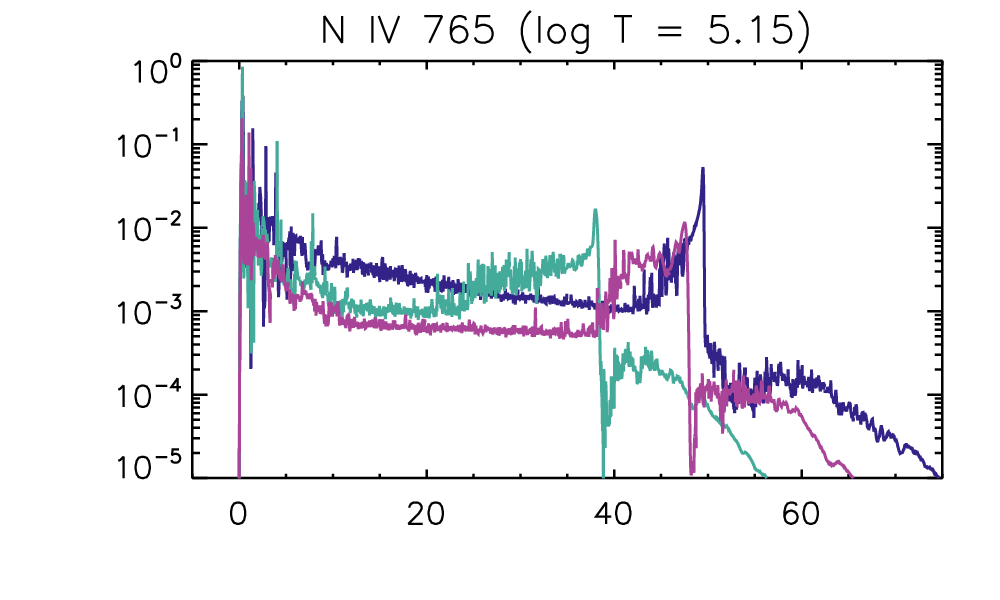}
    \includegraphics[width=0.32\textwidth]{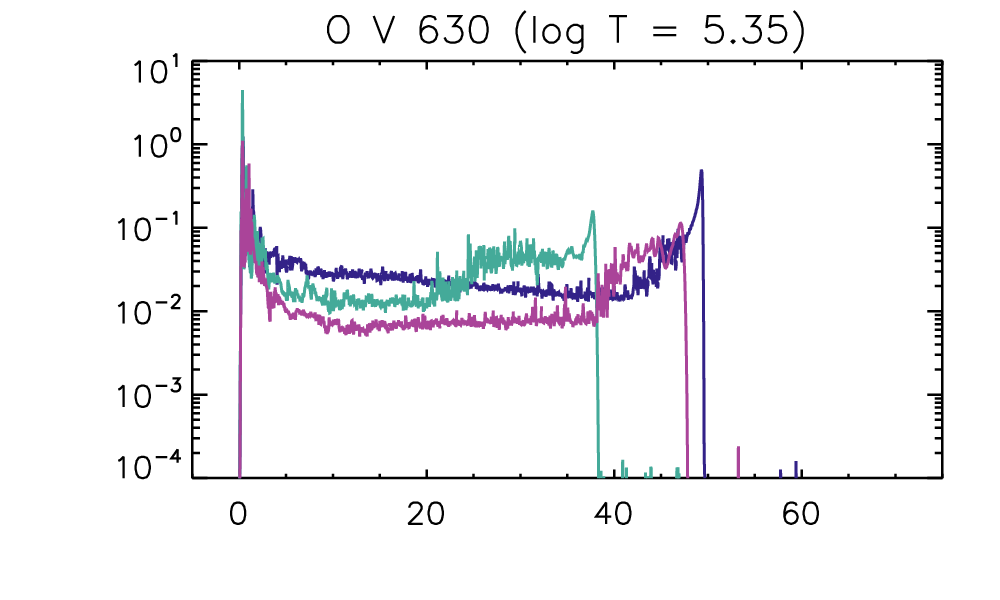}
    \includegraphics[width=0.32\textwidth]{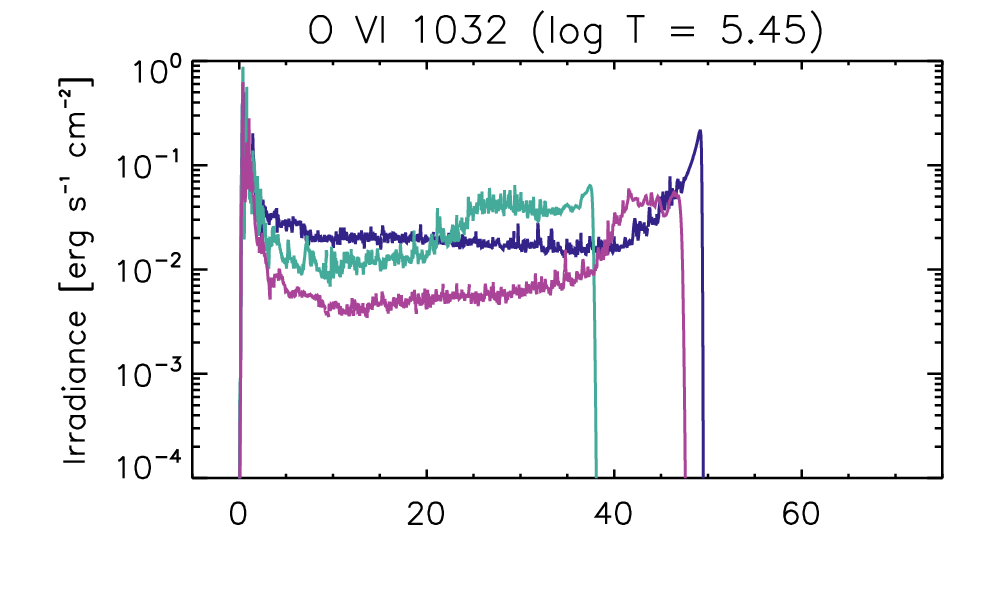}
    \includegraphics[width=0.32\textwidth]{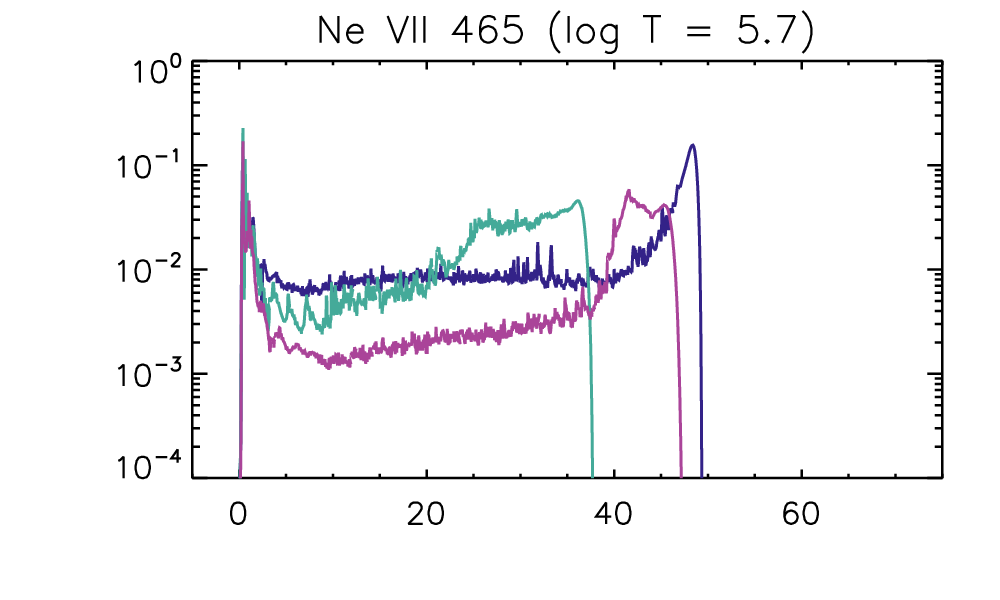}
    \includegraphics[width=0.32\textwidth]{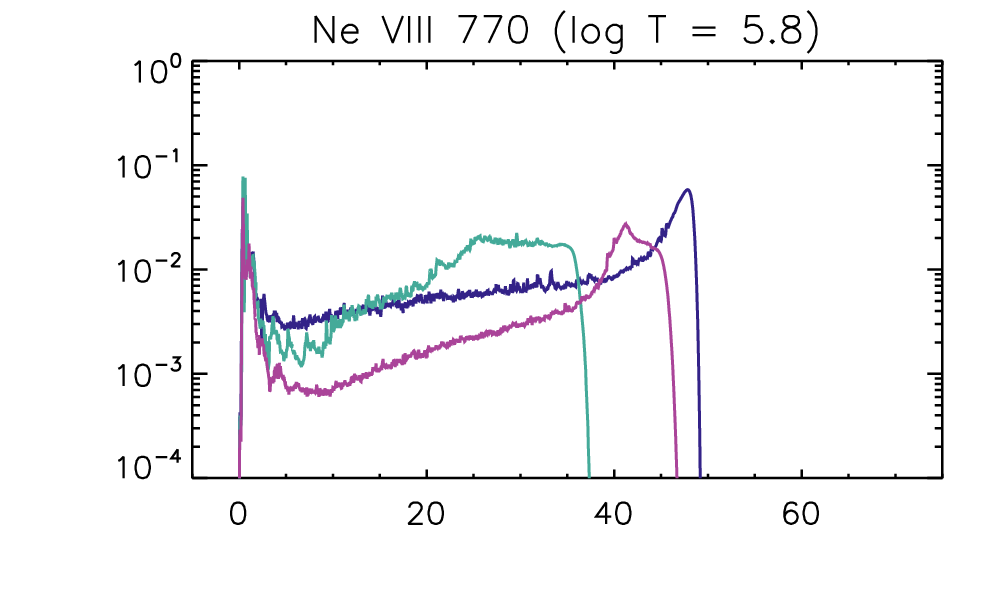}
    \includegraphics[width=0.32\textwidth]{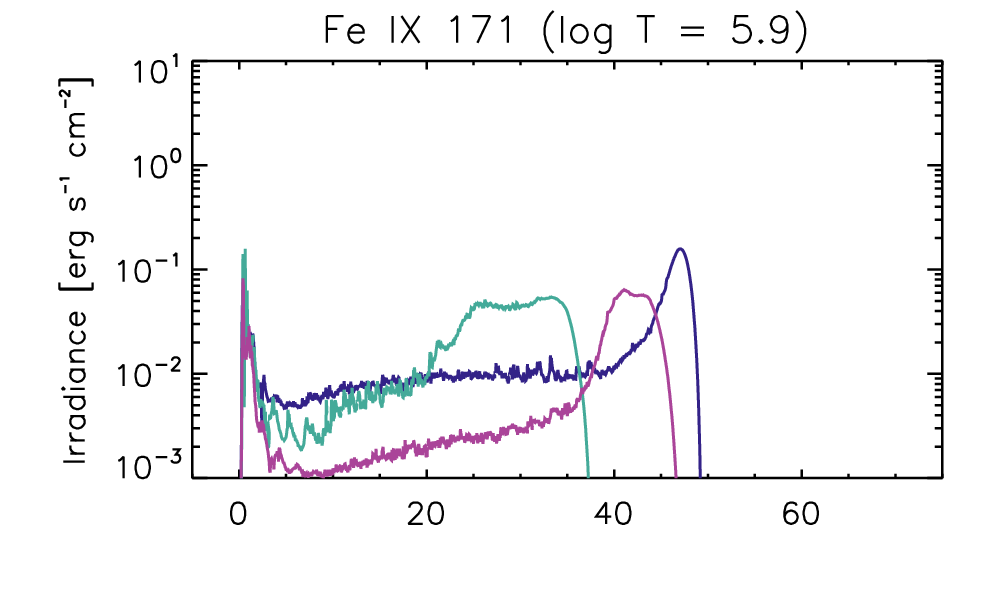}
    \includegraphics[width=0.32\textwidth]{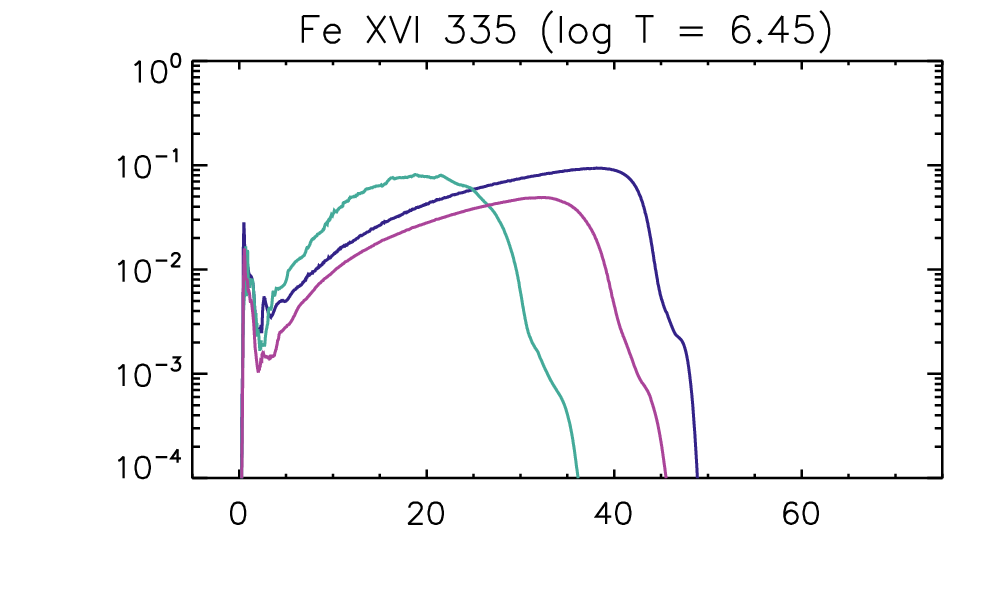}
    \includegraphics[width=0.32\textwidth]{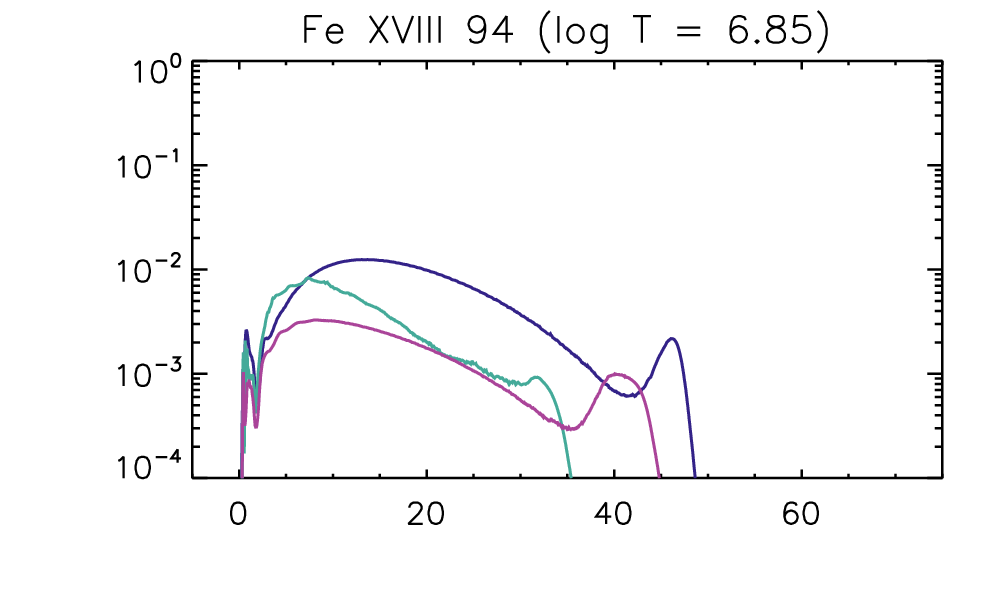}
    \includegraphics[width=0.32\textwidth]{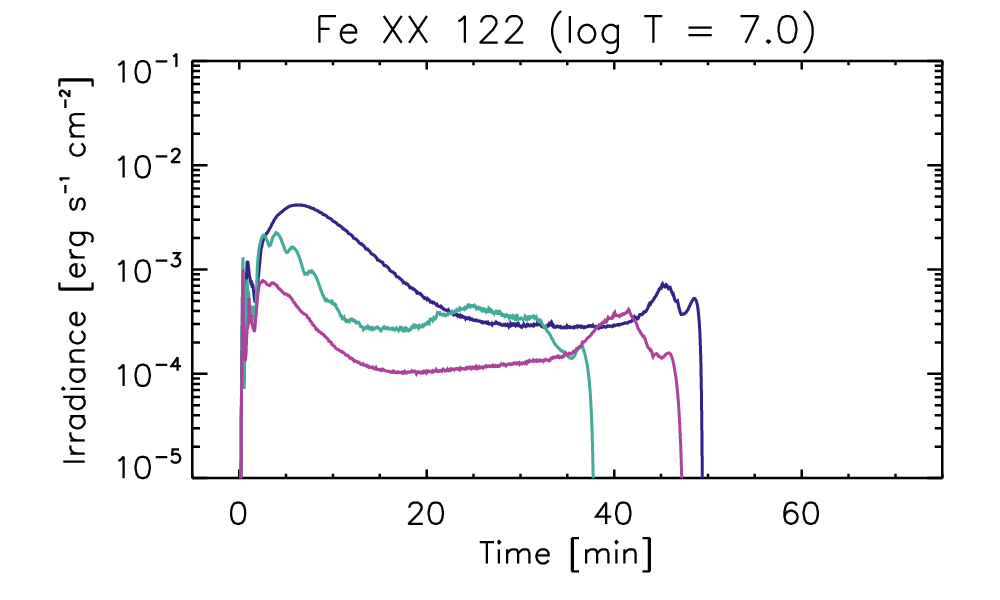}
    \includegraphics[width=0.32\textwidth]{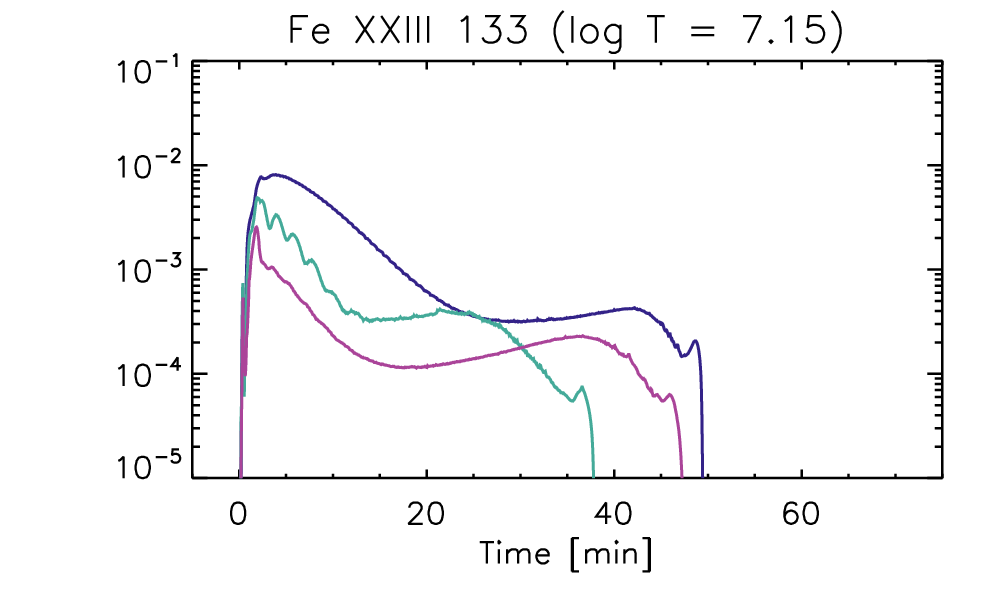}
    \includegraphics[width=0.32\textwidth]{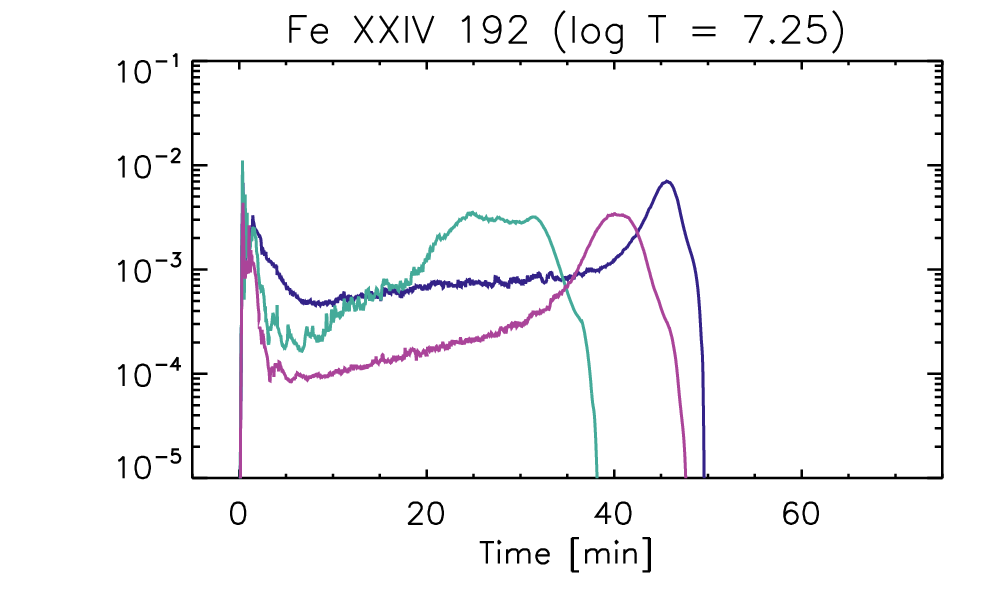}
    \caption{Similar to Figure \ref{fig:area_irr}, comparing three cases, gradual expansion (black), along with footpoint localized expansions with $s_{0} = 2.5$ Mm and $\sigma = 5$ Mm (blue), and $s_{0} = 1.5$ Mm and $\sigma = 3$ Mm (red).  The peak intensities are somewhat brighter in the gradual case because it has a higher density.  The cool lines brighten much later in the gradual expansion case because of the increased cooling times.  \label{fig:TR_irr}}
\end{figure*}

Finally, we once again synthesize the irradiance as might be seen by SDO/EVE for each case to understand how the dynamics affect the radiative output.  In Figure \ref{fig:TR_irr}, we show the synthesized irradiance for the same set of 12 lines, ranging in formation temperature from $\log T = 4.7$ to 7.25.  We once again normalize the total volume of each simulation.  Since the cooling time is reduced when the expansion occurs near the footpoint, the coolest lines peak in intensity earlier in the localized expansion cases.  The hot lines peak at similar times, but are somewhat brighter in the gradual expansion case as the density is higher (compare the apex densities in Fig. \ref{fig:TR_comp}).  

\section{Elliptical Loops}
\label{sec:elliptical}

In this section, we examine the assumption of semi-circular loops.  The gravitational acceleration parallel to the field line is given generally by:
\begin{equation}
    g_{\parallel} = g_{\sun}  \Big(\frac{R_{\sun}}{R_{\sun} + h \sin \theta} \Big)^{2} \cos \theta
\end{equation}
\noindent where $h$ is the height and $\theta$ is the angle relative to the center of the loop above the surface.  We derive the implementation of this for the more general case in Appendix \ref{app:gravity}, and then use that in simulations to understand how it affects the dynamics.

\subsection{Comparison of Elliptical Loops}

We examine a heating rate typical of solar flare simulations.  We assume that the loop is heated by an electron beam for 100\,s, with a peak heating rate of $10^{10.3}$ erg s$^{-1}$ cm$^{-2}$, low energy cut-off of $15$ keV, and spectral index $\delta$ of 5.  

\begin{figure*}
\centering
\includegraphics[width=0.32\linewidth]{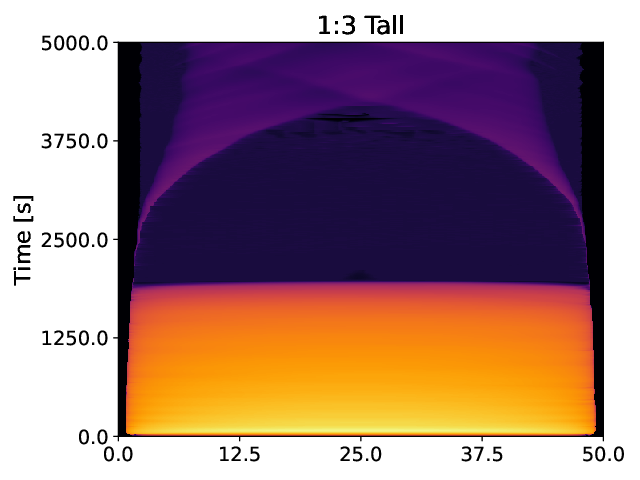}
\includegraphics[width=0.32\linewidth]{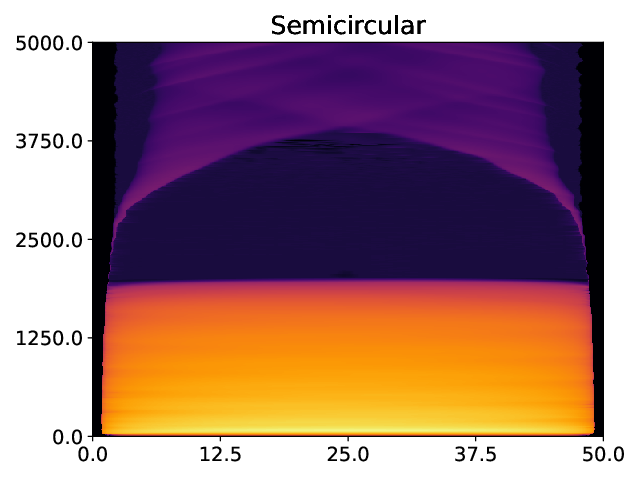}
\includegraphics[width=0.32\linewidth]{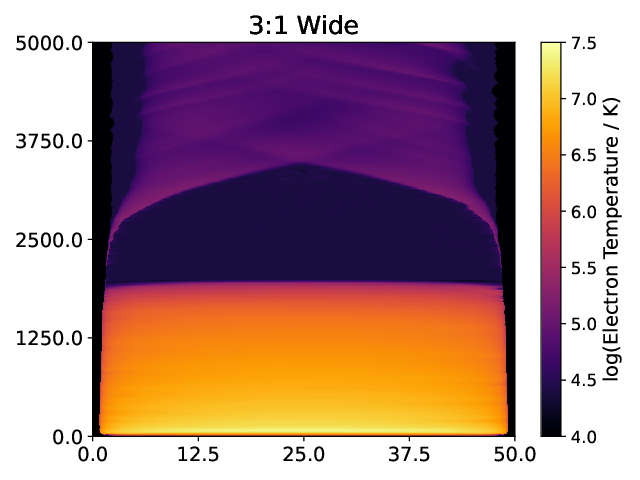}
\includegraphics[width=0.32\linewidth]{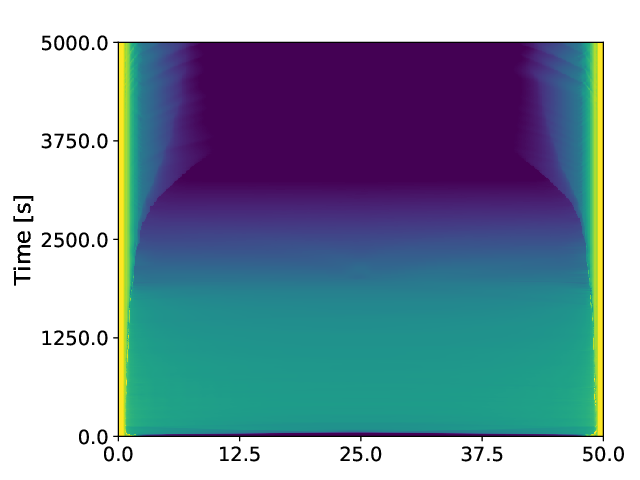}
\includegraphics[width=0.32\linewidth]{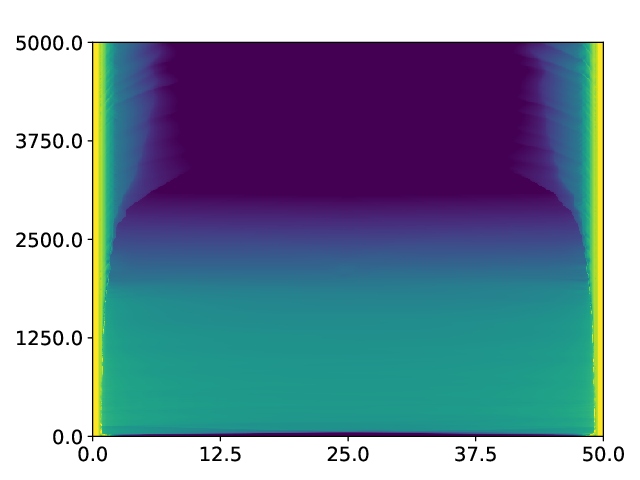}
\includegraphics[width=0.32\linewidth]{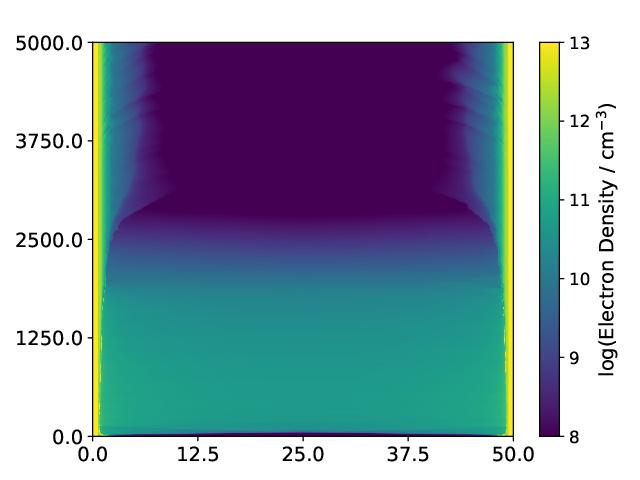}
\includegraphics[width=0.32\linewidth]{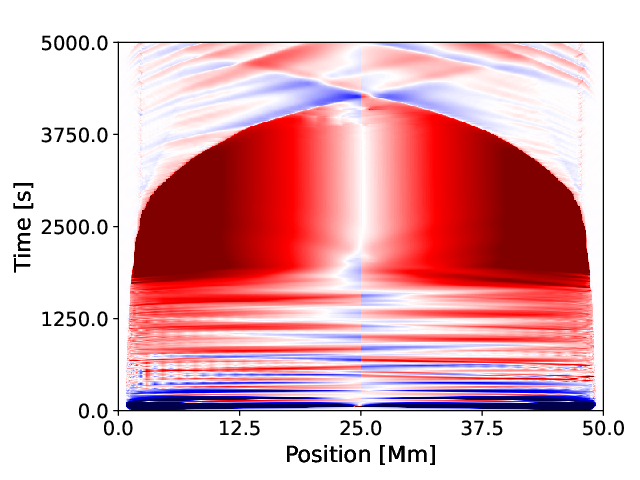}
\includegraphics[width=0.32\linewidth]{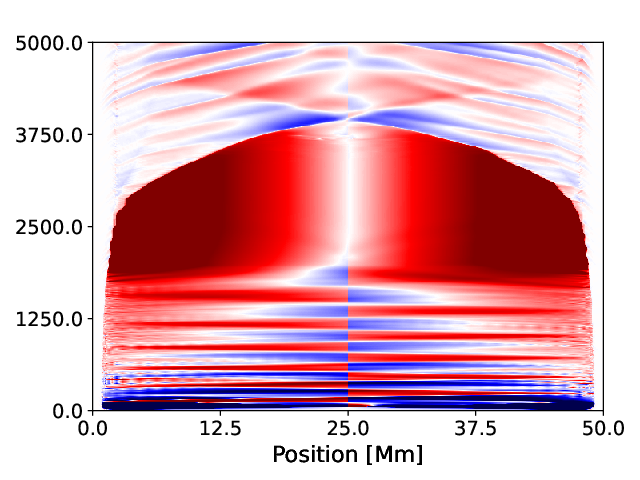}
\includegraphics[width=0.32\linewidth]{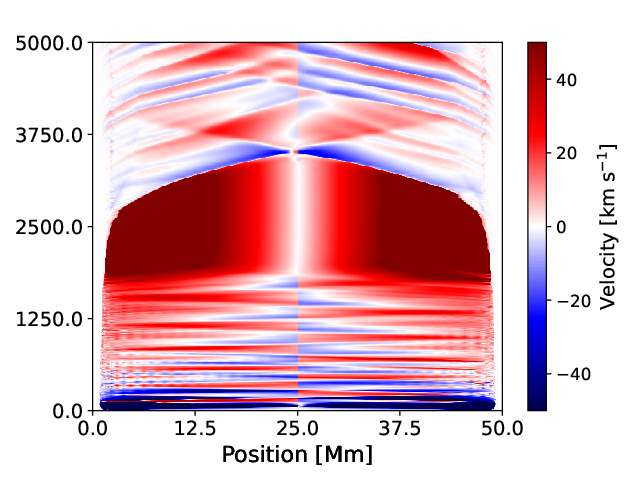}
\includegraphics[width=0.48\linewidth]{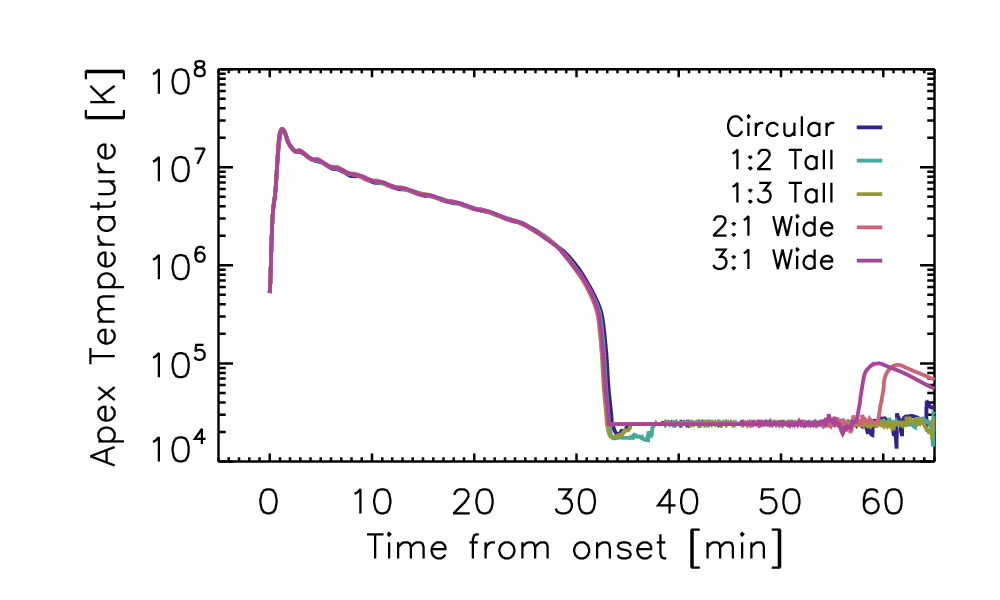}
\includegraphics[width=0.48\linewidth]{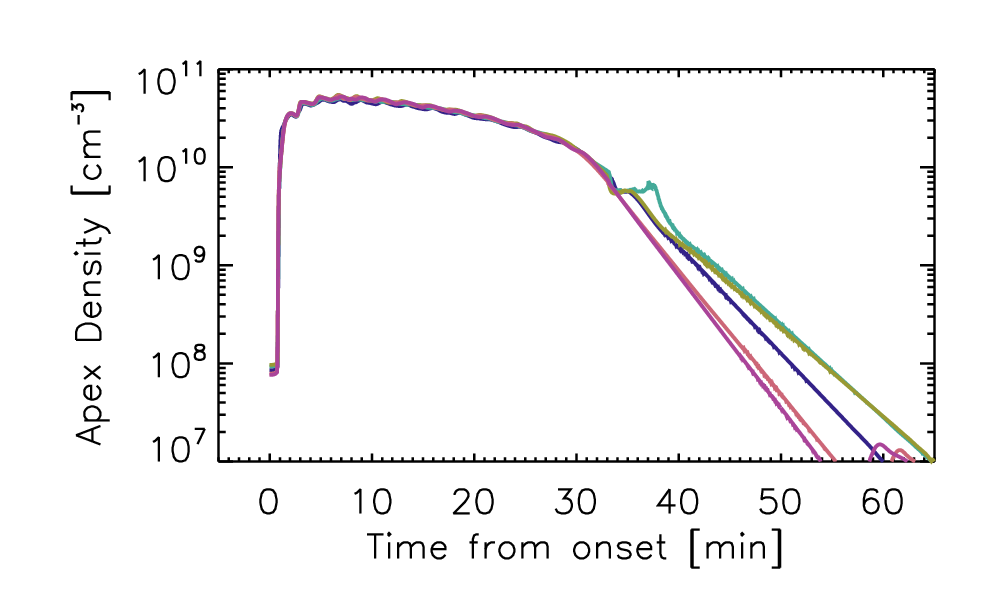}
\caption{The overall hydrodynamic evolution in a set of elliptical 50 Mm flaring loops with uniform area.  The left column shows a tall loop (1:3), the center a semicircular loop (1:1), and the right a wide loop (3:1).  The bottom row shows a comparison of the apex temperature and density in the three cases.  The draining time of the loops increases slightly with height, but the evolution is otherwise remarkably similar.  Ellipticity does not strongly impact the hydrodynamics in this case.  \label{fig:ell_hydro} } 
\end{figure*}

Figure \ref{fig:ell_hydro} shows a comparison of the hydrodynamic evolution of three elliptical loops.  The simulation parameters are all the same except for the gravitational acceleration.  The left column shows a tall loop, where the semi-major axis is oriented radially outwards from the solar surface and 3 times larger than the semi-minor axis.  The center column shows a semicircular loop.  The right column shows a wide loop, where the semi-major axis is oriented parallel to the solar surface, and also 3 times larger than the semi-minor axis.  The first three rows show the electron temperature, electron density, and bulk flow velocity along the loop (x-axis) as it evolves in time (y-axis).  Red flows indicate motion away from the apex and blue indicates motion towards the apex.  The bottom row shows a comparison of the electron temperatures and densities at the loop apex (including wide and tall loops with $a = 2b$).  The overall evolution of the temperatures, densities, and velocities are nearly identical in the three cases.  The only major difference is that the draining time increases somewhat with height, as might be expected for a slightly weaker gravitational acceleration. 

\begin{figure*}
    \centering
    \includegraphics[width=0.32\textwidth]{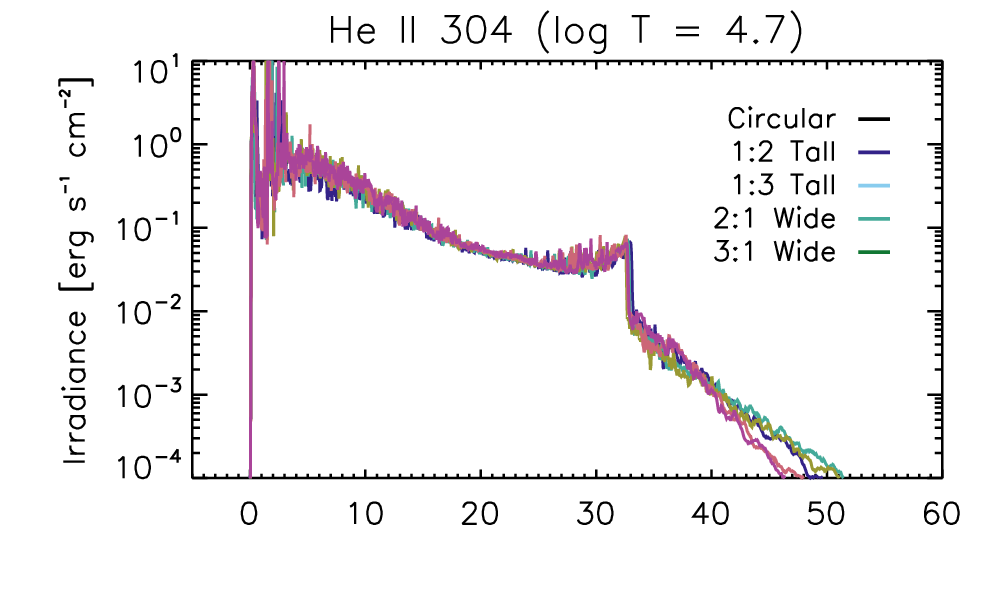}
    \includegraphics[width=0.32\textwidth]{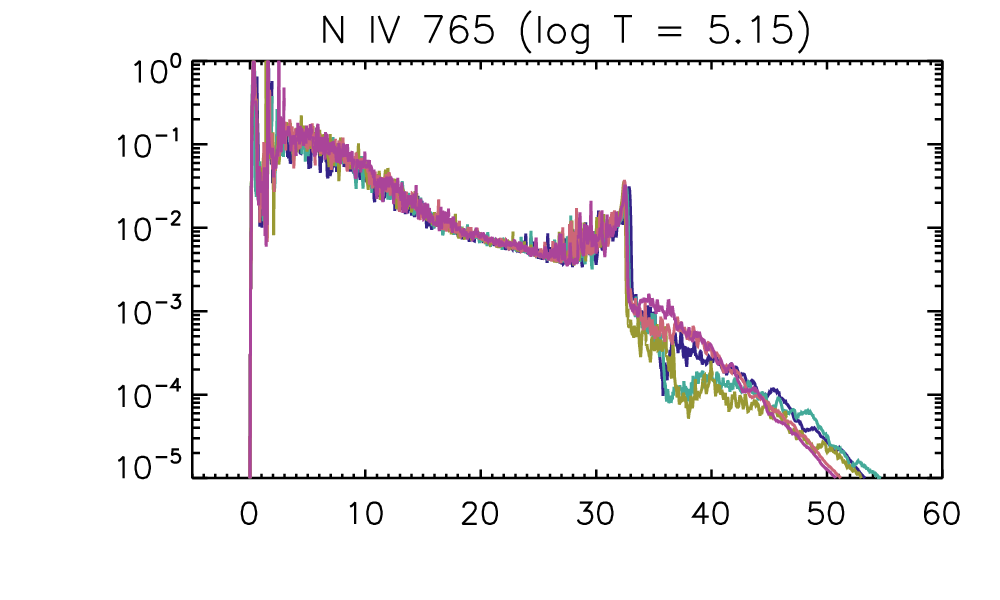}
    \includegraphics[width=0.32\textwidth]{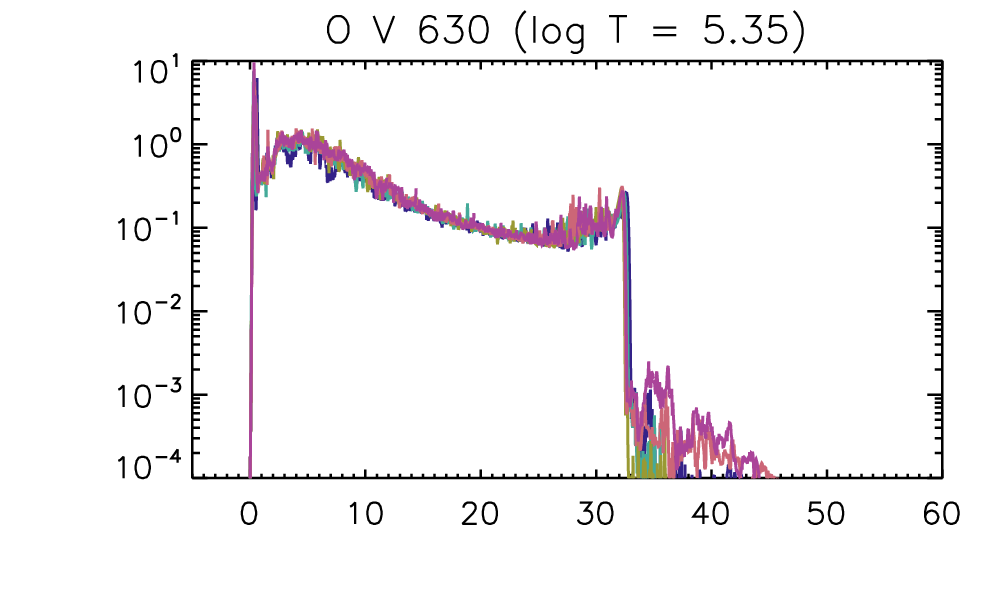}
    \includegraphics[width=0.32\textwidth]{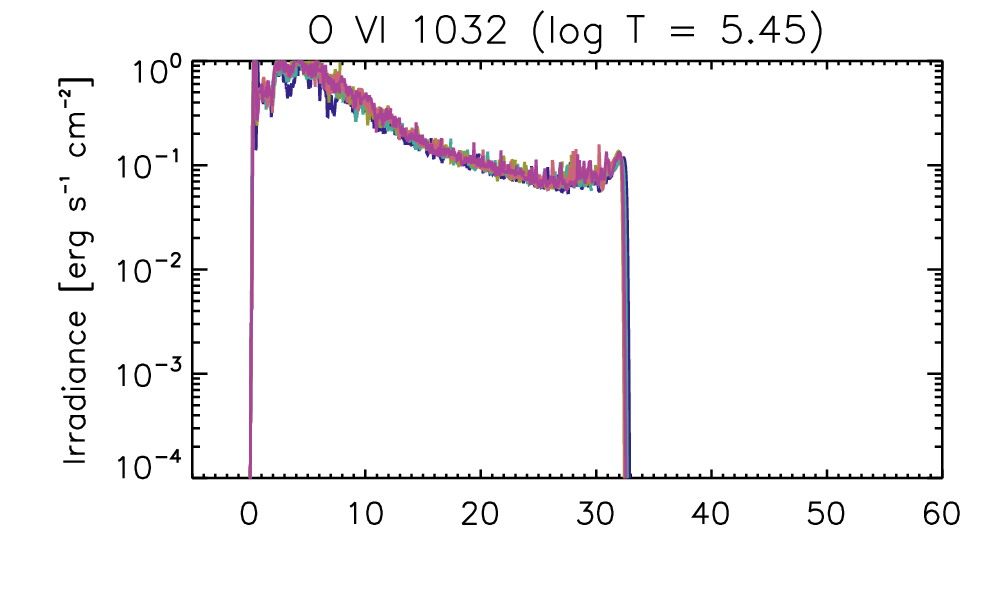}
    \includegraphics[width=0.32\textwidth]{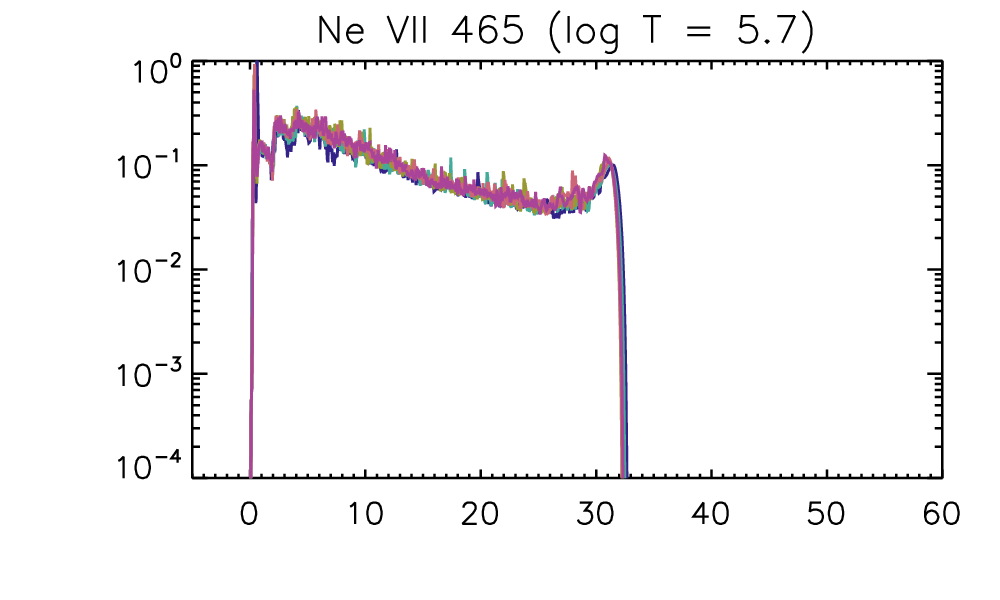}
    \includegraphics[width=0.32\textwidth]{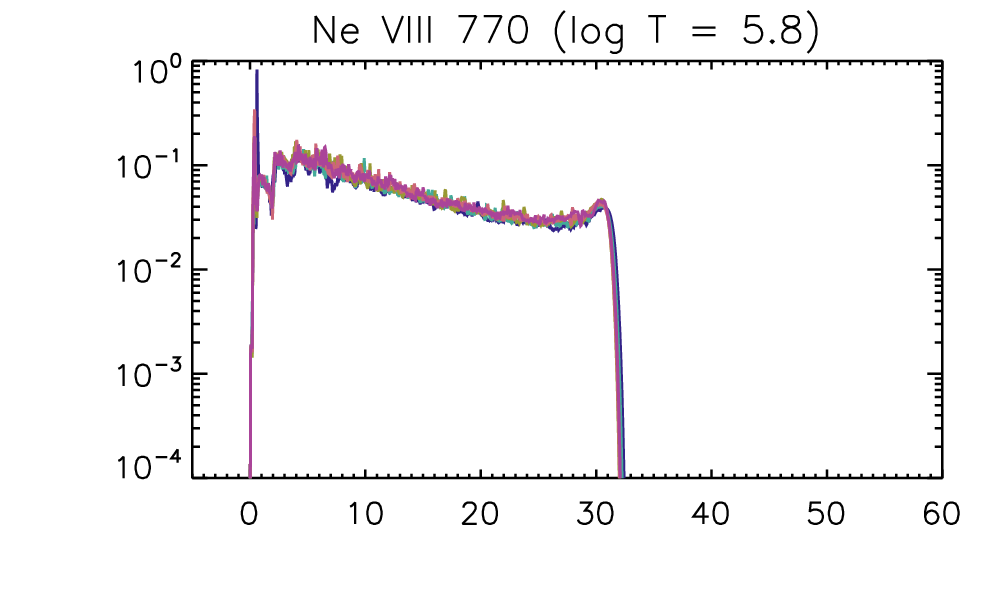}
    \includegraphics[width=0.32\textwidth]{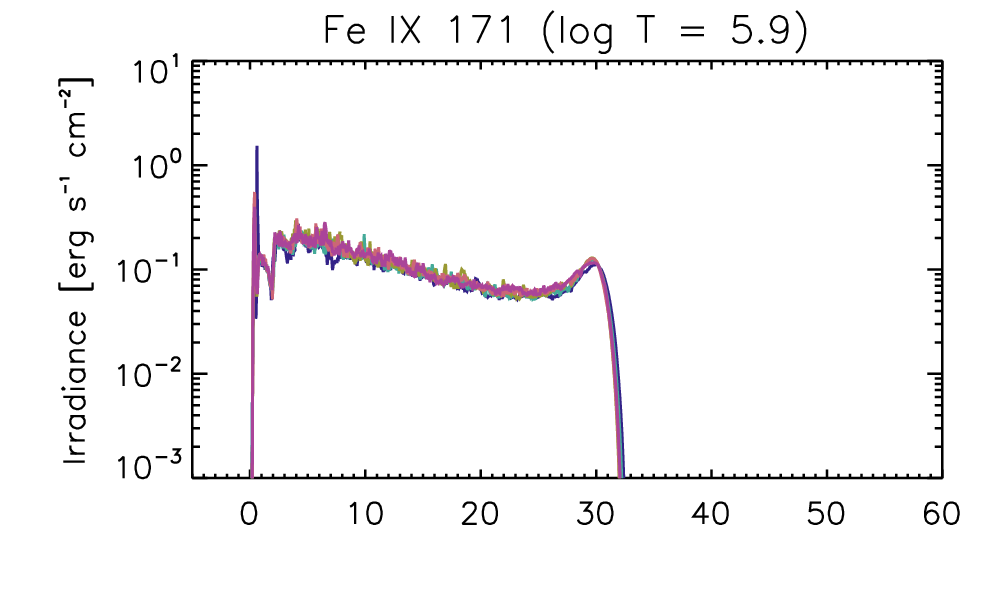}
    \includegraphics[width=0.32\textwidth]{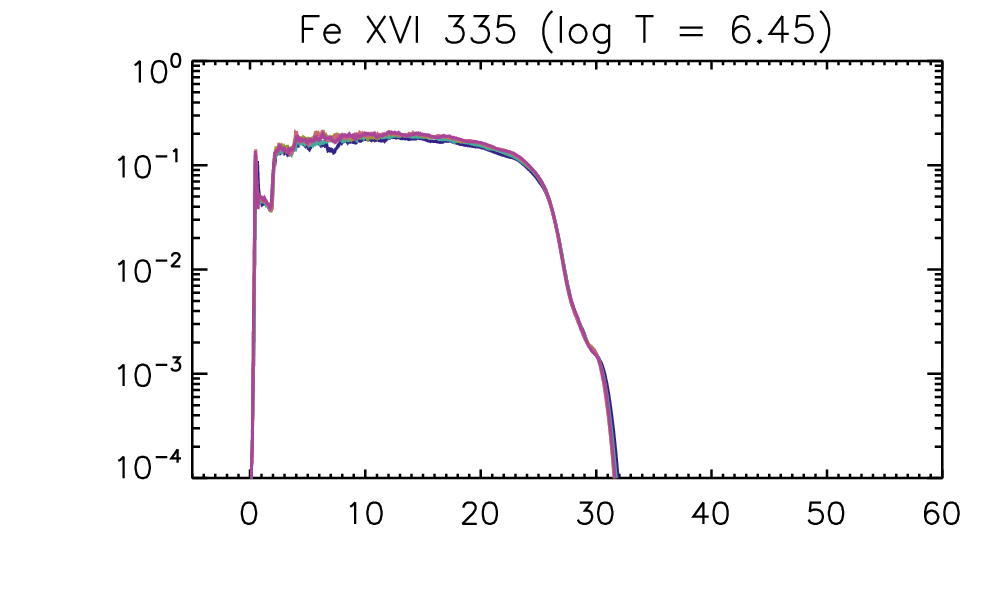}
    \includegraphics[width=0.32\textwidth]{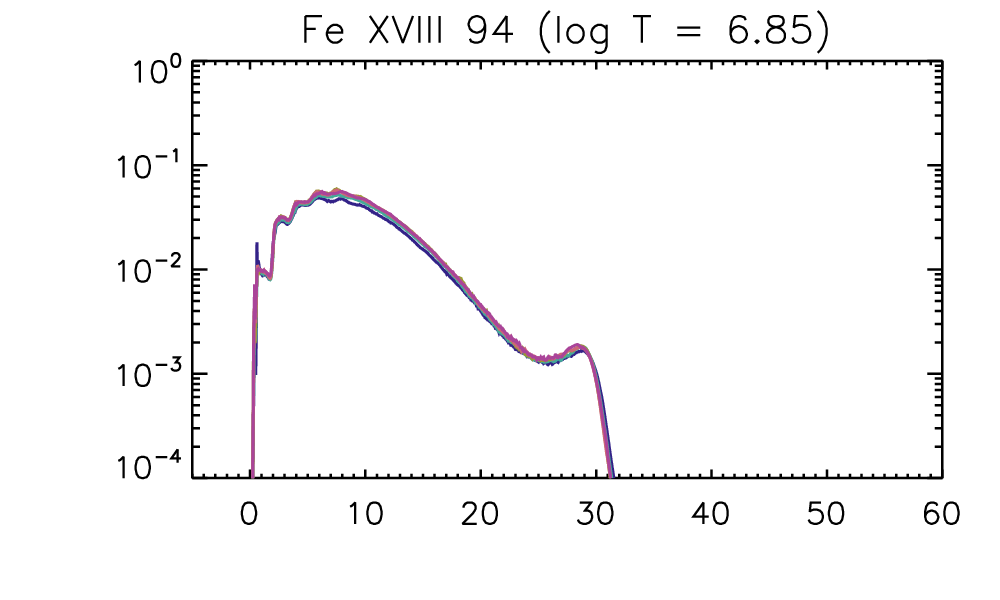}
    \includegraphics[width=0.32\textwidth]{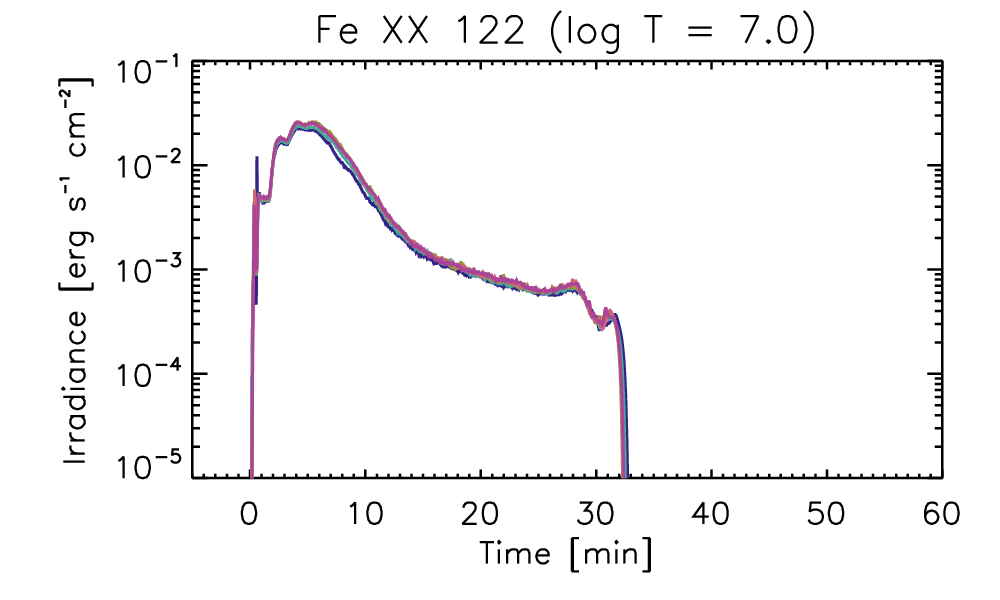}
    \includegraphics[width=0.32\textwidth]{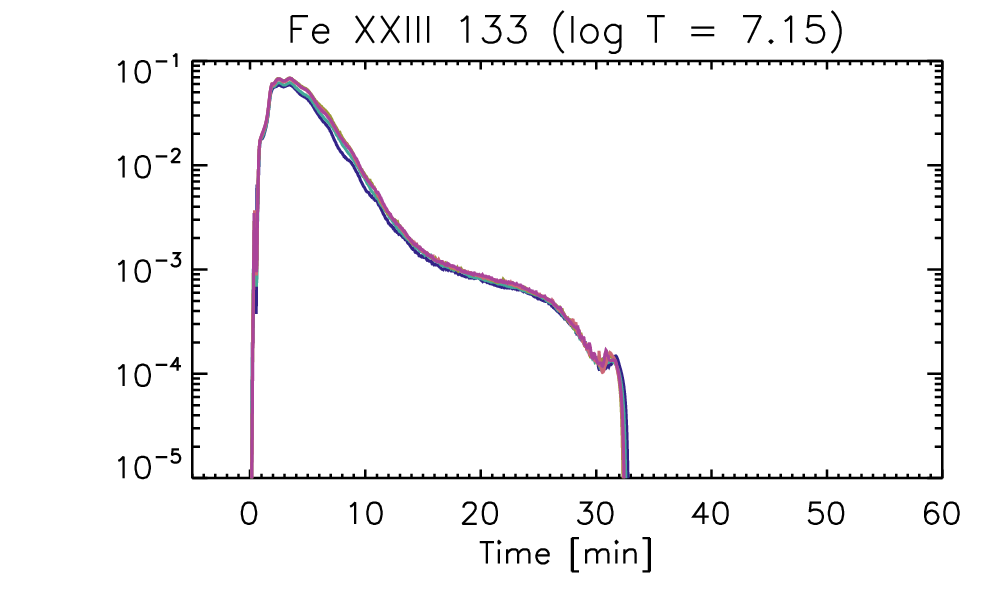}
    \includegraphics[width=0.32\textwidth]{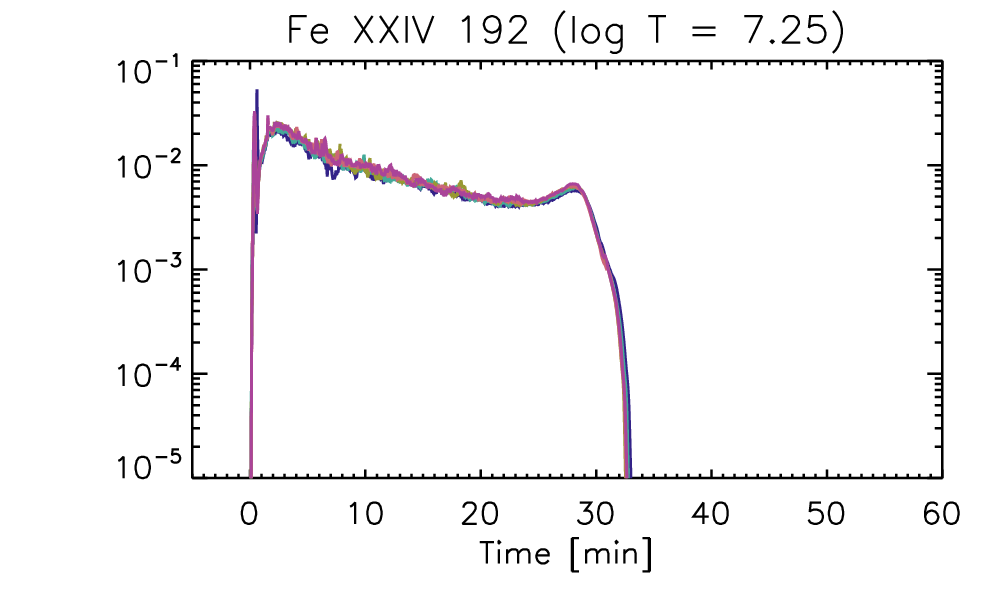}
    \caption{The synthetic irradiance for 12 lines as might be seen by SDO/EVE.  The loops were assumed elliptical, as annotated, with a uniform cross-sectional area.  The ellipticity of the loops has a negligible effect on the line intensities.  \label{fig:ell_irr}}
\end{figure*}

In Figure \ref{fig:ell_irr}, we also show the synthetic irradiance for 12 spectral lines, ranging in formation temperature from about 50 kK to 20 MK, as might be observed by SDO/EVE.  We show five cases with different ellipticities, for both tall and wide loops, corresponding to the gravitational acceleration profiles in Figure \ref{fig:gravity}.  Unsurprisingly, since the hydrodynamic evolution is not largely impacted, the line intensities are also not strongly impacted, once again indicating that the ellipticity is relatively unimportant.  

\begin{figure*}
\centering
\includegraphics[width=0.48\linewidth]{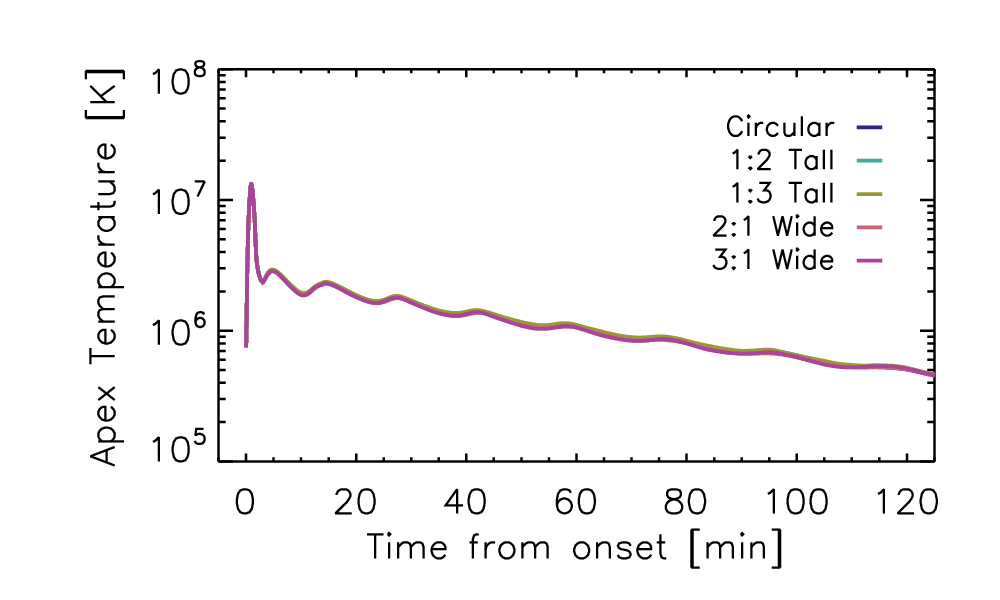}
\includegraphics[width=0.48\linewidth]{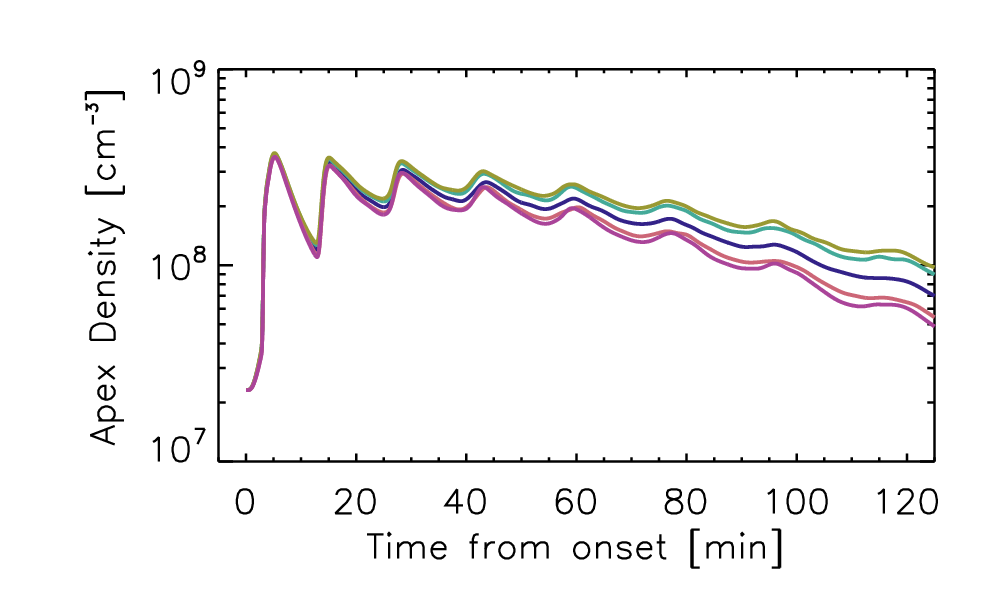}
\caption{The evolution of the apex temperatures and densities for elliptical loops of total length 150 Mm (uniform cross section), heated with a nanoflare level impulsive energy burst, lasting 100 s.  In this case, the gravitational scale height has been reduced, so there is a more noticeable divergence in the density evolution.  The effect is still relatively small, however.  \label{fig:L150} } 
\end{figure*}

Since these loops were short and hot, however, the gravitational scale height is large compared to the loops' heights.  To further check the importance of ellipticity, we therefore have additionally run five simulations of elliptical loops with total length 150 Mm, using a nanoflare strength heating: $\approx$ 0.7 erg s$^{-1}$ cm$^{-3}$, uniform over the loop, for 100 s, with a 50 s rise and 50 s decay.  In this case, the scale height will be reduced, so we expect that there should be more divergence in the evolution compared to the previous case.  Figure \ref{fig:L150} shows the evolution of the apex temperatures and densities in these five loops.  During the heating, and shortly thereafter, the five evolve identically, but as the loops cool and drain, the densities do begin to diverge.  The temperatures evolve similarly in all cases.  The divergence is still somewhat small, however.

\section{Elliptical, Expanding Loops}
\label{sec:ell_exp}

In general, we expect that coronal loops are both elliptical with expanding areas.  In this section, we combine the two geometrical effects to examine how this may affect the dynamics and emission.  Using the same elliptical loops in the previous section, we now assume that the cross-sectional area is not uniform.  Instead, we make the assumption that the magnetic field decreases with height above the solar surface $r$ as $B \propto \frac{1}{r^{2}}$, and since $A \propto \frac{1}{B}$, we have $A \propto r^{2}$.  We use a maximal expansion of precisely 10 from footpoint to apex in the circular loop.  Compared to the profiles used in Figure \ref{fig:area_comp}, this expansion profile is more gradual with distance along the loop.
\begin{figure*}
\centering
\includegraphics[width=0.48\linewidth]{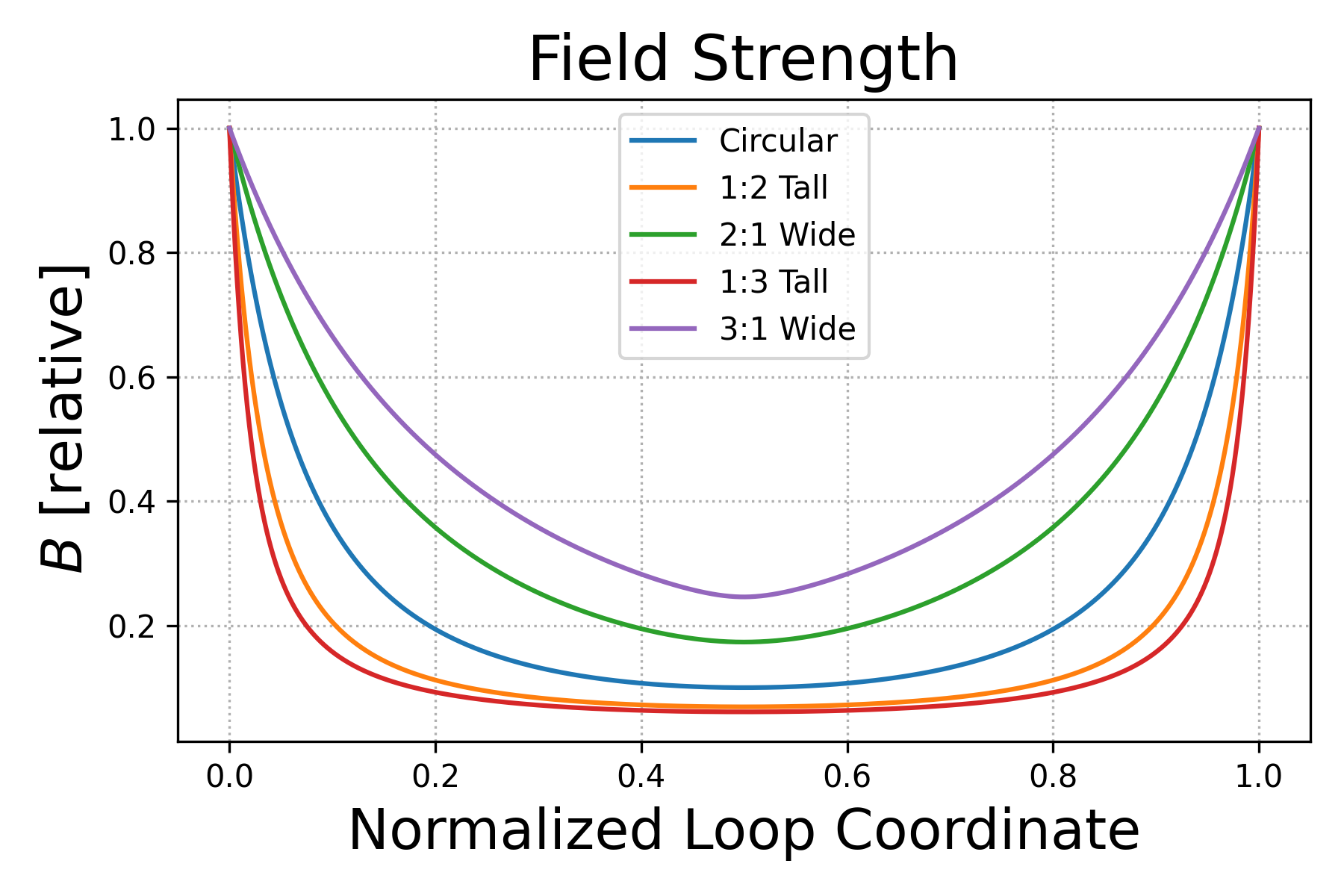}
\includegraphics[width=0.48\linewidth]{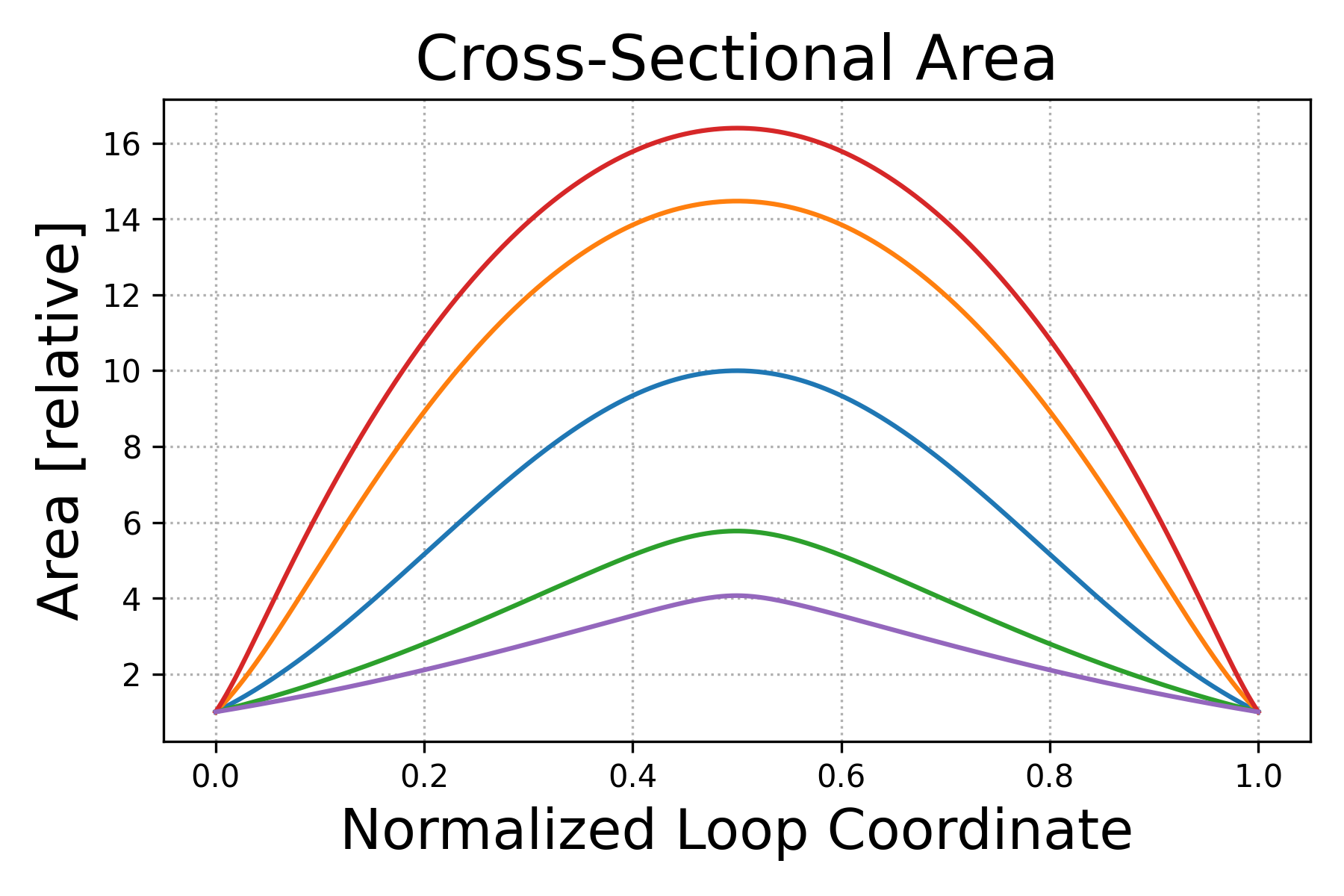}
\caption{The magnetic field variation, $B \propto \frac{1}{r^{2}}$, (left) and corresponding cross-sectional area (right) in the five loops examined in this section.  The circular case (blue) has an expansion of precisely 10 from footpoint to apex, while the taller loops (orange and red) have a larger expansion and wider loops (green and purple) a smaller expansion.  \label{fig:ee_area}} 
\end{figure*}

\begin{figure*}
\centering
\includegraphics[width=0.32\linewidth]{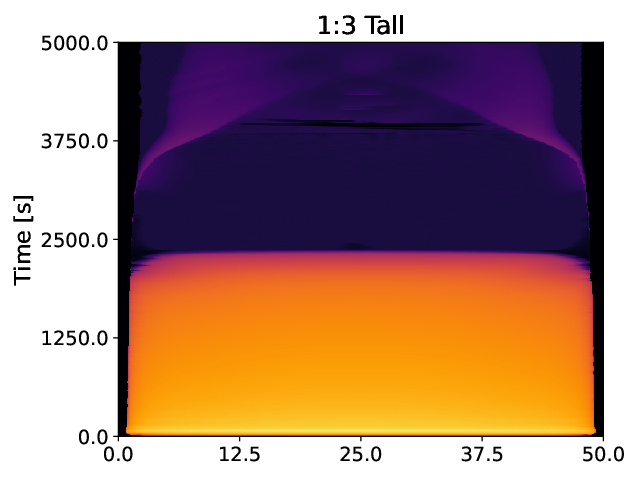}
\includegraphics[width=0.32\linewidth]{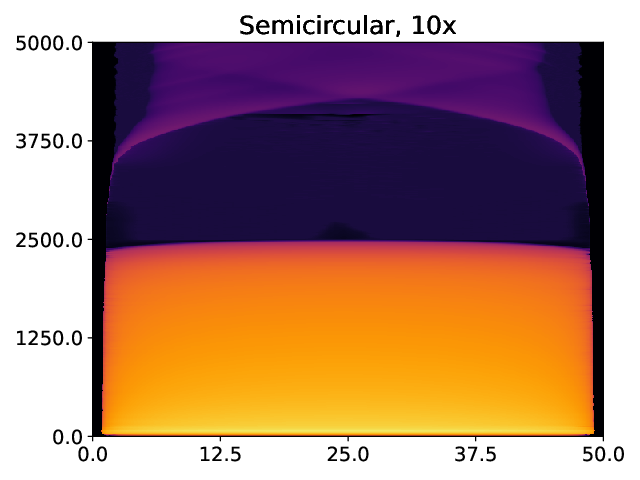}
\includegraphics[width=0.32\linewidth]{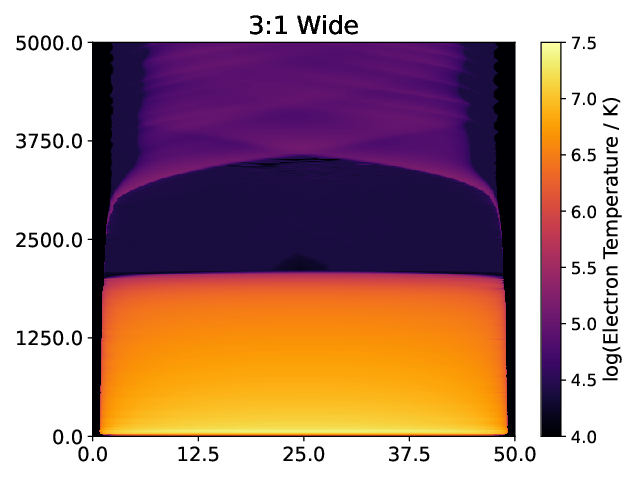}
\includegraphics[width=0.32\linewidth]{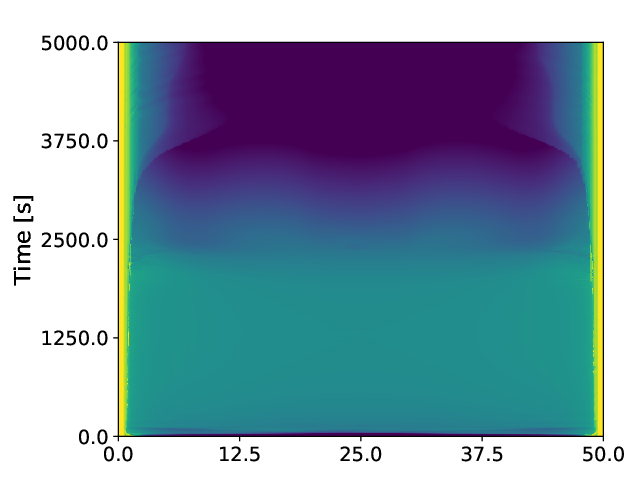}
\includegraphics[width=0.32\linewidth]{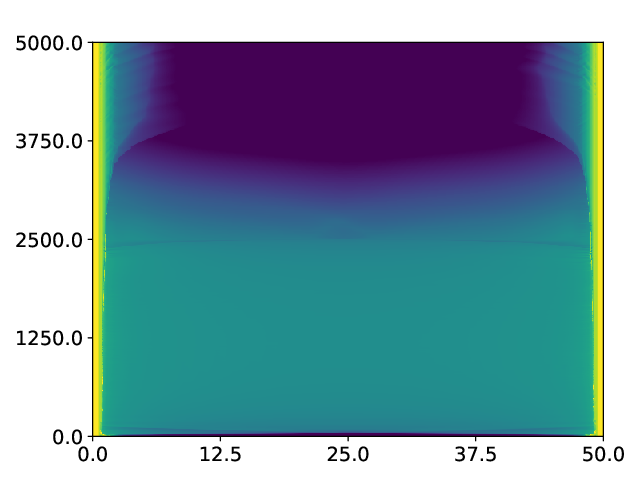}
\includegraphics[width=0.32\linewidth]{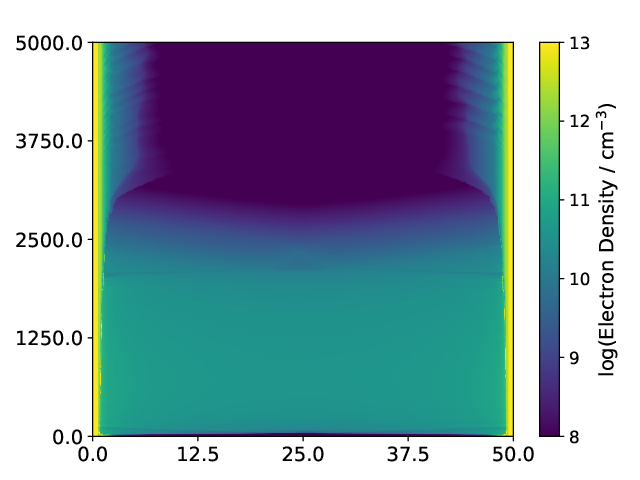}
\includegraphics[width=0.32\linewidth]{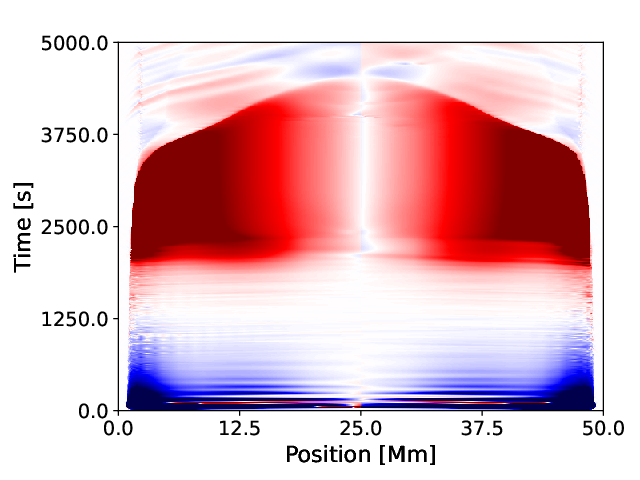}
\includegraphics[width=0.32\linewidth]{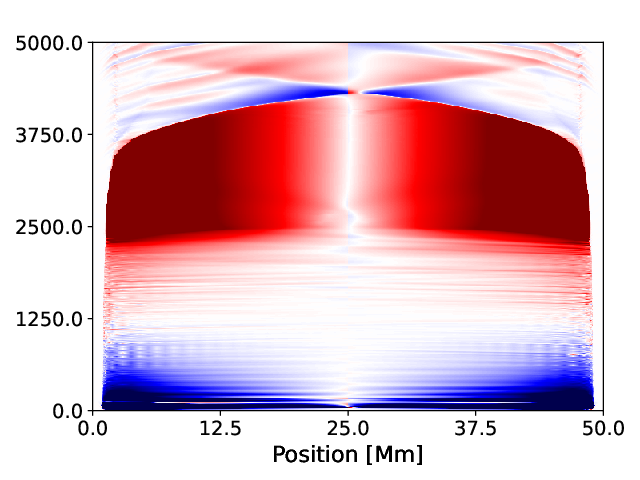}
\includegraphics[width=0.32\linewidth]{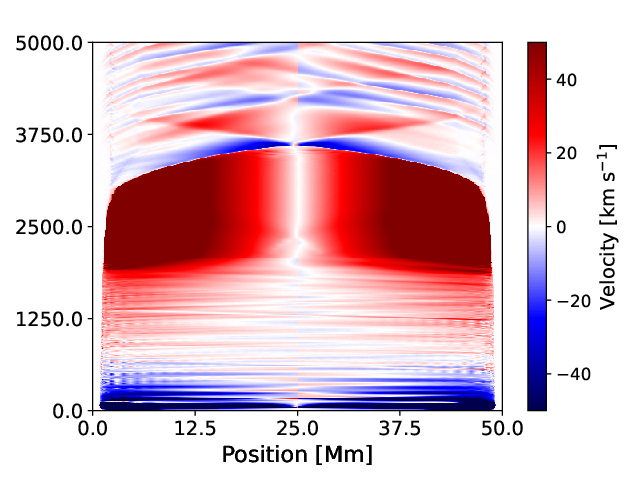}
\includegraphics[width=0.48\linewidth]{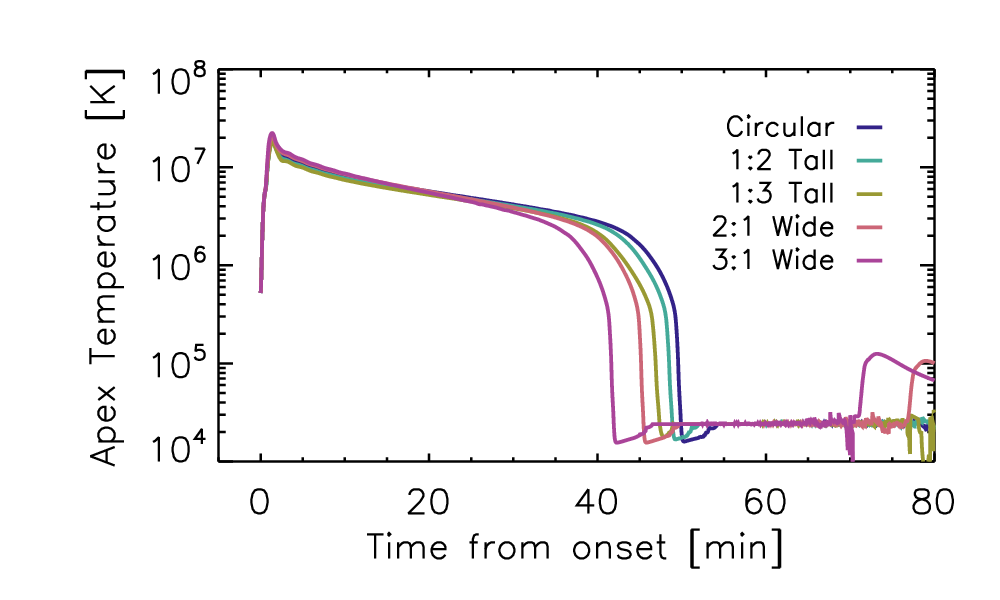}
\includegraphics[width=0.48\linewidth]{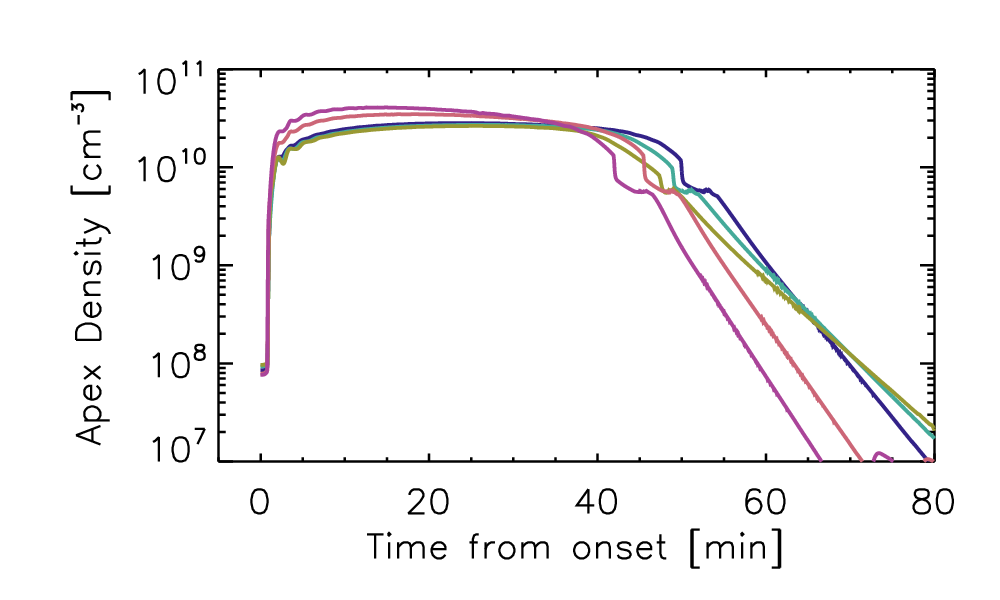}
\caption{The overall hydrodynamic evolution in a set of elliptical flaring loops with expanding area, corresponding to the gravity and area profiles in Figures \ref{fig:gravity} and \ref{fig:ee_area}.  The left column shows a tall loop (1:3), the center a circular loop (1:1), and the right a wide loop (3:1).  The wide loop drains and cools the fastest, while the circular loop cools the slowest and the tall loop drains the slowest.  The evaporative up-flows last longer in the tall case, and are also more confined near the footpoints, consistent with it having a somewhat larger area expansion. \label{fig:ee_hydro}} 
\end{figure*}

We examine simulations using both the gravity profiles in Figure \ref{fig:gravity} and area profiles in \ref{fig:ee_area}.  Figure \ref{fig:ee_hydro} shows the hydrodynamic evolution of these loops.  The left, center, and right columns show a tall loop (3:1), circular loop (1:1), and wide loop (1:3), respectively, while the bottom two plots show the evolution of the apex temperatures and densities.  Although the tall loop has the largest area expansion, the circular case has the longest cooling time.  Compared to the results in Section \ref{sec:area}, this indicates that the rate of expansion $\frac{dA}{ds}$ also affects the cooling time (see also \citealt{cargill2022}), since we have already seen that the ellipticity does not impact the cooling time.  The velocity profiles are also noticeably affected in these loops.  In taller loops, the speed of the evaporative upflows dampens with height quickly.  As the area gradually expands with height, the speed decreases due to the conservation of mass (Equation \ref{eqn:mass}), and since the taller loops expand more quickly, this effect is more pronounced than it is in shorter loops.  

\begin{figure*}
    \centering
    \includegraphics[width=0.32\textwidth]{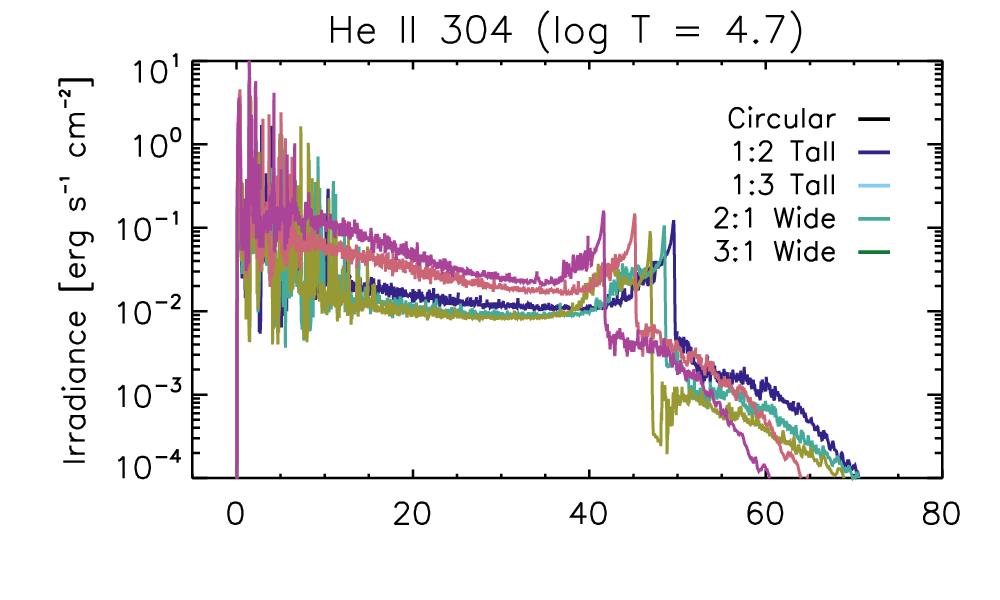}
    \includegraphics[width=0.32\textwidth]{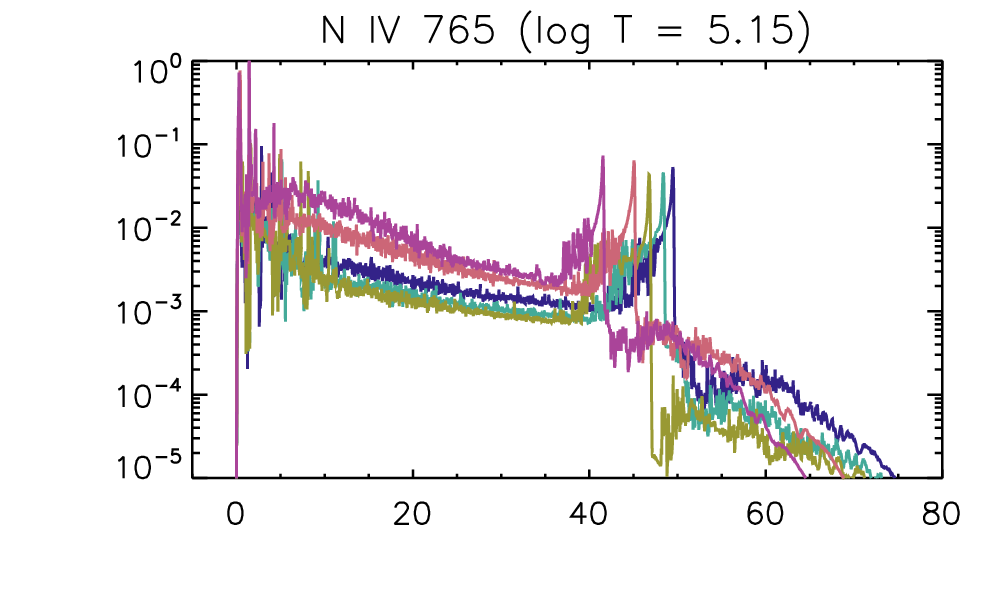}
    \includegraphics[width=0.32\textwidth]{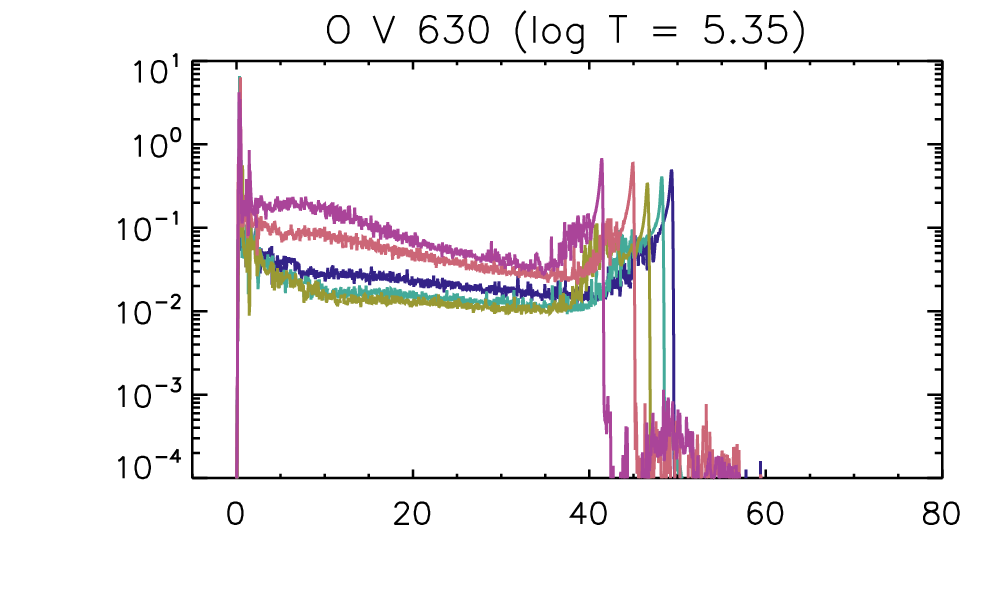}
    \includegraphics[width=0.32\textwidth]{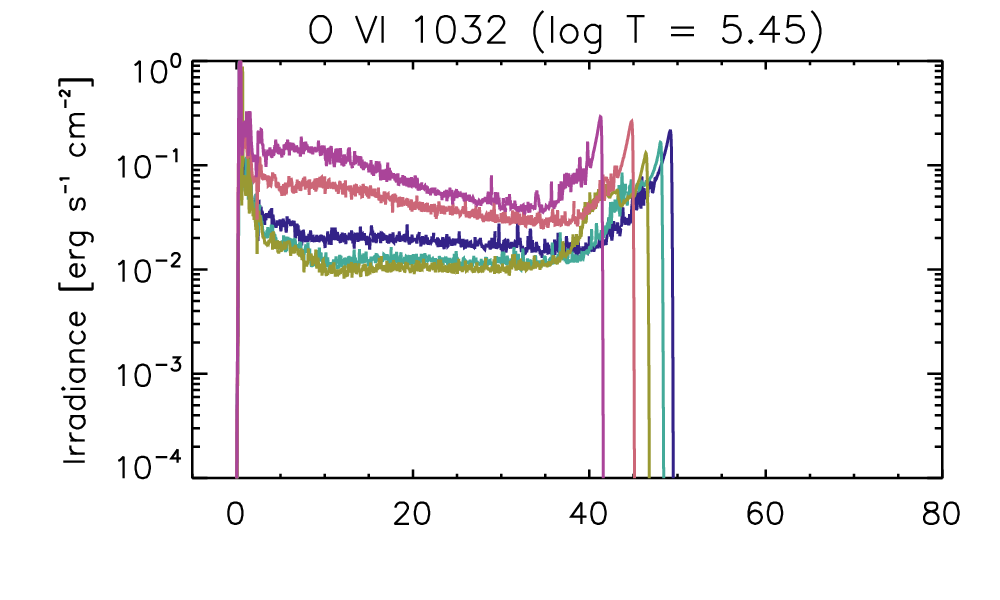}
    \includegraphics[width=0.32\textwidth]{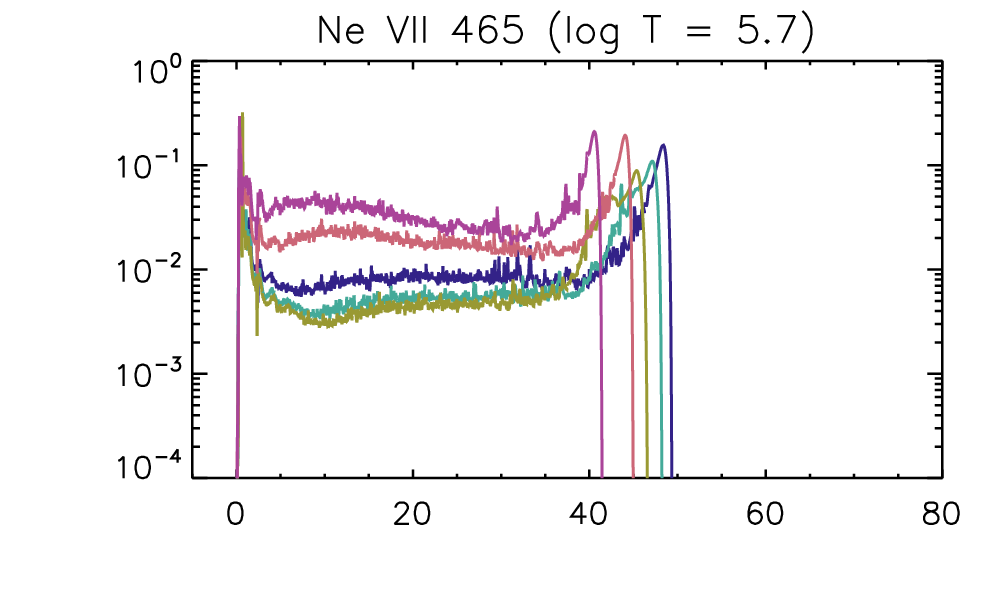}
    \includegraphics[width=0.32\textwidth]{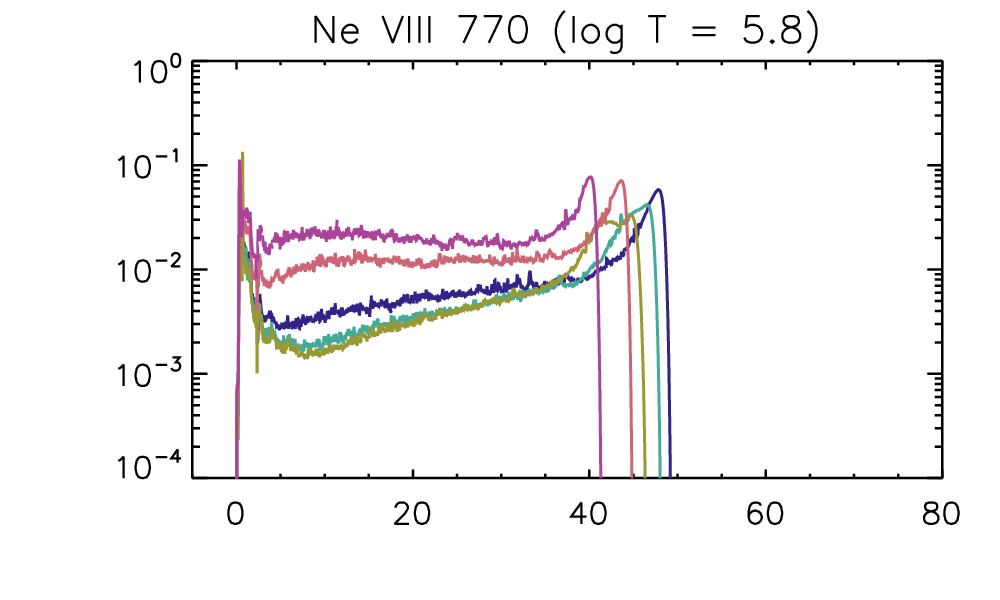}
    \includegraphics[width=0.32\textwidth]{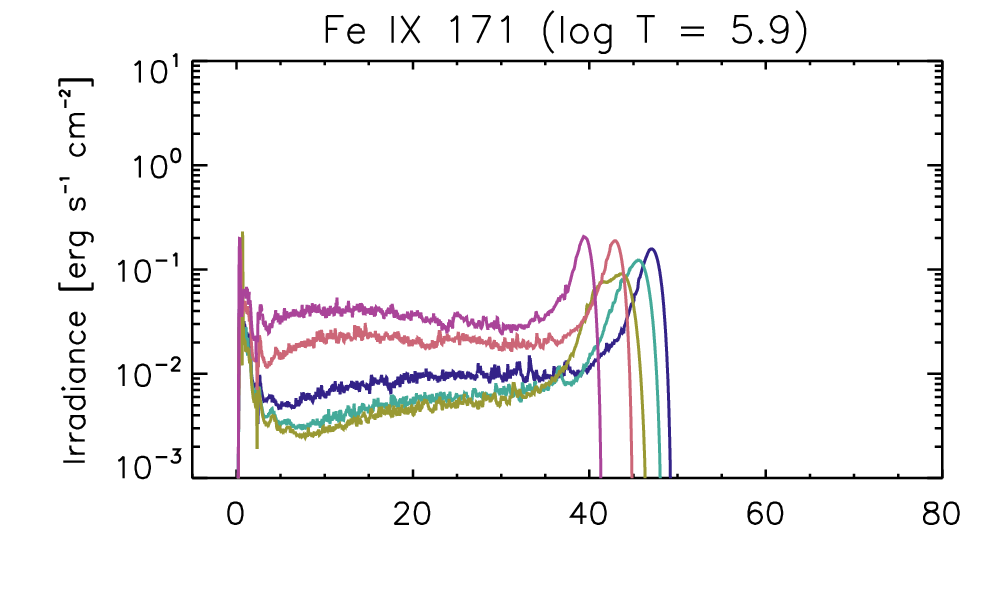}
    \includegraphics[width=0.32\textwidth]{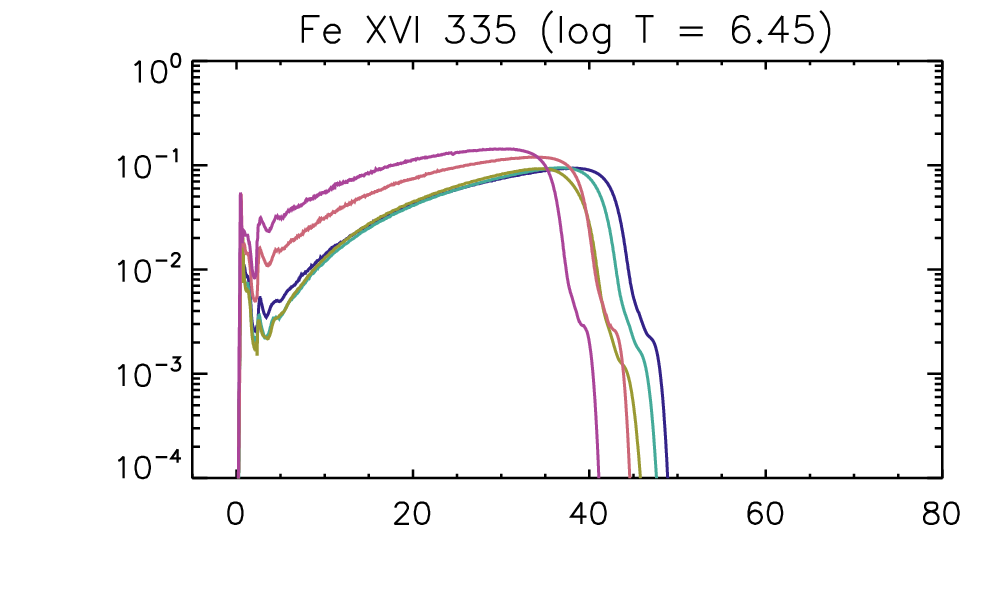}
    \includegraphics[width=0.32\textwidth]{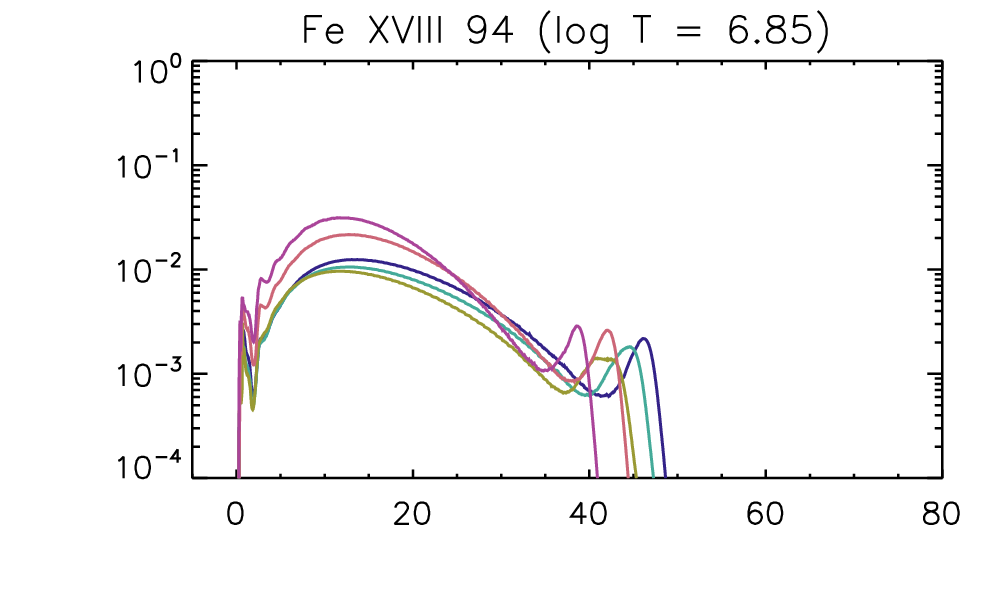}
    \includegraphics[width=0.32\textwidth]{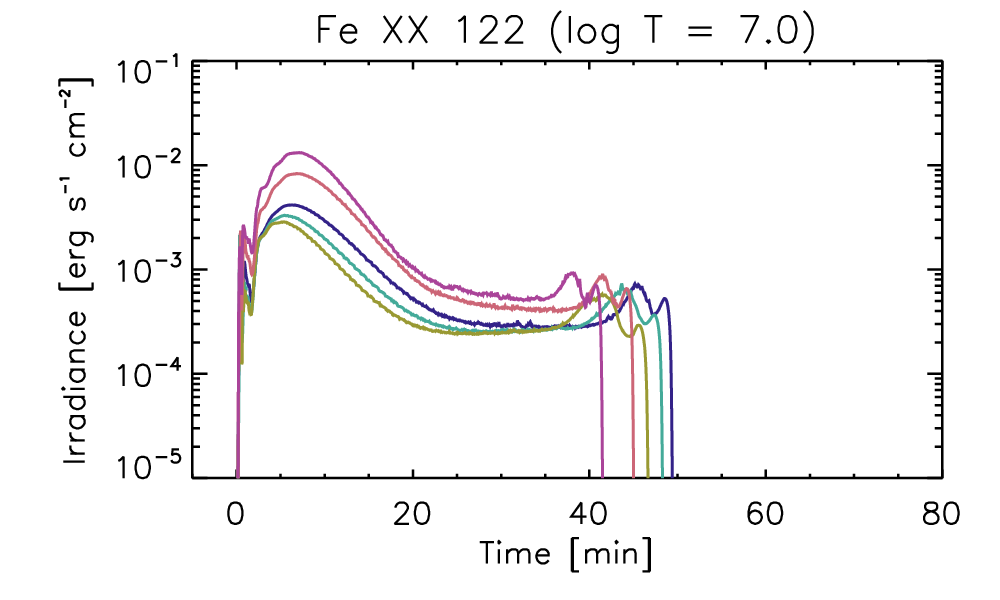}
    \includegraphics[width=0.32\textwidth]{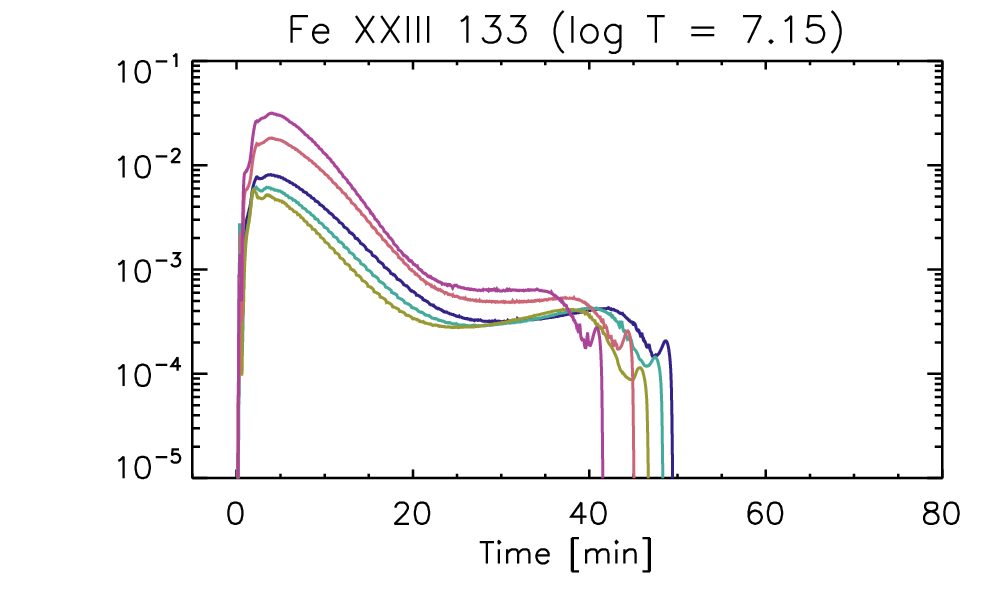}
    \includegraphics[width=0.32\textwidth]{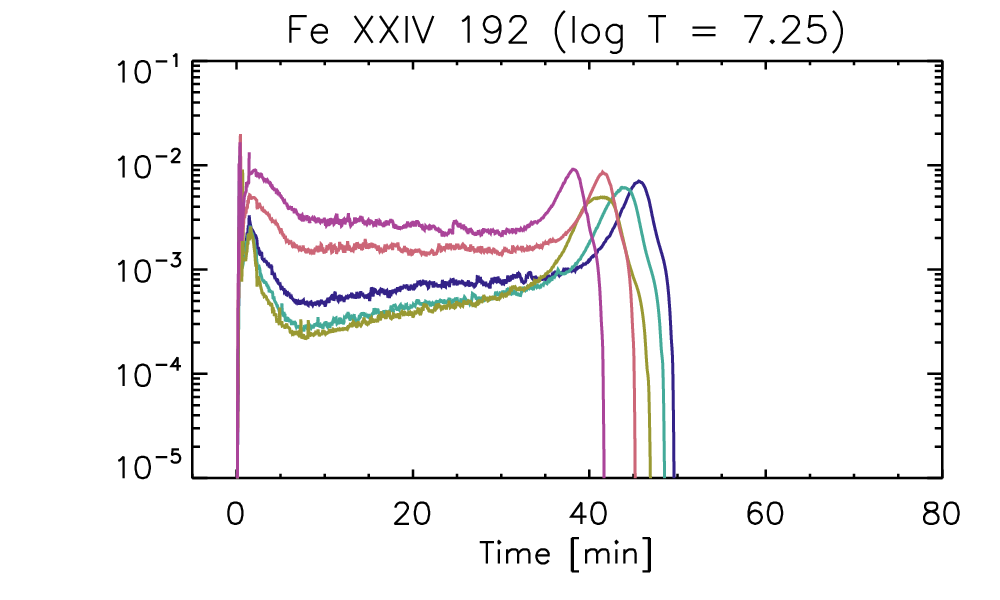}
    \caption{The synthetic irradiance for 12 spectral lines as might be seen by SDO/EVE, corresponding to the loops in Figure \ref{fig:ee_hydro}.  The cooling times are affected by the ellipticity here, and the tall loops, which have the largest volume, are typically the brightest.  The effect is relatively muted, though. \label{fig:ee_irr}}
\end{figure*}

We briefly examine the evolution of the irradiance of the spectral lines at various temperatures.  Figure \ref{fig:ee_irr} shows this comparison for the five loops.  As before, we have normalized the total volume of each loop to be equal.  In general, the tallest loop has the largest peak intensity because it has the largest coronal volume.  In the cooler lines, this peak occurs when the loop cools through the line's formation temperature.  However, before then, the short, wide loops have the highest densities (compare the apex densities in Figure \ref{fig:ee_hydro}), so they are brightest in transition region lines like \ion{O}{6} 1032\,\AA\ and \ion{Fe}{9} 171\,\AA.  In flaring lines like \ion{Fe}{23} 133\,\AA, however, short, wide loops are the brightest in peak intensity, since the intensity decreases with time as the loop cools.  

\begin{figure*}
\centering
\includegraphics[width=0.48\linewidth]{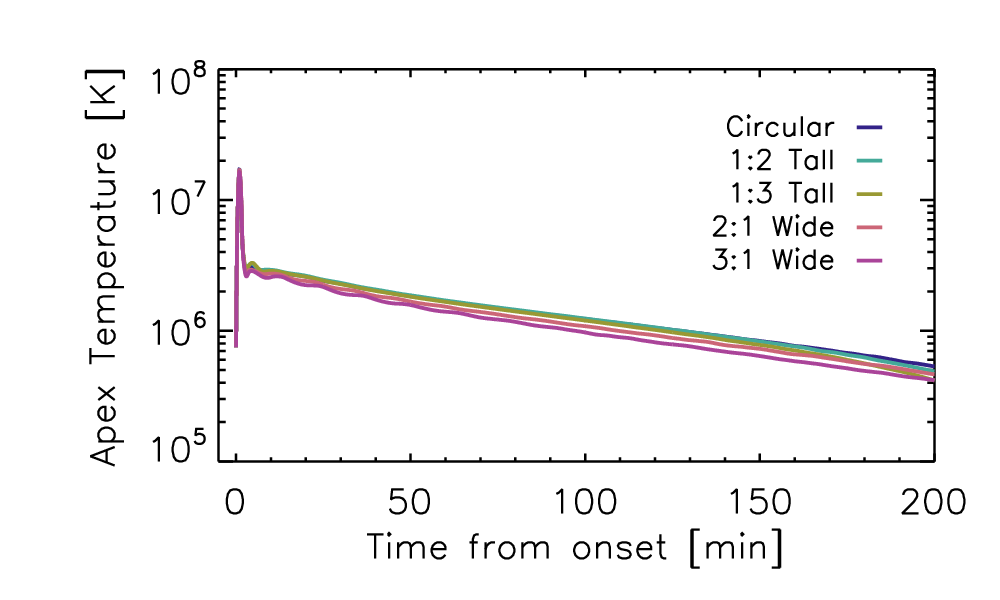}
\includegraphics[width=0.48\linewidth]{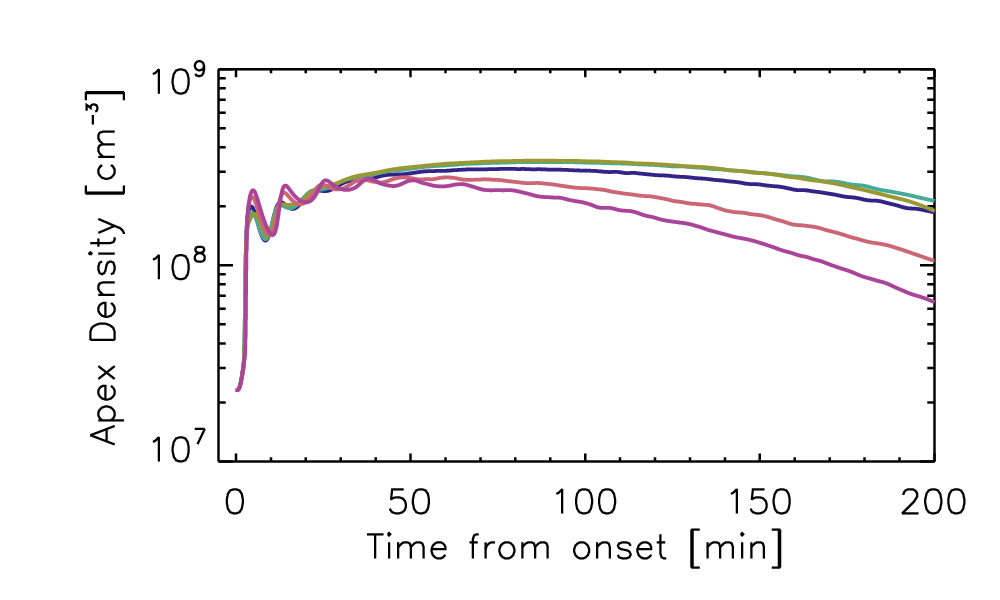}
\caption{The evolution of the apex temperatures and densities for expanding, elliptical loops of total length 150 Mm, where the circular case has an expansion of exactly 10, heated with a nanoflare level impulsive energy burst, lasting 100 s.  As with Figure \ref{fig:L150}, the scale heights are reduced, causing more noticeable divergence in the draining of the loops during the cooling phase.  The impulsive phase evolves similarly in each case.  \label{fig:L150_Rm10} } 
\end{figure*}

Once again, since we have focused on short, hot loops, we briefly examine longer loops heated with a smaller energy burst, so that the gravitational scale height is more comparable with the loop length.  In Figure \ref{fig:L150_Rm10}, we show the evolution of the apex temperatures and densities for expanding, elliptical loops of total length 150 Mm, where the circular case expands exactly by a factor of 10.  We once again use a nanoflare level of heating.  As before, we find that the impulsive phase in each loop evolves similarly, while there is a more noticeable divergence in the draining during the cooling phase.  Despite the slight variation in area expansion, the cooling times of all five loops are nearly identical in this case.

\section{Conclusions}
\label{sec:conclusions}

In this work, we have briefly examined two geometrical assumptions that are commonly used in hydrodynamic modeling of coronal loops.  It is often assumed in field-aligned simulations that coronal loops are semi-circular with constant cross-sectional area.  Observations of loop widths suggest that coronal loops only show minimal expansion, but the decrease of the magnetic field with height in the corona implies that there ought to be an expansion.  It is not clear how to resolve this discrepancy, but the comparison of simulations and observations may offer insight to better understand this problem.  On the other hand, many loops are very clearly not semi-circular, and are often very eccentric, but the impact of this on loop dynamics has been not been critically examined.  

We first examined the assumption of a non-uniform cross-sectional area in Section \ref{sec:area}.  This has been touched upon by a few authors \citep{emslie1992,mikic2013,froment2018,winebarger2018,reep2020, cargill2022} and found to be important for numerous reasons, such as impacting observed Doppler shifts or modifying the thermal conduction profiles.  However, this has not been universally adopted in loop simulations; uniform cross-sectional areas are commonly assumed.  The primary reason for this is that it is not at all apparent how the expansion profile varies with height, what its magnitude is, or what the rate of expansion with height is.  We have not attempted to reproduce any observations in this paper, but instead focus on simply understanding how an expansion would impact dynamics.  We find that an area expansion significantly increases the time for a loop to cool and drain, increases upflow durations and localizes upflows near the footpoints, and suppresses sound waves.  Importantly, the standard $T \sim n^{2}$ relation between cooling and draining does not generally hold with large area expansions, instead the cooling tends towards a static radiative cooling regime as the expansion grows.  Despite the changes in the hydrodynamics, we still did not find any coronal condensation events in any of the simulations (see also \citealt{reep2020}).  Synthetic line intensities are impacted both directly from the change in volume with height and indirectly through the effects on the density and ionization profiles.  Additionally, the location and rate of the expansion is important.  When the expansion is localized near the transition region, the cooling time and draining time are both reduced compared to a more gradual expansion of similar magnitude.  It is fundamentally important therefore to use a cross-sectional area expansion in loop simulations, both for understanding the hydrodynamics and the radiative output.  Many previous results ought to be critically reexamined in the context of expanding cross-sections.  Additionally, observations are required to further constrain the magnitude and rate of expansion.  

The assumption of semi-circular loops, that is, an eccentricity of 0, is almost universally used in the literature.  We have examined this assumption in Section \ref{sec:elliptical} by varying the gravitational acceleration parallel to the loop in accordance with both tall loops (vertical semi-major axis) and wide loops (horizontal semi-major axis).  We have found that the draining time of the loops is reduced in tall loops compared to the semi-circular case since gravity is weakened with height, whereas the draining time is increased in wide loops for the opposite reason.  The cooling times, peak temperatures, and peak densities are mostly unaffected.  As a result, the spectral line intensities are mostly unaffected.  The difference is more pronounced in longer loops or with weaker heating, which reduces the gravitational scale height.  This assumption is relatively unimportant during the heating phase or with short, dense loops, but becomes more important with long or tenuous loops during the cooling phase.

Finally, in Section \ref{sec:ell_exp}, we have combined the two geometrical assumptions, and examined elliptical loops with an area expansion that scales with height above the solar surface, such that the magnetic field $B \propto \frac{1}{r^{2}}$, or $A \propto r^{2}$.  In this case, we do find that the dynamics are somewhat affected.  A circular loop has the longest cooling time, which, when compared with the results of Section \ref{sec:area}, indicates that the rate of expansion $\frac{dA}{ds}$ is also an important factor.

We emphasize that a non-uniform cross-sectional area strongly impacts all of the hydrodynamic quantities, and therefore the radiation as well.  We note that in none of the simulations here do we find coronal condensation events characteristic of coronal rain, reiterating the result of \citet{reep2020}.  While the geometry does impact the basic quantities, it still appears that there needs to be some secondary heating term to produce rain.  In general, simulations of coronal loops must include area expansion to accurately simulate the dynamics, particularly the cooling of loops.  Observations, additionally, are required to constrain the magnitude, rate, and location of the expansion to better inform the simulations.  

\leavevmode \newline

\acknowledgments  
The authors were supported by a NASA Heliophysics Supporting Research Grant, number NNH19ZDA001N, and by the Office of Naval Research 6.1 Support Program.  This research benefited from discussions held at a meeting at the International Space Science Institute, in Bern, Switzerland, led by Drs. Vanessa Polito and Graham Kerr.  The authors also thank the referee for helpful comments that have improved this paper.

\appendix
\section{Observationally-Inferred Area Expansion Factors}
\label{app:area}
\begin{figure*}
\centering
\includegraphics[width=0.49\textwidth]{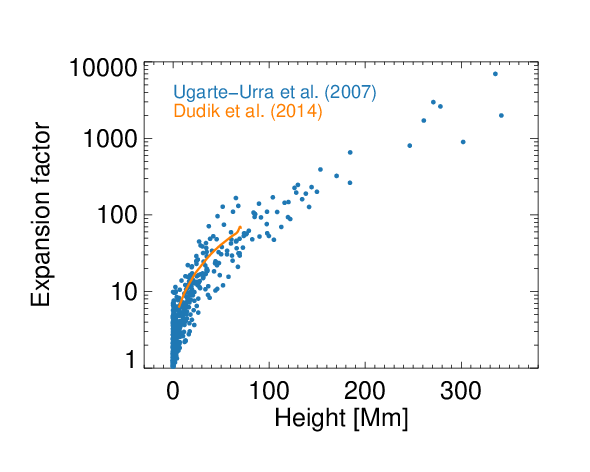}
\includegraphics[width=0.49\textwidth]{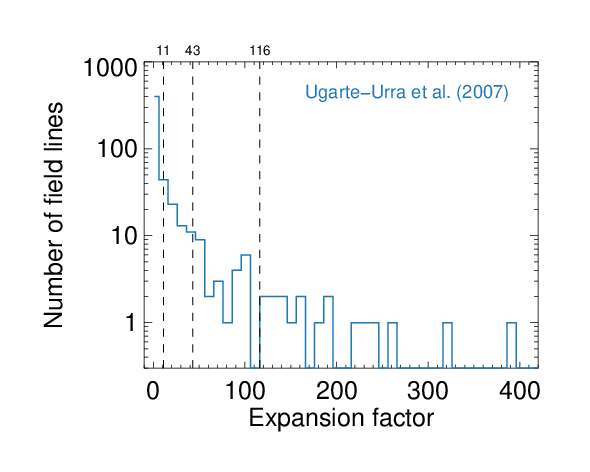}
\caption{Expansion factors of closed magnetic field lines in two potential
field extrapolations of a flare-active and a quiescent active region from
\citet{ugarte2007} and \citet{dudik2014}. The dashed lines show a reference
to expansion factors used in Section \ref{subsec:continuous}.
\label{fig:expansion_data}}
\end{figure*}

One consequence of the conservation of magnetic flux is that the cross-sectional
area of loops in the corona must expand as the magnetic field decreases in the
solar atmosphere. Figure~\ref{fig:expansion_data} shows the range of expansion
factors derived from magnetic field extrapolations of two active regions.  

Blue circles correspond to a current-free (potential) extrapolation of a MDI
\citep{scherrer1995} photospheric magnetogram of the flare-active active region
NOAA 9077 (2000 July 14, 09:36 UT). This extrapolation was part of a study of the
magnetic topology of 26 CME events \citep{ugarte2007}. The expansion factor was
calculated from the ratio between the maximum and minimum magnetic field strength
for 541 closed field lines within the Cartesian domain.

The orange curve corresponds to the potential magnetic field extrapolation of a
high-resolution {\it Hinode}/SOT magnetogram \citep{tsuneta2008} of the quiescent
active region NOAA 11482 (2012 May 18, 21:30 UT) presented in the investigation
of area expansion factors by \citet{dudik2014}. The expansion factor values are those described as ``height-averaged values of $\Gamma(Z)_{X,Y}$ Green's function extrapolation'' for closed coronal loops, in Figure 3 of that paper.

The right panel of Figure~\ref{fig:expansion_data} shows a histogram of the expansion factors in the NOAA 9077 dataset with a reference to the expansion factor values used in Section \ref{subsec:continuous} of the present paper: 1, 11, 43, 116. These numbers cover 96\% of the loops in this extrapolation.  See also the results of \citet{mok2005,mok2008} and \citet{asgari2013}, who found a similar range of expansion factors from magnetic field extrapolations.

\section{Gravitational Acceleration in Elliptical Loops}
\label{app:gravity}

\begin{figure}
\centering
\includegraphics[width=0.98\textwidth]{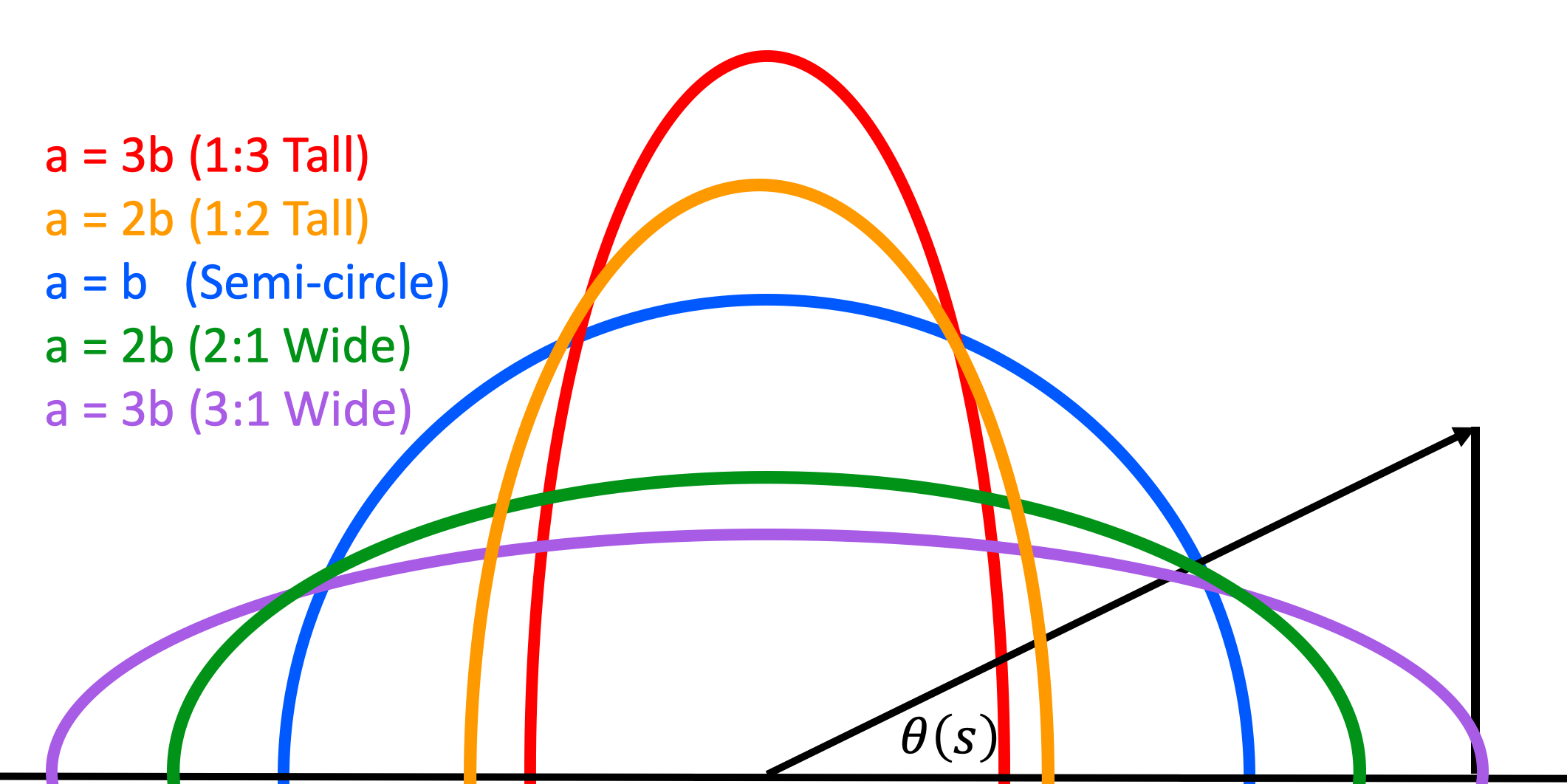}
\caption{A cartoon of the ellipse geometries assumed in this work.  We use two tall loops, with semi-major axis $a$ oriented radially outwards from the solar surface (red and orange), a semi-circular loop (blue), and two wide loops, with semi-major axis oriented parallel to the solar surface (green and purple).  (Not perfectly to scale.)
\label{fig:cartoon}}
\end{figure}

In order to examine how ellipticity affects loop hydrodynamics, we must modify the gravitational acceleration $g_{\parallel}(s)$ parallel to the loop coordinate $s$.  We consider five cases, shown in the cartoon in Figure \ref{fig:cartoon}.  In addition to the standard semi-circular case (blue), we use cases with semi-major axis $a$ oriented either radially outwards from the solar surface (``tall loops'') or parallel to it (``wide loops'').  

As a function of the angle $\theta$ relative to the ellipse center, the acceleration parallel to $s$ at a given height along a loop is
\begin{equation}
    g_{\parallel} = g_{\sun}  \Big(\frac{R_{\sun}}{R_{\sun} + h \sin \theta} \Big)^{2} \cos \theta
\end{equation}
\noindent where $g_{\sun}$ is the solar surface gravity, $R_{\sun}$ the solar radius, and $h$ the maximum height above the solar surface.  To determine the maximum height $h$ for a given loop length, we use the approximation for an ellipse's circumference given by \citet{ramanujan1914}\footnote{http://ramanujan.sirinudi.org/Volumes/published/ram06.pdf}:
\begin{equation}
    C = \pi \Big((a+b) + \frac{(a - b)^{2}}{10 (a + b) + \sqrt{a^{2} + 14ab + b^{2}})} +\epsilon \Big)
\end{equation}
\noindent where $a$ and $b$ are the semi-major and semi-minor axes, and $\epsilon$ is an error term of order $a k^{20}$ for $k$ the eccentricity.   If the semi-major axis $a$ is oriented vertically, then we can write the height $y = a \sin{\theta}$ and $x = b \cos{\theta}$ (if oriented horizontally, we would swap $a$ and $b$).  In order to then calculate the field-aligned gravitational acceleration, we must convert between $\theta$ and the loop coordinate $s$ used by \texttt{HYDRAD}.  From the Pythagorean theorem, we have 
\begin{align}
    ds &= \Bigg[\Big(\frac{dx}{d\theta}\Big)^{2} + \Big(\frac{dy}{d\theta}\Big)^{2} \Bigg]^{\frac{1}{2}}\ d\theta \nonumber \\
       &= \Big[b^{2} \sin^{2}\theta + a^{2} \cos^{2}\theta \Big]^{\frac{1}{2}}\ d\theta \nonumber \\
       &= a \Big( (1 - k^{2}) \sin^{2}\theta + \cos^{2}\theta\Big)^{\frac{1}{2}}\ d\theta
\end{align}
\noindent So we can integrate numerically
\begin{equation}
    s(\theta) = a \int_{\theta_{0}}^{\theta_{1}} \Big( (1 - k^{2}) \sin^{2}\theta + \cos^{2}\theta\Big)^{\frac{1}{2}}\ d\theta
\end{equation}
\noindent We can then use this to convert between $s$ and $\theta$, and since we know $g_{\parallel}(\theta)$, we have the gravitational acceleration as a function of loop coordinate $s$ for any specified eccentricity and loop length.  

\begin{figure}
    \centering
    \includegraphics[width=0.48\textwidth]{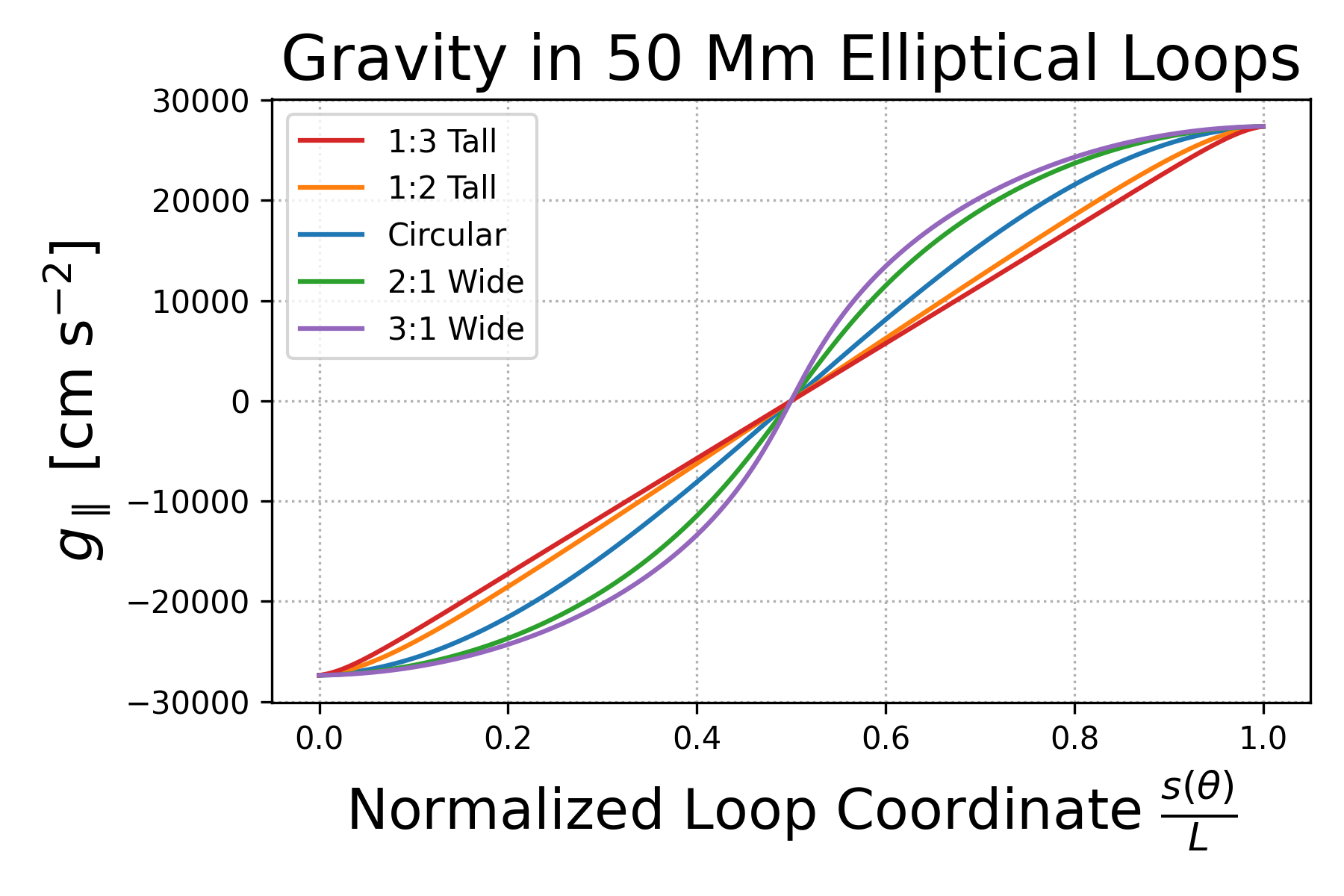}
    \includegraphics[width=0.48\textwidth]{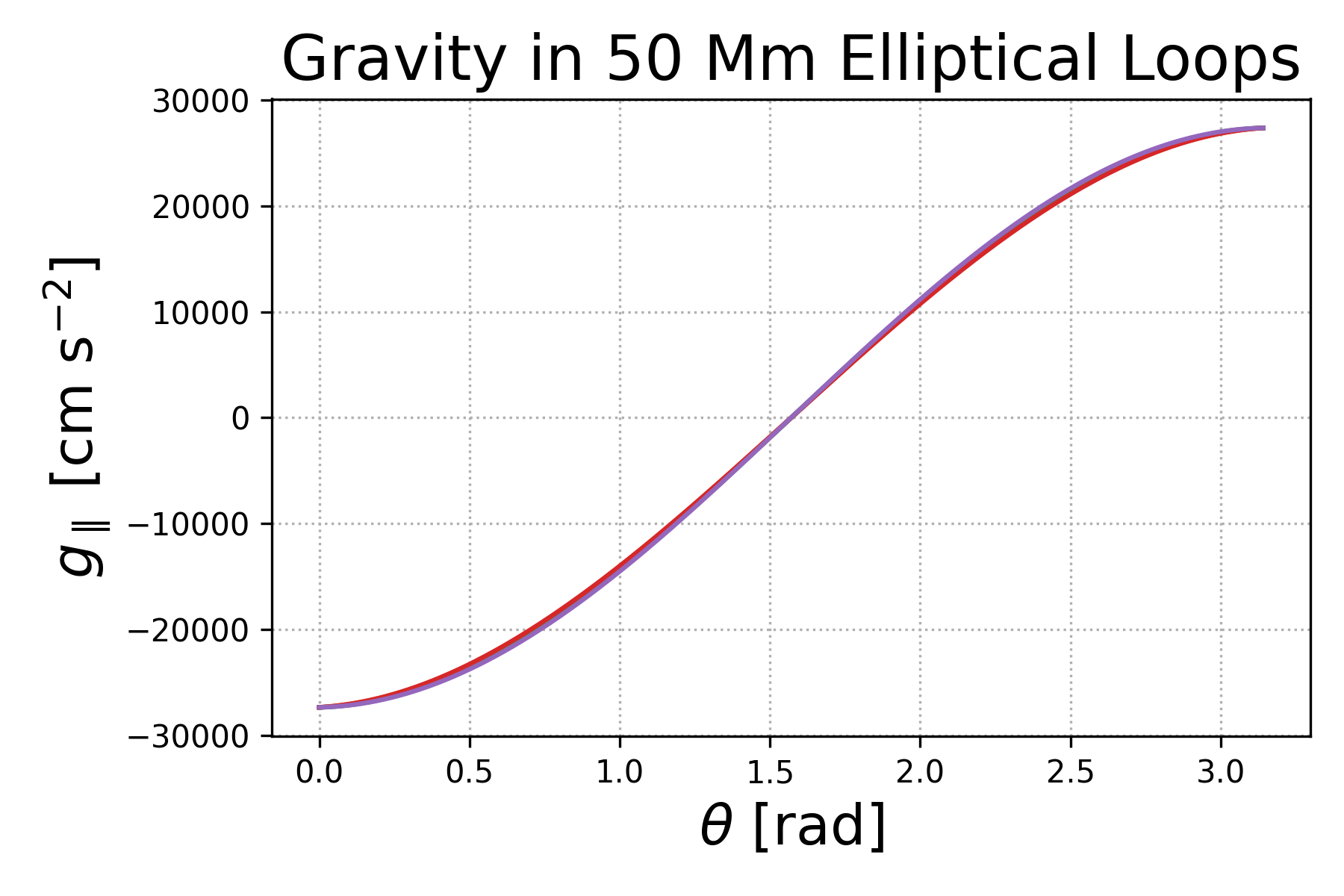}
    \includegraphics[width=0.48\textwidth]{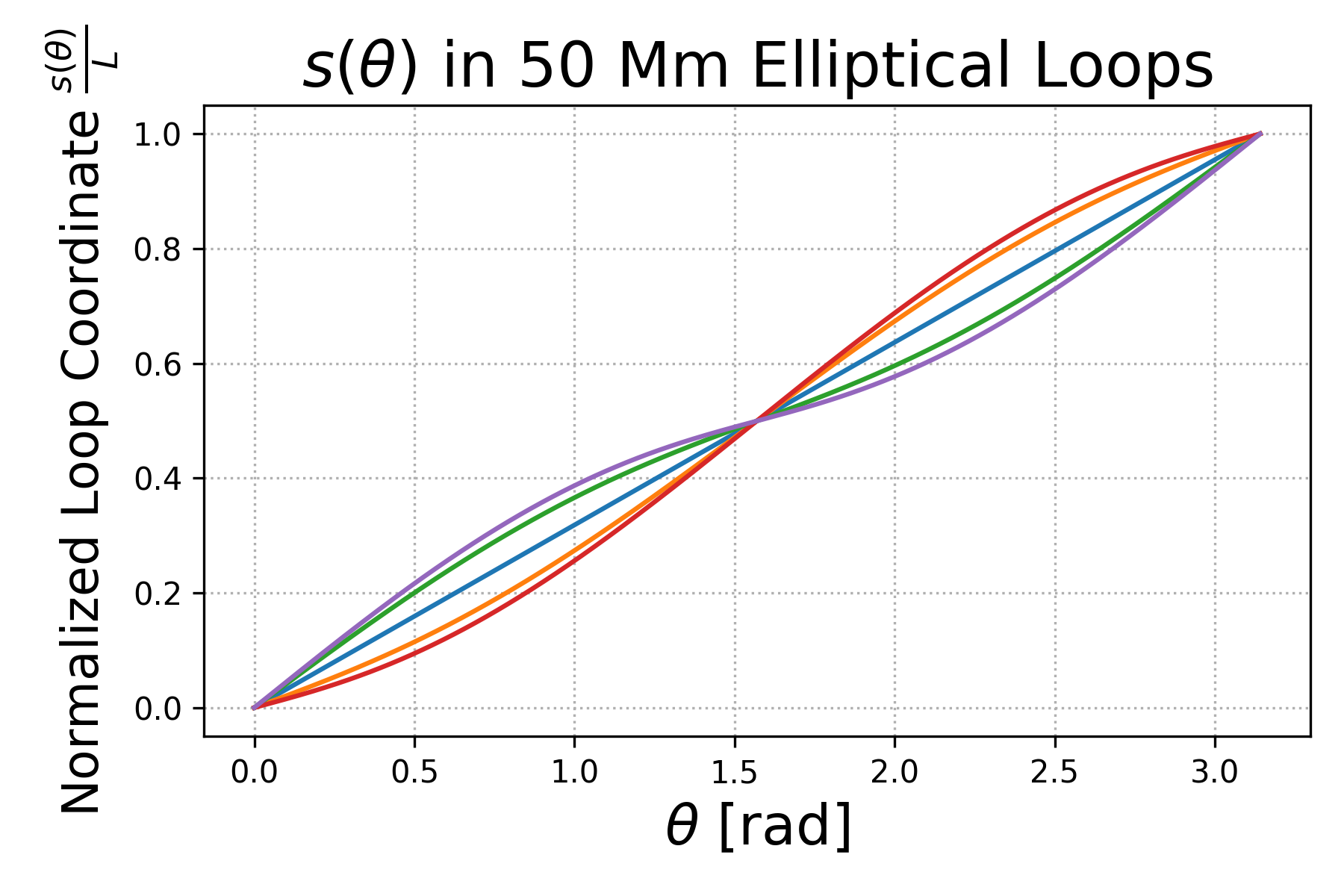}
    \caption{The parallel gravitational acceleration in 50 Mm semi-elliptical loops, as a function of position along the loop $s(\theta)$ (top left) and a function of $\theta$ (top right).  At bottom, the relation between $s$ and $\theta$ is shown explicitly for reference.  The blue line shows the semi-circular case.  The orange and red cases show tall loops, semi-major axis oriented radially outwards from the solar surface, $a = 2b$ and $a = 3b$ respectively.  The green and purple cases similarly show wide loops, semi-major axis oriented parallel to the solar surface, $a = 2b$ and $a = 3b$.  \label{fig:gravity}}
\end{figure}

Figure \ref{fig:gravity} shows an example calculation for 50 Mm semi-elliptical loops.  At top left, we show the acceleration along the loop coordinate $s(\theta)$, while the top right plot shows the acceleration as a function of $\theta$, and the bottom plot shows the conversion between $s$ and $\theta$ for clarity.  The blue line shows the acceleration for the semi-circular case, the most common assumption in loop models, which is a perfect sine wave.  The orange and red cases show tall loops, where the semi-major axis is oriented vertically, with $a = 2b$ and $a = 3b$ respectively.  The green and purple cases likewise show wide loops, semi-major axis oriented horizontally, $a = 2b$ and $a = 3b$ respectively.  It is clear that in tall (wide) loops the gravitational acceleration parallel to the field is reduced (increased) relative to the semi-circular case.

\bibliography{apj}
\bibliographystyle{aasjournal}

\end{document}